\documentclass[12pt]{article}

\usepackage{graphicx}

\def\rf#1{(\ref{eq:#1})}
\def\lab#1{\label{eq:#1}}

\def\br{\begin{eqnarray}}
\def\er{\end{eqnarray}}
\def\be{\begin{equation}}
\def\ee{\end{equation}}
\def\({\left(}
\def\){\right)}

\def\rlx{\relax\leavevmode}
\def\IR{\rlx\hbox{\rm I\kern-.18em R}}
\def\vp{\varphi}
\def\ve{\varepsilon}

\newcommand{\sbr}[2]{\left\lbrack\,{#1}\, ,\,{#2}\,\right\rbrack}

\def\IZ{\rlx\hbox{\sf Z\kern-.4em Z}}
\def\IR{\rlx\hbox{\rm I\kern-.18em R}}
\def\IC{\rlx\hbox{\,$\inbar\kern-.3em{\rm C}$}}
\def\one{\hbox{{1}\kern-.25em\hbox{l}}}

\begin{document}

\begin{titlepage}
\vspace*{-1cm}

\vskip 3cm

\vspace{.2in}
\begin{center}
{\large\bf The concept of quasi-integrability: a concrete example}
\end{center}

\vspace{.5cm}

\begin{center}
L. A. Ferreira~$^{\star}$, and Wojtek J. Zakrzewski~$^{\dagger}$

\vspace{.5 in}
\small

\par \vskip .2in \noindent
$^{(\star)}$Instituto de F\'\i sica de S\~ao Carlos; IFSC/USP;\\
Universidade de S\~ao Paulo  \\ 
Caixa Postal 369, CEP 13560-970, S\~ao Carlos-SP, Brazil\\

\par \vskip .2in \noindent
$^{(\dagger)}$~Department of Mathematical Sciences,\\
 University of Durham, Durham DH1 3LE, U.K.

\normalsize
\end{center}

\vspace{.5in}

\begin{abstract}

We use the deformed sine-Gordon models recently presented by Bazeia et
al \cite{Bazeia} 
to discuss possible definitions of quasi-integrability.
We present one such definition and use it to calculate an infinite
number of quasi-conserved 
quantities through a modification of the usual techniques of integrable
field theories.  Performing an expansion around the sine-Gordon theory
we are able to evaluate the charges and the anomalies of their
conservation laws in a perturbative power series in a small parameter 
 which  
describes the ``closeness'' to the integrable sine-Gordon model. Our
results indicate that in the case of the two-soliton scattering the
charges are conserved asymptotically, {\it i.e.} their values are the same
in the distant past and future, when the solitons are well separated. 

We back up our results with numerical simulations which also
demonstrate the existence 
of long lived breather-like and wobble-like states in these models.

\end{abstract} 
\end{titlepage}

\section{Introduction}
\label{sec:intro}
\setcounter{equation}{0}

Solitons and integrable field theories  play a central role in the
study of many non-linear phenomena. Indeed, it is perhaps correct to
say that many non-perturbative and
exact methods known in field theories are in one way or the other
related to solitons. The reason for that is twofold. On one hand, the
appearance of solitons in a given theory is often related to a high degree
of symmetries and so to the existence of a large number of conservation
laws.  On the other hand, in a large class of theories the solitons
possess a striking property. They become weakly coupled when the
interaction among the fundamental particles of the theory is strong,
and vice-versa. Therefore, the solitons are the natural candidates to
describe the relevant normal modes in the strong coupling (non-perturbative)
regime of the theory. Such relation between the strong and weak
coupling regimes have been observed in some (1+1) dimensional field theories,
as, for example, in the equivalence of the sine-Gordon and Thirring models
\cite{coleman}, as well as 
in four dimensional supersymmetric gauge theories where monopoles
(solitons) and fundamentals gauge particles exchange roles in the so-called
duality transformations \cite{duality}.   

The exact methods to study solitons in $(1+1)$-dimensional field
theories involve many algebraic and geometrical concepts, but the most
important ingredient is the so-called zero curvature condition or the
Lax-Zakharov-Shabat equation \cite{lax}. All theories known to possess
exact soliton solutions admit a representation of their equations of
motion as a zero curvature condition for a connection living in an
infinite dimensional Lie (Kac-Moody) algebra
\cite{leshouches,babelonbook}. In fact, in $(1+1)$ dimensions such zero
curvature condition is a conservation law, and the conserved quantities
are given by the eigenvalues of the holonomy of the flat connection
calculated on a spatial (fixed time) curve. On the other hand, many
techniques, like the dressing transformation method, for the
construction of exact solutions are based on the zero curvature
condition. In dimensions higher than two the soliton theory is not so
well developed, even though many exact results are known for some
$2$-dimensional theories such as the $CP^N$ models \cite{wojtekbook},
as well as in 
four dimensional gauge theories where  instanton and self-dual
monopoles are  the best examples \cite{mantonbook}. Some approaches
have been proposed 
for the study of integrable field theories in higher dimensions based on
generalizations of the two dimensional methods like the tetrahedron 
equations \cite{tetrahedron} and of the concept of zero curvature
involving connections in loop spaces \cite{afs}. 

Another important aspect of integrable field theories is that they
serve as good appro\-xi\-ma\-tions to many physical phenomena. In fact,
there is a vast literature exploring many aspects and applications of
perturbations around integrable models. In this paper we want to put
forward a technique that, so far as we know, has not been explored yet 
and which suggests 
that some non-integrable theories often possess many important properties of
fully integrable ones. We put forward and develop the concept of
quasi-integrability 
for theories that do not admit a representation of their equations of
motion in terms of the Lax-Zakharov-Shabat equation, but which can,
nevertherless, be 
associated with an almost flat connection in an infinite dimensional Lie
algebra.  In other words, we have an anomalous zero curvature
condition that leads to an infinite number of quasi-conservation
(almost conservation) 
laws. 

Moreover, in practice, in physical situations, like the scattering of solitons,
these charges are effectively conserved. The striking property we have
discovered is that as the scattering process takes place the charges do
vary in time. However, after the solitons have separated from each other the
charges return to the values they had prior to the scattering. Effectively
what we have is the asymptotic conservation of an infinite number of
non-trivial charges. There are still several aspects of this observation that have to be better understood but
we believe that if our results are indeed  robust  then such asymptotic
charges could play a role in many important properties of the theory like the
factorization of the S-matrix.

\begin{figure}[tbp]
    \centering
    \includegraphics[angle=270,width=0.5\textwidth]{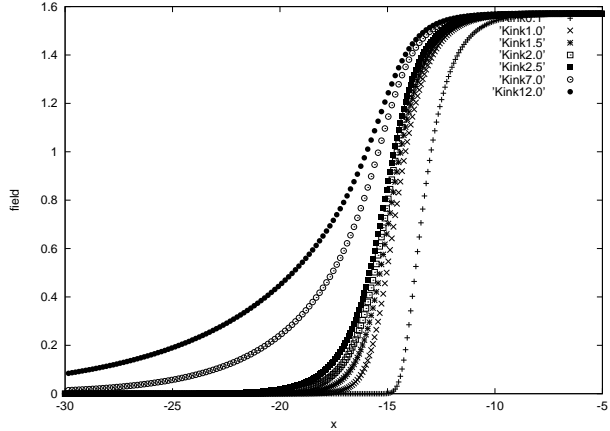}
    \begin{quote}
	\caption[AS]{\small Plots of solitons for various values on $n$ }
	\label{fig:onesoliton}
    \end{quote}
\end{figure}

\begin{figure}[tbp]
    \centering
    \includegraphics[width=0.5\textwidth]{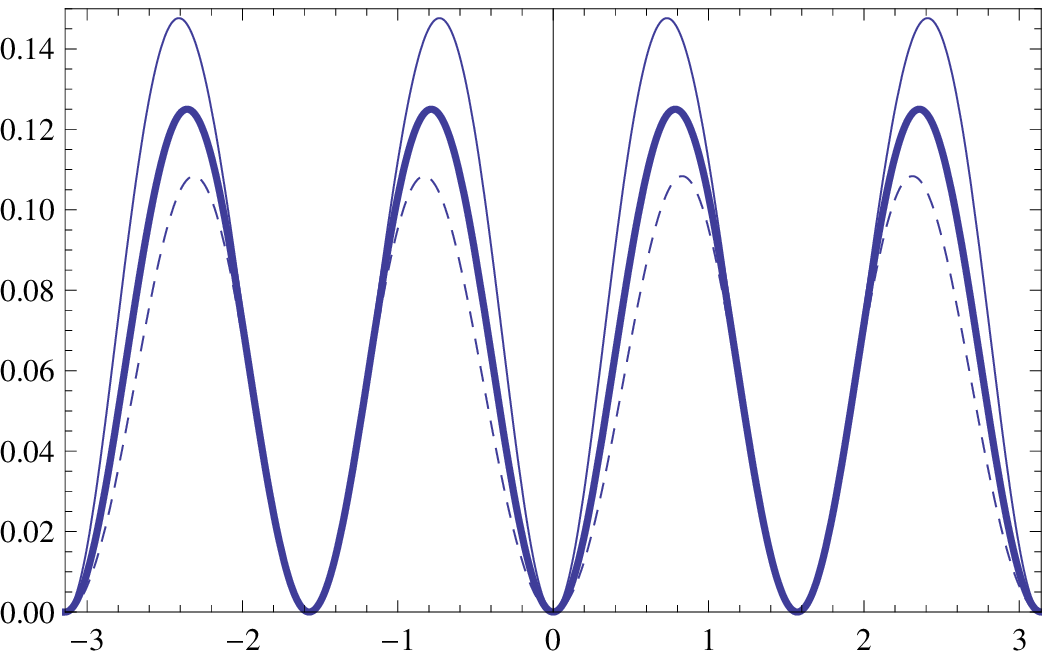}
    \begin{quote}
	\caption[AS]{\small Plots of the potential \rf{nicepotintroa}
          against $\vp$, for three values of the parameter $\ve$,
          where $n=2+\ve$, 
          namely $\ve 
          =-0.5$ (fine line), $\ve
          =0.0$ (thick line), and $\ve
          =0.5$ (dashed line). } 
	\label{fig:potentialintro}
    \end{quote}
\end{figure}

We introduce our concept of quasi-integrability through a concrete
example involving a real scalar field theory in $(1+1)$ dimensions
which is a special deformation of the sine-Gordon model. The scalar
field $\vp$ of our theory is subjected to the potential 
\be
V\(\vp,n\)=\frac{2}{n^2}\,\tan^2 \vp\left[1-\mid
  \sin\vp\mid^{n}\right]^2 
\lab{nicepotintroa}
\ee        
where $n$ is a real parameter which in the case $n=2$
reduces the potential to that of the sine-Gordon model,
{\it i.e.} $V\(\vp,2\)=\frac{1}{16}\left[1-\cos\(4\,\vp\)\right]$. 

This
potential \rf{nicepotintroa} is a slight modification of that
introduced by D. Bazeia et al, \cite{Bazeia}, in the sense that we
take the absolute value of $\sin\vp$ to allow $n$ to take real
 and not only integer values. 
 
 The potential \rf{nicepotintroa} has an infinite number of degenerate
 vacua that allow the existence of solutions with non-trivial topological
 charges. It is worth noticing that the positions of the vacua are
 independent of $n$ and so they are the same as in the
 sine-Gordon model, {\it i.e.} $\vp_{\rm vac.}= m\, \frac{\pi}{2}$, with $m$
 being any integer.

The model with the potential \rf{nicepotintroa}
 is fully topological ({\it i.e.} it satisfies its Bogomolnyi bound 
 for any $n$) and so its one soliton field configurations are known
 in an explicit form.
 They are given by:
 \be
\vp={\rm
  arcsin}\left[\frac{e^{2\,\Gamma}}{1+e^{2\,\Gamma}}\right]^{1/n},
\qquad\qquad \qquad \Gamma=\pm\,\frac{\(x-v\,t-x_0\)}{\sqrt{1-v^2}},
\lab{exactkinkintro}
\ee
where the velocity
$v$ is given in units of the speed of light, and the signs
correspond to the kink ($+$), and anti-kink ($-$), with topological
charges $+1$, and $-1$ respectively.
In Figure \ref{fig:onesoliton} we plot the fields of a one soliton
configuration for various values  
on $n$.  We see from this plot that the $n=2$ case does not appear to
be  very special; 
all soliton fields look very similar and the solitons for different
values of $n$ 
differ only in their slopes.

\begin{figure}[tbp]
    \centering
    \includegraphics[width=0.5\textwidth]{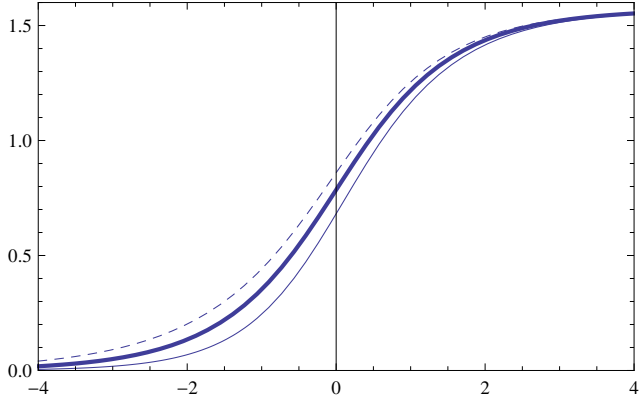}
    \begin{quote}
	\caption[AS]{\small Plots of the kink solutions \rf{exactkinkintro}
          against $x$, with $t=0$ and $x_0=0$, for three values of the
          parameter $\ve$, where $n=2+\ve$,  
          namely $\ve
          =-0.5$ (fine line), $\ve
          =0.0$ (thick line), and $\ve
          =0.5$ (dashed line). } 
	\label{fig:kinkintro}
    \end{quote}
\end{figure}

We are going to use this model to back up our discussion of
quasi-integrability and so next we look at 
$n=2+ \ve$, with $\ve$ small. 
 In Figure \ref{fig:potentialintro} we
 plot the potential \rf{nicepotintroa} for $\ve=-0.5; 0.0; 0.5$. 
 Of course, in this case 
 the kinks solutions are given by \rf{exactkinkintro} with $n$ replaced by
 $2+\ve$.
In Figure \ref{fig:kinkintro} we
plot the one kink solutions \rf{exactkinkintro} for the potentials
shown in Figure \ref{fig:potentialintro}, {\it i.e.} for $\ve=-0.5;
0.0; 0.5$. Note  that they connect the vacua $\vp_{\rm vac}=0$ to $\vp_{\rm
  vac.}=\frac{\pi}{2}$, as $x$ goes from $-\infty$ to $+\infty$, with the
slope of the kink increasing as the value of $\ve$ decreases.  

In this paper we study the concept of quasi-integrability in the context of the
theory \rf{nicepotintroa} from the analytical and numerical points
of view\footnote{Preliminary results of our 
approach have already been given in \cite{Torun}}. Our approach and the main results of this paper can be summarised as follows (more details are given in the following sections):

We first consider a real scalar field theory with a very general potential
  $V\(\vp\)$, and construct a connection $A_{\mu}$ based on the $sl(2)$
  loop algebra which, as a consequence of its equations of motion,
  satisfies an anomalous zero curvature condition. Using a
  modification of the methods employed in integrable field 
  theories we construct and infinite number of quasi-conserved charges
  for such a theory, {\it i.e.} 
\be
\frac{d\,Q^{(2n+1)}}{d\,t}=-\frac{1}{2}\,\alpha^{(2n+1)}\(t\)
\qquad\qquad\qquad n=0,\pm 1,\pm 2,\ldots,
\lab{quasichargesintro}
\ee
where the anomalies $\alpha^{(2n+1)}\(t\)$ are non-zero due to
the non-flatness of the connection $A_{\mu}$.  
The charges $Q^{(\pm 1)}$ are in fact conserved, {\it i.e.} $\alpha^{(\pm
  1)}=0$, and linear combinations 
of them correspond to the energy and momentum.  

We then restrict ourselves to the case of the potential
\rf{nicepotintroa} and set up a perturbative expansion around the
sine-Gordon theory. We expand all the quantities, equations of motion, field
$\vp$, charges and anomalies, in powers of the parameter $\ve$,
related to $n$
appearing in \rf{nicepotintroa} by $n=2+\ve$. For instance, we have 
\be
Q^{(2n+1)}=Q^{(2n+1)}_0+\ve\, Q^{(2n+1)}_1 + O\(\ve^2\),\qquad\qquad
\alpha^{(2n+1)}=\ve\, \alpha^{(2n+1)}_1+O\(\ve^2\).
\ee

The anomalies vanish in the lowest order (order zero) in $\ve$ because they
correspond to
the sine-Gordon theory which is integrable and so all the charges
$Q^{(2n+1)}_0$ are conserved during the dynamics of all field configurations.

 In this paper we concentrate our attention on the evaluation of
the first non-trivial charge and its anomaly, namely $Q^{(3)}$ and
$\alpha^{(3)}$, but our calculations can easily be extended to the
other charges. We considered the case of the scattering of two
kinks and also of a kink/anti-kink in the theory \rf{nicepotintroa},
where the solitons are far apart in the distant past and future, and
collide when $t\sim 0$. We found that the
first order anomaly $\alpha^{(3)}_1$, vanishes when integrated over
the whole time axis. Therefore, from  \rf{quasichargesintro} we see that
\be
Q^{(3)}_1\(t=+\infty\)=Q^{(3)}_1\(t=-\infty\).
\lab{conservcharge2solintro}
\ee
Consequently, the scattering of the solitons happen in a way
that, to first order in $\ve$ at least, the charge is asymptotically
conserved. That is a very important result, and if one can extend it
to higher orders and higher charges one would prove that effectively
the scattering of solitons in the theory \rf{nicepotintroa} takes place in the sdame way as
if the theory were   
a truly integrable theory. We have also analyzed the first order
charge $Q^{(3)}_1$ for the breather solution of the theory
\rf{nicepotintroa}, and found that even though the charge is not conserved, it
oscillates around a fixed value. In other words, the first order
anomaly vanishes when integrated over a period $\frac{\pi}{\nu}$, and so from
\rf{quasichargesintro} we find that 
\be
Q^{(3)}_1\(t\)=Q^{(3)}_1\(t+\frac{\pi}{\nu}\),
\lab{conservchargebreatherintro}
\ee
where $\nu$ is the angular frequency of the breather. That means that the
period of the charge is half of that of the breather. 

We have performed many numerical simulations of the full theory
\rf{nicepotintroa} using a fourth order Runge-Kuta method, and using
various lattice grids to make sure that the results are not contaminated by
numerical artifacts. We have found reliable results with lattice grids 
of at least 3001 points, where the kinks were of size $\sim 5$
points. Most of the simulations in this paper were performed with lattices
of 10001 points ({\it i.e.} well within this reliability). 
The main results we have found are the following:  We have found that if $\mid
\ve \mid$  does not get very close to unity the kinks and the kink/anti-kink
scatter without destroying themselves and preserve their original
shapes, given in \rf{exactkinkintro}.  For small values of $\mid
\ve \mid$ the anomaly $\alpha^{(3)}$ integrates to zero for large
values of the time interval, and so the charge $Q^{(3)}$ is
asymptotically conserved within our numerical errors. This is an
important confirmation of our analytical result described above, and
is valid for the full charge and not only for its first order
approximation as in \rf{conservcharge2solintro}.

One of the important discoveries of our numerical simulations is that
the theory \rf{nicepotintroa} also possesses very long lived breather
solutions for $\ve \neq 0$, which correspond to non-integrable
models. These long-lived breathers were obtained by starting 
the simulations with a field configuration corresponding to  a kink
and an anti-kink. 
As they get close to each other they interact and readjust their profiles and 
some radiation is emitted in this process. We absorbed this
radiation at the boundaries of the grid and the system stabilized to a
breather-like configuration. For $n=2$ the resultant field configuration
was the exact (and analytically known) breather while for
$\ve$ small those breather-like fields lived for
millions of units of time. As one changed  $\ve$ and made it come close to unity
the quasi-breathers radiated more and for even larger values they
eventually died. 
 We also looked at the anomalies for
such breather-like configurations and have found a good agreement with our
analytical results described above. The  anomaly, integrated in time, 
does oscillate and for small values of $\ve$ the charge is periodic
in time. That is, again, in agreement with the analytical result
\rf{conservchargebreatherintro}. Notice however, that the numerical
result is stronger in the sense that it corresponds to the full charge and
not only to its first order approximation as in
\rf{conservchargebreatherintro}. 

We have also performed similar numerical simulations of wobbles
\cite{wojtekwobble} which correspond to configurations of a breather
and a kink. Again, such configurations were obtained by starting the
simulation with two kinks and an anti-kink. As the three solitons
interact and adjust their profiles they radiate energy and this
radiation had been 
absorbed at the boundaries of the grid. Eventually the system 
has evolved to a breather and a kink and for small values of $\ve$ 
the resultant configuration was quite stable, 
 and for $n=2$ it agreed with the analytically known configuration of a wobble.
 Again, we believe that this is a very interesting result which shows that
non-integrable theories can  support such kinds of solutions. 

Our results open up the way to investigate large classes of models
which are not really exactly integrable  but which possess
properties which are very similar to those of integrable field theories. 
We believe that they will have applications in many non-linear
phenomena of physical interest. 

The paper is organized as follow: in section \ref{sec:quasizc} we
introduce the quasi-zero curvature condition, based on the $sl(2)$
loop algebra, for a real scalar field
theory subject to a generic potential, and construct an infinite
number of quasi-conserved quantities. In section \ref{sec:expansion}
we perform the expansion of the theory \rf{nicepotintroa} around the
sine-Gordon model, and evaluate the first non-trivial charge and its
corresponding anomaly. The numerical simulations, involving the two
solitons scattering, breathers and a wobble, are presented in
section \ref{sec:numerical}. In section \ref{sec:summary} we present our
conclusions; the details of  the $sl(2)$
loop algebra, charge calculations and $\ve$-expansion are presented in
the appendices.

\section{The quasi zero curvature condition}
\label{sec:quasizc}
\setcounter{equation}{0}

We shall consider Lorentz invariant field theories in
$(1+1)$-dimensions with a real scalar field $\vp$ and equation of
motion given by 
\be
\partial^2 \varphi + \frac{\partial\, V\(\varphi\)}{\partial\, \varphi}=0,
\lab{eqmot}
\ee
where $V\(\varphi\)$ is the scalar potential. Thus we want to study the
integrability properties of such theory using the techniques of
integrable field theories \cite{lax,leshouches,babelonbook}. We then
start by trying to set up a zero 
curvature representation of the equations of motion \rf{eqmot}, and so
we introduce the Lax potentials  as 
\br
A_{+}&=& \frac{1}{2}\left[ \(\omega^2 \, V -m\)\, b_{1}
  -i\,\omega\, \frac{d\,V}{d\,\varphi}\,F_1\right], 
\nonumber\\
A_{-}&=& \frac{1}{2}\, b_{-1} - \frac{i}{2}\,
\omega\, \partial_{-}\varphi\, F_0. 
\lab{potentials}
\er
Our Lax potentials live on the so-called $sl(2)$ loop algebra with
generators $b_{2n+1}$ and $F_n$, with $n$ integer; their commutation
relations are given in Appendix \ref{sec:appendix-algebra}.  The
parameters $\omega$ and $m$ are constants, and they play a special
role in our analysis. Note, that the dynamics governed by \rf{eqmot}
does not depend upon them, but since they appear in \rf{potentials}
they will play a role in the quasi-conserved quantities that we will construct
through the Lax equations. In the expression above we have used light cone
coordinates 
$x_{\pm}=\frac{1}{2}\(t\pm
x\)$, where $\partial_{\pm}= \partial_t\pm \partial_x$,  and   
$\partial_{+}\partial_{-}=\partial^2_t-\partial^2_x \equiv \partial^2$.

The curvature of the connection \rf{potentials} is given by 
\be
F_{+-}\equiv \partial_{+}A_{-}-\partial_{-}A_{+}+\sbr{A_{+}}{A_{-}}= X
\, F_1 -\frac{i\,\omega}{2}\left[\partial^2 \varphi + \frac{\partial\,
    V}{\partial\, \varphi} \right]\,F_0
\lab{zc}
\ee
with
\be
X = \frac{i\,\omega}{2}\,  \partial_{-}\varphi\,
\left[\frac{d^2\,V}{d\,\varphi^2}+\omega^2\, V-m\right].
\lab{xdef}
\ee

As in the case of the sine-Gordon model where the potential is given by
\be
V_{\rm SG}=\frac{1}{16}\left[1-\cos\(4\,\vp\)\right]
\ee
we find that $X$, given by \rf{xdef}, vanishes when we take
$\omega=4$ and $m=1$. Then the  curvature \rf{zc}
vanishes when the equations of motion \rf{eqmot} hold. The vanishing
of the curvature allows us to use several powerful techniques to construct
conserved charges and exact solutions. We want to analyze what can be
said about the conservation laws for potentials when $X$ does not
vanish but can be considered small. 

In general, the conserved charges can be constructed using the fact that
the path ordered integral of the connection along a curve $\Gamma$,
namely
$P\,\exp\left[\int_{\Gamma}d\sigma\,A_{\mu}\,\frac{d\,x^{\mu}}{d\,\sigma}\right]$,
  is path independent when the connection is flat
  \cite{leshouches,babelonbook,ours}.  Here, we will use a
  more refined version of this technique and try to gauge transform
  the connection into the abelian subalgebra generated by the
  $b_{2n+1}$. We follow the usual procedures of integrable field
  theories discussed for instance in \cite{olive1,olive2,aratyn}. An
  important ingredient of the method is that our $sl(2)$ loop algebra
  ${\cal G}$ is graded,  with $n$
  being the grades determined by the grading operator $d= T_3+
  2\,\lambda \frac{d\;}{d\lambda}$ (see appendix
  \ref{sec:appendix-algebra} for details)  
\be
{\cal G}=\sum_{n}{\cal G}_n\; ; \qquad\qquad \sbr{{\cal G}_m}{{\cal
    G}_n}\subset {\cal G}_{m+n}\; ; \qquad\qquad \sbr{d}{{\cal
    G}_n}= n \, {\cal G}_n.
\ee
We perform a gauge transformation 
\be
A_{\mu}\rightarrow a_{\mu}=g\, A_{\mu}\,g^{-1}-\partial_{\mu}g\,
g^{-1}
\lab{gaugeminus}
\ee
with the group element $g$ being an exponentiation of generators lying in
the positive grade subspace generated by the $F_n$'s, {\it i.e.}, 
\be
g={\rm exp}\left[\sum_{n=1}^{\infty} \zeta_n\, F_n\right]
\lab{gplusdef}
\ee
with $\zeta_n$ being parameters to be determined as we will
explain below. Under \rf{gaugeminus} the curvature \rf{zc} is transformed as
\be
F_{+-}\rightarrow
g\,F_{+-}\,g^{-1}=\partial_{+}a_{-}-\partial_{-}a_{+}+\sbr{a_{+}}{a_{-}}=
X \, g\, F_1\,g^{-1},
\lab{newcurvature}
\ee
where we have used the equations of motion \rf{eqmot} to drop the term
proportional to $F_0$ in \rf{zc}. The component $A_{-}$ of the
connection \rf{potentials} has terms with grade $0$ and
$-1$. Therefore, under \rf{gaugeminus} it is transformed into
$a_{-}$ which has terms with grades ranging from $-1$ to
$+\infty$. Decomposing $a_{-}$ into grades we get from \rf{gaugeminus}
and \rf{gplusdef} that
\br
a_{-}&=&\frac{1}{2}\,b_{-1}
\lab{determinezeta}\\
&-&\frac{1}{2}\,\zeta_1\,\sbr{b_{-1}}{F_1}- \frac{i}{2}\,
\omega\, \partial_{-}\varphi\, F_0\nonumber\\
&-&\frac{1}{2}\,\zeta_2\,\sbr{b_{-1}}{F_2}
+\frac{1}{4}\,\zeta_1^2\,\sbr{\sbr{b_{-1}}{F_1}}{F_1}
- \frac{i}{2}\,
\omega\, \partial_{-}\varphi\, \zeta_1\,\sbr{F_1}{F_0}
-\partial_{-}\zeta_1\,F_1
\nonumber\\
&\vdots&\nonumber\\
&-&\frac{1}{2}\,\zeta_n\,\sbr{b_{-1}}{F_n}+ \ldots. \nonumber
\er

Next we note that one can choose the parameters $\zeta_n$ recursively
by requiring that the component in the direction on $F_{n-1}$ 
cancels out in $a_{-}$. Thus we can put that $\zeta_1= \frac{i}{2}\,
\omega\, \partial_{-}\varphi$, and so on. In the appendix
\ref{sec:appendix-gauge} we give the first few $\zeta_n$'s obtained
that way. In consequence, the component $a_{-}$ is rotated into the abelian
subalgebra generated by the $b_{2n+1}$. Note that this procedure
has not used the equations of motion \rf{eqmot}. We then have
\be
a_{-}=\frac{1}{2}\,b_{-1}+\sum_{n=0}^{\infty}a_{-}^{(2n+1)}\, b_{2n+1}
\ee
and the first three components are
\br
a_{-}^{(1)}&=&-\frac{1}{4} \omega^2  (\partial_{-}\varphi)^2
\lab{3aminus},\\
a_{-}^{(3)}&=&-\frac{1}{16} \omega^4
(\partial_{-}\varphi)^4-\frac{1}{4} \omega^2 
\partial_{-}^3\varphi \partial_{-}\varphi,\nonumber\\
a_{-}^{(5)}&=&-\frac{1}{32} \omega^6
  (\partial_{-}\varphi)^6-\frac{7}{16} \omega^4  
\partial_{-}^3\varphi 
(\partial_{-}\varphi)^3-\frac{11}{16} \omega^4 
(\partial_{-}^2\varphi)^2 
(\partial_{-}\varphi)^2-\frac{1}{4} \omega^2 
\partial_{-}^5\varphi 
\partial_{-}\varphi.\nonumber
\er
With the $\zeta_n$'s determined this way we perform the transformation of the
$A_{+}$ component of the connection \rf{potentials}. Since the
$\zeta_n$'s are polynomials of $x_{-}$-derivatives of $\vp$ (see
appendix \ref{sec:appendix-gauge}) and since there will be terms involving
$x_{+}$-derivatives of $\zeta_n$'s, we use the equations of motion to
eliminate terms involving $\partial_{+}\partial_{-}\vp$. Due to the nonvanishing of the 
anomaly term $X$ in \rf{zc} we are not able to transform $a_{+}$ into
the abelian subalgebra generated by the $b_{2n+1}$. We find that
$a_{+}$ is of the form
\be
a_{+}=\sum_{n=0}^{\infty}a_{+}^{(2n+1)}\, b_{2n+1}
+\sum_{n=2}^{\infty} c_{+}^{(n)}\,F_n
\ee
where
\br
a_{+}^{(1)}&=&\frac{1}{2} \, \left[\omega^2\, V-m\right]
\lab{3aplus},\\
a_{+}^{(3)}&=&\frac{1}{4}\omega^2\, 
\partial_{-}^2\varphi \, \frac{d\, V}{d\, \varphi}\, 
-\frac{1}{2} i \omega \partial_{-}\varphi X,
\nonumber\\
a_{+}^{(5)}&=&
-\frac{3}{8} i \omega^3 (\partial_{-}\varphi)^3 
X+\frac{5}{16} \omega^4 \partial_{-}^2\varphi 
(\partial_{-}\varphi)^2 \, \frac{d\, V}{d\, \varphi}\,
-\frac{1}{2} i \omega \partial_{-}\varphi 
\partial_{-}^2 X+\frac{1}{2} i \omega 
\partial_{-}^2\varphi \partial_{-}X
\nonumber\\
&-& \frac{1}{2} i \omega 
\partial_{-}^3\varphi X+\frac{1}{4} \omega^2\,
\partial_{-}^4\varphi \frac{d\,V}{d\,\varphi}
\nonumber
\er
with $X$ given in \rf{xdef}, and $V$ being the potential (see
\rf{eqmot}). 
See appendix \ref{sec:appendix-gauge} for more details, including the
terms involving $c_{+}^{(n)}$. 

The next step is to decompose the curvature \rf{newcurvature} into the
component lying in the abelian subalgebra generated by $b_{2n+1}$ and one
lying in the subspace generated by $F_n$. Since the equation of motion
\rf{eqmot} has been imposed, it turns out that the terms 
proportional to the $F_n$'s in the combination
$-\partial_{-}a_{+}+\sbr{a_{+}}{a_{-}}-X\,g\,F_1\,g^{-1}$,  exactly
cancel out. We are then left with terms in the direction of the $b_{2n+1}$
only. Therefore, the transformed curvature \rf{newcurvature} leads to
 equations of the form
\be
\partial_{+}a_{-}^{(2n+1)}-\partial_{-}a_{+}^{(2n+1)}=
\beta^{(2n+1)}\qquad\qquad n=0,1,2,\ldots
\lab{quasiconserv}
\ee
with $\beta^{(2n+1)}$ being linear in the anomaly $X$ given in
\rf{xdef}, and the first three of them being given by
\br
\beta^{(1)}&=&0,
\lab{anomalydensity}\\
\beta^{(3)}&=&i \omega \;
\partial_{-}^2\varphi\; X,\nonumber\\
\beta^{(5)}&=&i \omega \;\left[\frac{3}{2}  \omega^2 (\partial_{-}\varphi)^2 
\partial_{-}^2\varphi + 
\partial_{-}^4\varphi\right]\; X.
\nonumber
\er 
Working with the $x$ and $t$ variables we have that \rf{quasiconserv}
takes the form $\partial_{t}a_{x}^{(2n+1)}-\partial_{x}a_{t}^{(2n+1)}=-\frac{1}{2}\,
\beta^{(2n+1)}$ and so we find that
\be
\frac{d\,Q^{(2n+1)}}{d\,t}=-\frac{1}{2}\,\alpha^{(2n+1)}+
a_{t}^{(2n+1)}\mid_{x=-\infty}^{x=\infty} 
\ee
with
\be
Q^{(2n+1)}\equiv \int_{-\infty}^{\infty}dx\,a_{x}^{(2n+1)},\qquad\qquad\qquad
\alpha^{(2n+1)}\equiv \int_{-\infty}^{\infty}dx\,\beta^{(2n+1)}.
\lab{chargeanomalydef}
\ee

As we are interested in finite energy solutions of the theory
\rf{eqmot} we are concerned with field
configurations satisfying the boundary conditions
\be
\partial_{\mu}\vp \rightarrow 0\; ; \qquad\qquad V\(\vp\)\rightarrow \mbox{\rm
  global minimum}\qquad\qquad {\rm as} \qquad  x\rightarrow \pm \infty.
\lab{bc}
\ee
Therefore from \rf{potentials} we see that
\be
A_{+}\rightarrow \frac{1}{2}\,\(\omega^2\,V_{\rm vac.}-m\)\,b_1,
\qquad\qquad \qquad A_{-}\rightarrow \frac{1}{2}\,b_{-1}\qquad\qquad
{\rm as} \qquad  x\rightarrow \pm \infty, 
\ee
where $V_{\rm vac.}$ is the value of the potential at the global
minimum which, in general, is taken to be zero. As we have seen the
parameters $\zeta_n$ of the gauge \rf{gaugeminus} and \rf{gplusdef}
are polynomials in $x_{-}$-derivatives of the field $\vp$ (see
appendix \ref{sec:appendix-gauge}). Therefore,
for finite energy solutions we see that $g \rightarrow 1$ as
$x\rightarrow \pm \infty$, and so
\br
a_{t}^{(-1)} &\rightarrow& \frac{1}{4},
\nonumber\\
a_{t}^{(1)} &\rightarrow & \frac{1}{4}\, \(\omega^2\,V_{\rm
  vac.}-m\)\qquad\qquad\qquad {\rm as} \qquad
x\rightarrow \pm \infty,
\lab{bcat}\\
a_{t}^{(2n+1)} &\rightarrow & 0 \qquad\qquad n=1,2,\ldots 
\nonumber
\er
We can also investigate this behaviour more explicitly by analyzing  
\rf{3aminus}, \rf{3aplus}, \rf{xdef} and \rf{bc}.
Consequently, for finite energy solutions satisfying \rf{bc}, we  have that
\be
\frac{d\,Q^{(1)}}{d\,t}= 0\; , \qquad\qquad\qquad
\frac{d\,Q^{(2n+1)}}{d\,t}=-\frac{1}{2}\,\alpha^{(2n+1)}\qquad
n=1,2,\ldots
\lab{quasiconserv2}
\ee
Of course, the theory \rf{eqmot} is invariant under  space-time
translations and so its energy momentum tensor is conserved. The
conserved charge $Q^{(1)}$ is in fact a combination of the energy and
momentum of the field configuration. In section \ref{sec:expansion} we
will analyze the anomalies $\alpha^{(2n+1)}$ for a concrete
perturbation of the sine-Gordon model, and we will show that even
though the charges are not exactly conserved they lead to very
important consequences for the dynamics of the soliton solutions. 

A result that we can draw for general potentials, thus, is the
following. For static finite energy solutions the
charges $Q^{(2n+1)}$ are obviously time independent, and as a
consequence of \rf{quasiconserv2}  one sees
that the anomalies vanish, {\it i.e.} $\alpha^{(2n+1)}=0$.  Under a
$(1+1)$-dimensional Lorentz transformation where $x_{\pm }\rightarrow
\gamma^{\pm 1}\, x_{\pm}$ one finds that the connection \rf{potentials}
does not really transform as a vector. However, consider the internal
transformation 
\be
A_{\mu}\rightarrow \gamma^d\, A_{\mu}\,\gamma^{-d}
\lab{nicelorentz}
\ee
where $d$ is the grading operator introduced in
\rf{gradingop}. Then, one notices that $A_{\mu}$, given in
\rf{potentials},  transforms as a
vector under the combination of the external Lorentz transformation
and the internal transformation \rf{nicelorentz}. For the same reasons
the transformed connection $a_{\mu}$, defined in \rf{gaugeminus}, is
also a vector under the combined transformations. Consequently, the
anomalies $\beta^{(2n+1)}$, introduced in \rf{quasiconserv}, are
pseudo-scalars under the same combined transformation. Therefore, in
any Lorentz reference frame  the integrated anomalies
$\alpha^{(2n+1)}$, defined in \rf{chargeanomalydef}, satisfy
\be
\alpha^{(2n+1)}=0 \qquad\qquad \mbox{\rm for any static or a travelling 
  finite energy solution}
\lab{nicetheorem}
\ee
where by a {\em travelling solution} we mean any solution that can be put
at rest by a Lorentz boost. Even though this result may look trivial,
it can perhaps shed some light on the nature of the anomalies
$\alpha^{(2n+1)}$. In fact, as we will see in our concrete example of
section \ref{sec:expansion}, the anomalies vanish in multi-soliton
solutions when the solitons they describe are far apart and so 
when they are not in interaction with each
other. The anomalies seem to be turned on only when the interaction takes place
among the solitons.   

\subsection{A second set of quasi conserved charges}
\label{sec:secondsetcharges}

Note that we can also construct a second set of quasi conserved charges for the theories
\rf{eqmot} using another zero curvature representation of their
equations of motion. The new Lax potentials are obtained from
\rf{potentials} by interchanging $x_{+}$ with $x_{-}$, and by reverting
the grades of the generators. Then we introduce the Lax potentials 
\br
{\tilde A_{-}}&=& \frac{1}{2}\left[ \(\omega^2\,V-m\)\, b_{-1}
-i\,\omega\,  \frac{d\,V}{d\,\varphi}\,F_{-1}\right], 
\nonumber\\
{\tilde A_{+}}&=& \frac{1}{2}\, b_{1} - \frac{i}{2}\,
\omega\, \partial_{+}\varphi\, F_0.
\lab{pluspotentials}
\er
In this case using the commutation relations of appendix
\ref{sec:appendix-algebra} we observe that the curvature of such a connection is 
\be
{\tilde F}_{+-}\equiv \partial_{+}{\tilde A_{-}}-\partial_{-}{\tilde
  A_{+}}+\sbr{{\tilde 
    A_{+}}}{{\tilde A_{-}}}= {\tilde X} \, F_{-1}
+\frac{i}{2}\,\omega\,\left[\partial^2\vp+\frac{\partial V}{\partial
    \vp}\right]\, F_0
\ee
with
\be
{\tilde X} = -\frac{i}{2}\,
\omega\, \partial_{+}\varphi\,\left[\frac{d^2\,V}{d\,\varphi^2}+\omega^2\,
  V-m\right].
\lab{ydef}
\ee
The construction of the corresponding charges follows the same procedure as in
section \ref{sec:quasizc}. We perform the gauge transformation 
\be
{\tilde A}_{\mu}\rightarrow {\tilde a}_{\mu}={\tilde g}\, {\tilde
  A}_{\mu}\,{\tilde g}^{-1}-\partial_{\mu}{\tilde g}\, {\tilde g}^{-1} 
\lab{gaugeplus}
\ee
with the group element being 
\be
{\tilde g}={\rm exp}\left[\sum_{n=1}^{\infty} \zeta_{-n}\,
  F_{-n}\right]
\lab{gaugeplus2}
\ee
and analogously to the case of section \ref{sec:quasizc}, we choose
the $\zeta_{-n}$'s to cancel the $F_{-n}$'s components of ${\tilde
  a}_{+}$. We then have  
\be
\partial_{+}{\tilde a}_{-}-\partial_{-}{\tilde a}_{+}+\sbr{{\tilde
    a}_{+}}{{\tilde a}_{-}}= {\tilde X} \, {\tilde g}\,
F_{-1}\,{\tilde g}^{-1}
\lab{newcurvaturetilde}
\ee
where we have used the equation of motion \rf{eqmot} to cancel the
component of ${\tilde F}_{+-}$ in the direction of $F_0$. The details
of the calculations are given in the appendix
\ref{sec:appendix-gauge2}. The transformed connection takes the form
\br
{\tilde a}_{+}&=&\frac{1}{2}\,b_{1}
+\sum_{n=0}^{\infty}{\tilde a}_{+}^{(-2n-1)}\, b_{-2n-1},
\nonumber\\
{\tilde a}_{-}&=&\sum_{n=0}^{\infty}{\tilde a}_{-}^{(-2n-1)}\, b_{-2n-1}
+\sum_{n=2}^{\infty} {\tilde c}_{+}^{(-n)}\,F_{-n}.
\nonumber
\er
The transformed curvature \rf{newcurvaturetilde} leads to
 equations of the form
\be
\partial_{+}{\tilde a}_{-}^{(-2n-1)}-\partial_{-}{\tilde a}_{+}^{(-2n-1)}=
{\tilde \beta}^{(-2n-1)}\qquad\qquad n=0,1,2,\ldots
\lab{quasiconservtilde}
\ee
with ${\tilde \beta}^{(2n+1)}$ being linear in the anomaly ${\tilde X}$, given in
\rf{ydef}, and the first three are given by
\br
{\tilde \beta}^{(-1)}&=&0,\nonumber\\
{\tilde \beta}^{(-3)}&=&i \omega\, \partial_{+}^2\varphi\, {\tilde X},
\nonumber\\
{\tilde \beta}^{(-5)}&=&i \omega\,\left[\frac{3}{2}  \omega ^2
  (\partial_{+}\varphi)^2 
    \partial_{+}^2\varphi + 
\partial_{+}^4\varphi\right]\, {\tilde X}.
\nonumber
\er
Following the same reasoning as in section \ref{sec:quasizc}, we find
that for finite energy solutions we have the quasi conservation laws
\be
\frac{d\,{\tilde Q}^{(-1)}}{d\,t}= 0\; , \qquad\qquad\qquad
\frac{d\,{\tilde Q}^{(-2n-1)}}{d\,t}=-\frac{1}{2}\,{\tilde
  \alpha}^{(-2n-1)}\qquad 
n=1,2,\ldots
\lab{quasiconserv2tilde}
\ee
with
\be
{\tilde Q}^{(-2n-1)}\equiv \int_{-\infty}^{\infty}dx\,{\tilde
  a}_{x}^{(-2n-1)},\qquad\qquad\qquad 
{\tilde \alpha}^{(-2n-1)}\equiv \int_{-\infty}^{\infty}dx\,{\tilde
  \beta}^{(-2n-1)}. 
\lab{chargeanomalydeftilde}
\ee

\section{The expansion around the sine-Gordon model}
\label{sec:expansion}
\setcounter{equation}{0}

The construction of quasi conserved charges of  section
\ref{sec:quasizc} was performed for a very general potential, and no
estimates were done on how small the anomaly of the zero curvature
condition really is. We now turn to the problem of evaluating the
anomalies $\alpha^{(2n+1)}$, introduced in \rf{chargeanomalydef}, and
to discuss the usefulness of the quasi conservation laws
\rf{quasiconserv2}. In order to do that we choose a specific
potential which is a perturbation of the sine-Gordon potential and
that preserves its main features like infinite degenerate vacua and the
existence of soliton-like solutions.
So we consider the potential given in \rf{nicepotintroa}
and we put $n=2+\ve$ {\it i.e.} we take
\be
V\(\vp,\ve\)=\frac{2}{\(2+\ve\)^2}\,\tan^2 \vp\left[1-\mid
  \sin\vp\mid^{2+\ve}\right]^2.
\lab{nicepot}
\ee

In order to analyze the role of zero curvature anomalies we shall
expand the equation of motion \rf{eqmot} for the potential
\rf{nicepot}, as well as the 
solutions, in powers of $\ve$. We then write   
\be
\vp= \vp_0+\vp_1\,\ve +\vp_2\,\ve^2+\ldots
\ee
and
\br
\frac{\partial\,V}{\partial\,\vp} &=& 
\frac{\partial\,V}{\partial\,\vp} \mid_{\ve=0} +
\left[\frac{d\,}{d\,\ve}\(\frac{\partial\,V}{\partial\,\vp}
  \)\right]_{\ve=0}\, \ve +\ldots\nonumber\\
&=& 
\frac{\partial\,V}{\partial\,\vp} \mid_{\ve=0} +
 \left[\frac{\partial^2 V}{\partial \ve\partial\vp}+
\frac{\partial^2 V}{\partial \vp^2}\,\frac{\partial \vp}{\partial
  \ve}\right]_{\ve=0} \, \ve +\ldots\nonumber
\er
Using the results of appendix \ref{sec:appendix-expansion}, where we
give the detailed  calculations of such expansion, we have that the
order zero field $\vp_0$ must satisfies the sine-Gordon equation,
{\it i.e.} 
\be
\partial^2\vp_0+\frac{1}{4}\,\sin\(4\,\vp_0\)=0.
\lab{eqforphi0}
\ee
On the other hand the first order field $\vp_1$ has to satisfy the
equation  
\be
\partial^2\vp_1+\cos\(4\,\vp_0\)\, \vp_1=
 \sin ( \vp_0)\,\cos ( \vp_0) 
\left[2\,\sin^2\vp_0\, \ln\( \sin^2 (\vp_0)\)
+\cos^2 ( \vp_0) \right].
\lab{eqforphi1}
\ee   
We shall consider here only the anomalies for the charges constructed in section
\ref{sec:quasizc} (the analysis for the charges constructed in section
\ref{sec:secondsetcharges} is very similar). We  expand the anomaly
$X$ introduced in \rf{xdef} as 
\be
X=X_0+X_1\,\ve +X_2\,\ve^2+\ldots
\ee
and we also expand the parameters
\br
\omega &=& \omega_0+\omega_1\, \ve +\omega_2\, \ve^2 +\ldots\nonumber\\
m&=&m_0+m_1\,\ve +m_2\,\ve^2+\ldots.
\er
Then we find that
\be
X_0= \frac{i\,\omega_0}{2}\,  \partial_{-}\varphi_0\,
\left[\frac{d^2\,V}{d\,\varphi^2}\mid_{\ve=0}+\omega^2_0\,
  V\mid_{\ve=0}-m^2_0\right].
\ee
Using the results of appendix \ref{sec:appendix-expansion} we find
that $X_0$ vanishes by an appropriate choice of parameters, {\it i.e.}
\be
X_0=0 \qquad\qquad {\rm when} \qquad\qquad \omega_0=4\qquad {\rm
  and}\qquad m_0=1.
\ee
With such a choice the first order contribution to $X$ reduces to (again
using the results of appendix \ref{sec:appendix-expansion})
\br
X_1&=& i\,2\,  \partial_{-}\varphi_0\,
\left[-6\, \sin^2\vp_0\,\ln\( \sin^2 \vp_0\) -\cos^2\vp_0 -m^2_1 
\right. \nonumber\\
&+&\left.
8\,\(\frac{\omega_1}{2}-1\)\,\sin^2\vp_0\,\cos^2\vp_0\right]
\lab{x1def}
\er
and so  we see that $X_1$ does not depend upon $\vp_1$. 

Since the anomalies $\alpha^{(2n+1)}$, introduced in
\rf{chargeanomalydef}, are linear in $X$ and since $X_0=0$, it
follows that their zero order contribution vanishes, as it should
since sine-Gordon is integrable. Thus we write our anomalies as
\be
\alpha^{(2n+1)} =
\alpha^{(2n+1)}_1\,\ve+\alpha^{(2n+1)}_2\,\ve^2+\ldots
\lab{anomalyexpansion}
\ee
and the first order contribution to the first two of them are
(remember that $\alpha^{(1)}=0$)
\br
\alpha^{(3)}_1&=&
i\,\omega_0\,\int_{-\infty}^{\infty}dx\,X_1\,\partial_{-}^2\vp_0,
\lab{firstorderanomaly}\\
\alpha^{(5)}_1&=&i\,\omega_0\,\int_{-\infty}^{\infty}dx\,X_1\,
\left[
\frac{3}{2}\,\omega^2_0\,\(\partial_{-}\vp_0\)^2 \partial_{-}^2\vp_0 +
\partial_{-}^4\vp_0\right]
\nonumber 
\er  
with $X_1$ given in \rf{x1def}. Thus, the first order anomalies
do not depend on the first order field $\vp_1$. The first order
charges, however, do depend upon $\vp_1$. To see this we expand the charges
as 
\be
Q^{(2n+1)}= Q^{(2n+1)}_0+Q^{(2n+1)}_1\,\ve+Q^{(2n+1)}_2\,\ve^2+\ldots
\lab{chargeexpansion}
\ee
Then we find that $Q^{(2n+1)}_0$ are conserved and correspond to the
charges of the sine-Gordon model, and involve $\vp_0$ only. As an
example we present the first
charge at first order 
\br
Q^{(3)}_1&=&\int_{-\infty}^{\infty}dx\, \left[
8\,\(\partial_{-}\vp_0\)^3\(\omega_1\,\partial_{-}\vp_0+4\,\partial_{-}\vp_1\)
+\partial_{-}^3\vp_0\,\(\omega_1\,\partial_{-}\vp_0
+2\,\partial_{-}\vp_1\)
\right.\nonumber\\
  &+& \left. 2\,\partial_{-}^3\vp_1\,\partial_{-}\vp_0
+\frac{1}{4}\,\sin\(4\,\vp_0\)\,\(\omega_1\partial_{-}^2\vp_0
+2\partial_{-}^2\vp_1\)
-2\,\partial_{-}^2\vp_0\,\partial_{+}\partial_{-}\vp_1-i\,X_1\partial_{-}\vp_0
\right]\nonumber
\er
which, indeed, does depend on $\vp_1$. 

We can now evaluate the anomaly, to first order, for some  physical
relevant solutions of the theory \rf{eqmot} with the potential 
given by \rf{nicepot}. As we have stressed this earlier the first order anomaly depends
only upon the zero order field $\vp_0$ which is an exact solution of the
sine-Gordon equation \rf{eqforphi0}.

\subsection{Anomaly for the kink}
 First we look at the case of one kink. 
The kink solution is given by \rf{exactkinkintro} with $n=2+\ve$ and
it is an exact 
solution of \rf{eqmot} for $V$ given by \rf{nicepot}. The first order
anomaly depend upon the kink solution of the sine-Gordon equation
\rf{eqforphi0} which is given by 
\be
\vp_0=\arctan\(e^x\).
\lab{phi0kink}
\ee
Inserting this expression into \rf{firstorderanomaly} and
\rf{x1def} we find that
\be
\alpha^{(3)}_1=\alpha^{(5)}_1=
\int_{-\infty}^{\infty}dx\,\frac{\sinh x}{\cosh^4 x}\, 
\left[6\,e^x\,\ln\(\frac{1}{2}\,\frac{e^x}{\cosh x}\)+e^{-x}\right].
\ee
This expression can be integrated explicitly using the fact that
\br
\frac{d\;}{d\,x}\left[
\frac{2 \sinh (x)+e^x \left(e^{2
   x}-3\right)\ln\( \frac{1}{2}\,\frac{e^x}{\cosh x}\)}{2\, 
\cosh^3\( x\)}\right]=
\frac{\sinh x}{\cosh^4 x}\, 
\left[6\,e^x\,\ln\(\frac{1}{2}\,\frac{e^x}{\cosh x}\)+e^{-x}\right]
\nonumber
\er
and so 
\be
\int_{-\infty}^{\infty}dx\,\frac{\sinh x}{\cosh^4 x}\, 
\left[6\,e^x\,\ln\(\frac{1}{2}\,\frac{e^x}{\cosh x}\)+e^{-x}\right]=0.
\ee
Therefore, the first order anomalies vanish,
{\it i.e.} $\alpha^{(3)}_1=\alpha^{(5)}_1=0$, agreeing with the general
result shown in \rf{nicetheorem}.

\subsection{Anomalies for the 2-soliton solutions}

\subsubsection{The soliton/anti-soliton scattering}

Let us consider a 2-soliton solution corresponding, for $\eta =1$, to
a soliton   moving to 
the right with speed $v$ and located at $x=-L$ at $t=0$, and  an 
anti-soliton moving to the left with speed $v$ and located at $x=L$ at $t=0$. 
For $\eta=-1$ the roles of soliton and anti-soliton are interchanged. 
The solution at order zero in the $\ve$-expansion is given by a
solution of the sine-Gordon equation \rf{eqforphi0} given by
\be
\vp_0={\rm ArcTan}\left[\frac{\eta\, v \,\cosh y_1}{\sinh \tau_1}\right]
\lab{phi0solitonantisoliton}
\ee
with
\br
y_1= \frac{x}{\sqrt{1-v^2}},\qquad\qquad\qquad 
\tau_1= \frac{v\,t-L}{\sqrt{1-v^2}}+\eta \ln v.
\er
Putting this expression  into \rf{firstorderanomaly} and
\rf{x1def} we find  that the first anomaly at first order is
\br
&\alpha^{(3)}_1&=\frac{8\, v^2}{(1-v^2)^{3/2}}\, \sinh\tau_1\,\cosh\tau_1\, 
\int_{-\infty}^{\infty}dx\, \frac{1}{\Lambda_1^3}\, 
\left[v\(\(3+v^2\)\Omega_1+4\,v^2\)\,\cosh^2 y_1-2\,v\,\Omega_1\right]
\times \nonumber\\
&\times&\left[-6\, \frac{v^2\,\cosh^2 y_1}{\Lambda_1}\,
\ln\( \frac{v^2\,\cosh^2 y_1}{\Lambda_1}\) -\frac{\sinh^2 \tau_1}{\Lambda_1} -m^2_1 
+
8\,\(\frac{\omega_1}{2}-1\)\,\frac{v^2\,\cosh^2 y_1}{\Lambda_1}\,
\frac{\sinh^2 \tau_1}{\Lambda_1}\right],\nonumber\\
\lab{firstanomaly2sol}
\er
where we have introduced
\be
\Lambda_1= \sinh^2\tau_1 + v^2\, \cosh^2 y_1,\qquad\qquad
\Omega_1=\sinh^2\tau_1 - v^2\, \cosh^2 y_1.
\ee
Note that $\alpha^{(3)}_1$ given in \rf{firstanomaly2sol}, is an odd
function of $\tau_1$ due to the term $\sinh\tau_1$ in front of the
integral. All other terms involving $\tau_1$ in
\rf{firstanomaly2sol} appear as $\cosh\tau_1$ or $\sinh^2\tau_1$, and
so are even in $\tau_1$.  Consequently we see that 
\be
\int_{-\infty}^{\infty} dt\, \alpha^{(3)}_1 =0.
\ee
We point out that this result is independent of the values of
$\omega_1$ and $m_1$ which appear in the expression for
$\alpha^{(3)}_1$. 
Note that, from \rf{quasiconserv2}, \rf{anomalyexpansion} and
\rf{chargeexpansion},   we have that
\be
\frac{d\,Q^{(3)}_1}{d\,t}=-\frac{1}{2}\,\alpha^{(3)}_1
\lab{timederq31}
\ee
and so
\be
Q^{(3)}_1\(t=\infty\)=Q^{(3)}_1\(t=-\infty\).
\ee
Thus, in the scattering of the soliton and anti-soliton the
charge at first order is conserved asymptotically. From the physical
point of view that is as effective as in the case of the integrable
sine-Gordon theory. The solitons have to scatter preserving higher
charges (at least in first order approximation).

\subsubsection{The soliton/soliton scattering}

Next we consider a 2-soliton solution corresponding, for $\eta =1$, to
a soliton   
moving to 
the right with speed $v$ and located at $x=-L$ at $t=0$, and  another  
soliton moving to the left with speed $v$ and located at $x=L$ at $t=0$. 
For $\eta=-1$ the roles of soliton and anti-soliton are interchanged. 
The solution, at order zero in the $\ve$-expansion, is again given by a
solution of the sine-Gordon equation \rf{eqforphi0}, namely
\be
\vp_0={\rm ArcTan}\left[\frac{-\eta\, \cosh \tau_2}{v \,\sinh y_2}\right]
\lab{phi0solitonsoliton},
\ee
where
\br
y_2= \frac{x}{\sqrt{1-v^2}}+\eta \ln v,\qquad\qquad\qquad 
\tau_2= \frac{v\,t-L}{\sqrt{1-v^2}}.
\er
Following the same procedure as in the case of soliton/anti-soliton
solution, by putting \rf{phi0solitonsoliton}  into \rf{firstorderanomaly} and
\rf{x1def} we find that the first anomaly, at first order, is
\br
\alpha^{(3)}_1&=& \frac{8\,v^2}{\(1-v^2\)^{3/2}}\,\sinh\tau_2\,\cosh\tau_2\,
\int_{-\infty}^{\infty}dx\,\frac{1}{\Lambda_2^3}\,
\left[v\,\left(\left(v^2+3\right)\Omega_2-4 v^2\right)\,\sinh^2y_2+
2\,v \,\Omega_2\right]\nonumber\\
&\times&
\left[-6\, \frac{\cosh^2 \tau_2}{\Lambda_2}\,\ln\( \frac{\cosh^2
    \tau_2}{\Lambda_2}\) -\frac{v^2\,\sinh^2 y_2}{\Lambda_2} -m^2_1  
+
8\,\(\frac{\omega_1}{2}-1\)\,\frac{\cosh^2
  \tau_2}{\Lambda_2}\,\frac{v^2\,\sinh^2 y_2}{\Lambda_2}\right] 
\nonumber\\
\lab{firstanomaly2sol2}
\er
with
\be
\Lambda_2= \cosh^2\tau_2 + v^2\, \sinh^2 y_2,\qquad\qquad\qquad
\Omega_2=\cosh^2\tau_2 - v^2\, \sinh^2 y_2.
\ee
Again, one notices that $\alpha^{(3)}_1$ given in
\rf{firstanomaly2sol2} is odd in 
$\tau_2$. Indeed, except for the factor $\sinh\tau_2$ in front of the
integral, all other terms are even in $\tau_2$ since they involve
only   $\cosh\tau_2$. Consequently, we again have 
\be
\int_{-\infty}^{\infty} dt\, \alpha^{(3)}_1 =0
\ee
and  such result is independent of the values of $\omega_1$
and $m_1$. Again, using \rf{timederq31} we see that
$Q^{(3)}_1\(t=\infty\)=Q^{(3)}_1\(t=-\infty\)$. So, the solitons
scatter preserving higher charges asymptotically, like in the case of
soliton/anti-soliton scattering discussed above.

\subsection{Anomalies for breathers}

As we show in the next section the
theory \rf{eqmot} with potential \rf{nicepot} has long lived
breather-like solutions. Hence, next we evaluate the anomaly, to
first order, for 
such a solution. For that we need the solution for the zero order field
$\vp_0$ which is a breather solution for the sine-Gordon equation
\rf{eqforphi0}, {\it i.e.} 
\be
\vp_0=\arctan\(\frac{\sqrt{1-\nu^2}}{\nu}\,\frac{\sin\(\nu\,t\)}
{\cosh\(\sqrt{1-\nu^2}\,x\)}\), 
\lab{phi0breather}
\ee
where $\nu$ is the frequency of the breather ($0 < \nu < 1$), and
we have chosen to 
express it in its Lorentz rest frame. Putting this  configuration
into  \rf{firstorderanomaly} and
\rf{x1def} we find that the first anomaly, to first order, is
\be
\alpha^{(3)}_1=-4\,\nu^3\,\(1-\nu^2\)\,\sin\(2\,\nu\,t\)\,\left[
I_1\(\nu,t\)-m^2_1\,I_2\(\nu,t\)+8\,\(\frac{\omega_1}{2}-1\)\,I_3\(\nu,t\)
\right]
\lab{anomaly31breather}
\ee
with
\br
I_1\(\nu,t\)&=&\int_{-\infty}^{\infty}dx\,
\left[ 2\,\(1-\nu^2\)\,\sinh^2\(\sqrt{1-\nu^2}\,x\)\,\Omega
\right.\\
&+&\left.\cosh^2\(\sqrt{1-\nu^2}\,x\)\,\(\(1-2\,\nu^2\)\,\Omega
-4\,\nu^2\,\(1-\nu^2\)\)\right]
\times\nonumber\\
&\times&\frac{1}{\Lambda^3}\,
\left[-6\, \frac{\(1-\nu^2\)\,\sin^2\(\nu\,t\)}{\Lambda}\,
\ln\( \frac{\(1-\nu^2\)\,\sin^2\(\nu\,t\)}{\Lambda}\) 
-\frac{\nu^2\,\cosh^2\(\sqrt{1-\nu^2}\,x\)}{\Lambda} \right]
\nonumber
\er
and
\br
I_2\(\nu,t\)&=&\int_{-\infty}^{\infty}dx\,
\left[ 2\,\(1-\nu^2\)\,\sinh^2\(\sqrt{1-\nu^2}\,x\)\,\Omega
\right.\nonumber\\
&+&\left.\cosh^2\(\sqrt{1-\nu^2}\,x\)\,\(\(1-2\,\nu^2\)\,\Omega
-4\,\nu^2\,\(1-\nu^2\)\)\right]\,\frac{1}{\Lambda^3}\,
\er
and
\br
I_3\(\nu,t\)&=&\int_{-\infty}^{\infty}dx\,
\left[ 2\,\(1-\nu^2\)\,\sinh^2\(\sqrt{1-\nu^2}\,x\)\,\Omega
\right.\nonumber\\
&+&\left.\cosh^2\(\sqrt{1-\nu^2}\,x\)\,\(\(1-2\,\nu^2\)\,\Omega
-4\,\nu^2\,\(1-\nu^2\)\)\right]
\times\nonumber\\
&\times&\frac{1}{\Lambda^3}\,
\left[
\frac{\nu^2\,\(1-\nu^2\)\,\sin^2\(\nu\,t\)\,\cosh^2\(\sqrt{1-\nu^2}\,x\)}
{\Lambda^2}\right],
\er
where we have denoted
\br
\Lambda &=&
\nu^2\,\cosh^2\(\sqrt{1-\nu^2}\,x\)+\(1-\nu^2\)\,\sin^2\(\nu\,t\),
\nonumber\\
\Omega&=&
\nu^2\,\cosh^2\(\sqrt{1-\nu^2}\,x\)-\(1-\nu^2\)\,\sin^2\(\nu\,t\).
\er
Note that the time dependence of the integrals $I_j\(\nu,t\)$,
$j=1,2,3$, 
comes only through the factor  
$\sin^2\(\nu\,t\)=\frac{1}{2}\,\left[1-\cos\(2\,\nu\,t\)\right]$. Since
these integrals are multiplied by the factor $\sin \(2\,\nu\,t\)$ in
\rf{anomaly31breather}, we conclude that $\alpha^{(3)}_1$ is periodic
in time with period $T\equiv \frac{\pi}{\nu}$. In addition, we observe
that $I_j\(\nu,t\)=I_j\(\nu,-t\)$, and so
$\alpha^{(3)}_1\(t\)=-\alpha^{(3)}_1\(-t\)$, due to the overall factor
$\sin \(2\,\nu\,t\)$ in \rf{anomaly31breather}. Consequently, we have
that
\be
\int_{t}^{t+T}dt^{\prime}\,\alpha^{(3)}_1\(t^{\prime}\)= 
\int_{-T/2}^{T/2}dt^{\prime}\,\alpha^{(3)}_1\(t^{\prime}\)=0,
\ee 
where we have used the fact that
$\int_{t}^{t+T}=\int_{t}^{-T/2}+\int_{-T/2}^{T/2}+\int_{T/2}^{t+T}$,
and so the first and third integrals  cancel  due to the fact that
$\alpha^{(3)}_1\(t\)=\alpha^{(3)}_1\(t+T\)$. Consequently, we find
from \rf{timederq31} that the charge (to first order) is periodic in time
\be
Q^{(3)}_1\(t\)=Q^{(3)}_1\(t+\frac{\pi}{\nu}\).
\ee

\section{Numerical support}
\label{sec:numerical}
\setcounter{equation}{0}

To check our results on the anomaly we have decided to perform various
simulations 
of the Bazeia at al model - studying two kinks, a kink-antikink, and a
system involving two  
kinks and an antikink.

In all our numerical work the time evolution was simulated by the
fourth order Runge - Kuta method. 
 We used various lattice grids (to make sure that our results
were not contaminated by any numerical artefacts, the issue here was
the size of the lattice 
and the lattice step). We found that to have reliable results the lattice grid 
(given that the kinks were of size $\sim \pm5$) had to stretch 
to, at least, $\pm 50$. Hence most of our work was performed using
even larger grids and  
the results given in this paper were obtained in simulations in which
the lattice 
contained 10001 equally spaced points and stretched from -75 to 75. At
the edge of the grid (in practice from -71 to -75 and from +71 to +75)
we absorbed the kinetic 
energy of the fields.  During the scattering process there was some
radiation sent out 
towards the edges of the grid and it is this radiation that our
procedure absorbed 
(so that we would not have any reflection of the radiation from the
boundaries). Thus 
our procedure had the effect of simulating an infinite
grid in which we looked only at the fields in a finite region. 
Thus, due to this absorption, the total energy seen in our simulations
would decrease but this decrease could be associated with the system
radiating some energy towards the boundaries and the energy seen by us
corresponded to 
the energy of the system that we have tried to describe.

\subsection{Kink-kink interactions}
\label{sec:kink-kink}
\setcounter{equation}{0}

First we looked at the interaction between 2 kinks.
To study this we placed two kinks at some distance from each other and
then performed a simulation to see what happens. 

When we performed this simulation with static kinks we have found
that the kinks repel. To decide what happens during the scattering we decided 
to plot the positions of the kinks as a function of time. There are several 
possible definitions of `the position of the kink' but, physically the most sensible
one, involves looking at the energy density of each kink, with the position being
defined at the location of the maximum of this density. This is the definition
we have used in our analysis.

In fig.4   we present the trajectories of two kinks (given the definition
of the position as mentioned above),  initially at rest, 
as seen in simulations for {$n=2$} and       {$n=1.9$}.
The kinks were initially placed at   {$d=\pm 7.0$} and so far away
from any boundary.

\begin{figure}[tbp]
    \centering
	\includegraphics[angle=0,width=4.5cm]{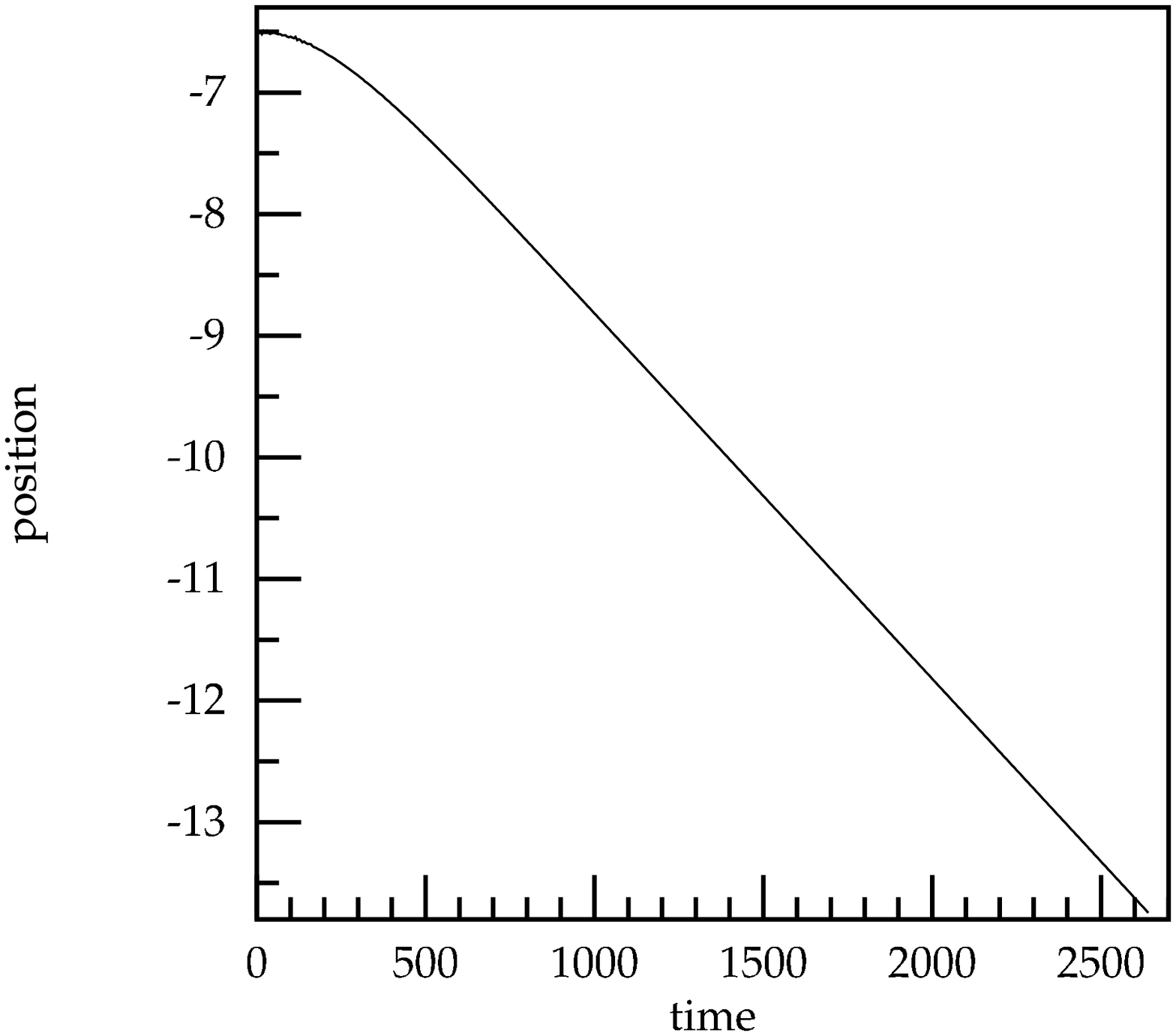}
	 \includegraphics[angle=0,width=4.5cm]{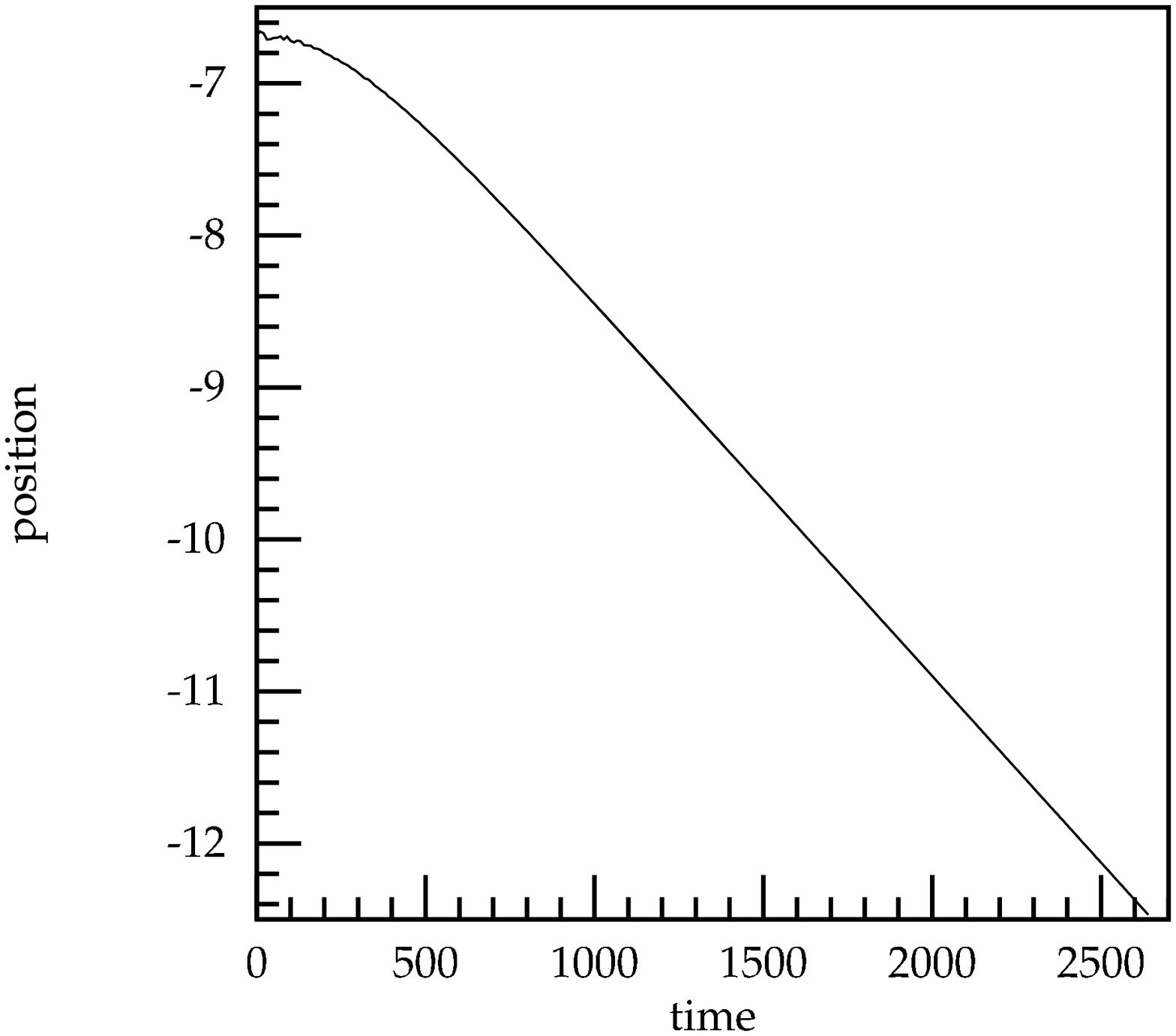}
    \begin{quote}
	\caption[AS]{  Trajectories: a)       {$n=2$}, b)       {$n=1.9$},} 
	\label{fig:Fig4}
    \end{quote}
\end{figure}

It is clear from these plots that the kinks repel. We have repeated
our simulations 
 for various values of 
$n$ and each time the situation was the same.
 Looking at the plots of the trajectories
we do not see much difference between $n=2$ and $n=1.9$.

Next we sent the kinks towards each other with some velocities. 
In the fig. 5  we present the trajectories of kinks sent towards each 
other with velocity       {$v=0.5$}  for  the cases corresponding to     {$n=2$} and       {$n=1.9$}.
Initially the kinks were placed at       {$d=\pm 15.5$}

\begin{figure}[tbp]
    \centering
	\includegraphics[angle=0,width=4.5cm]{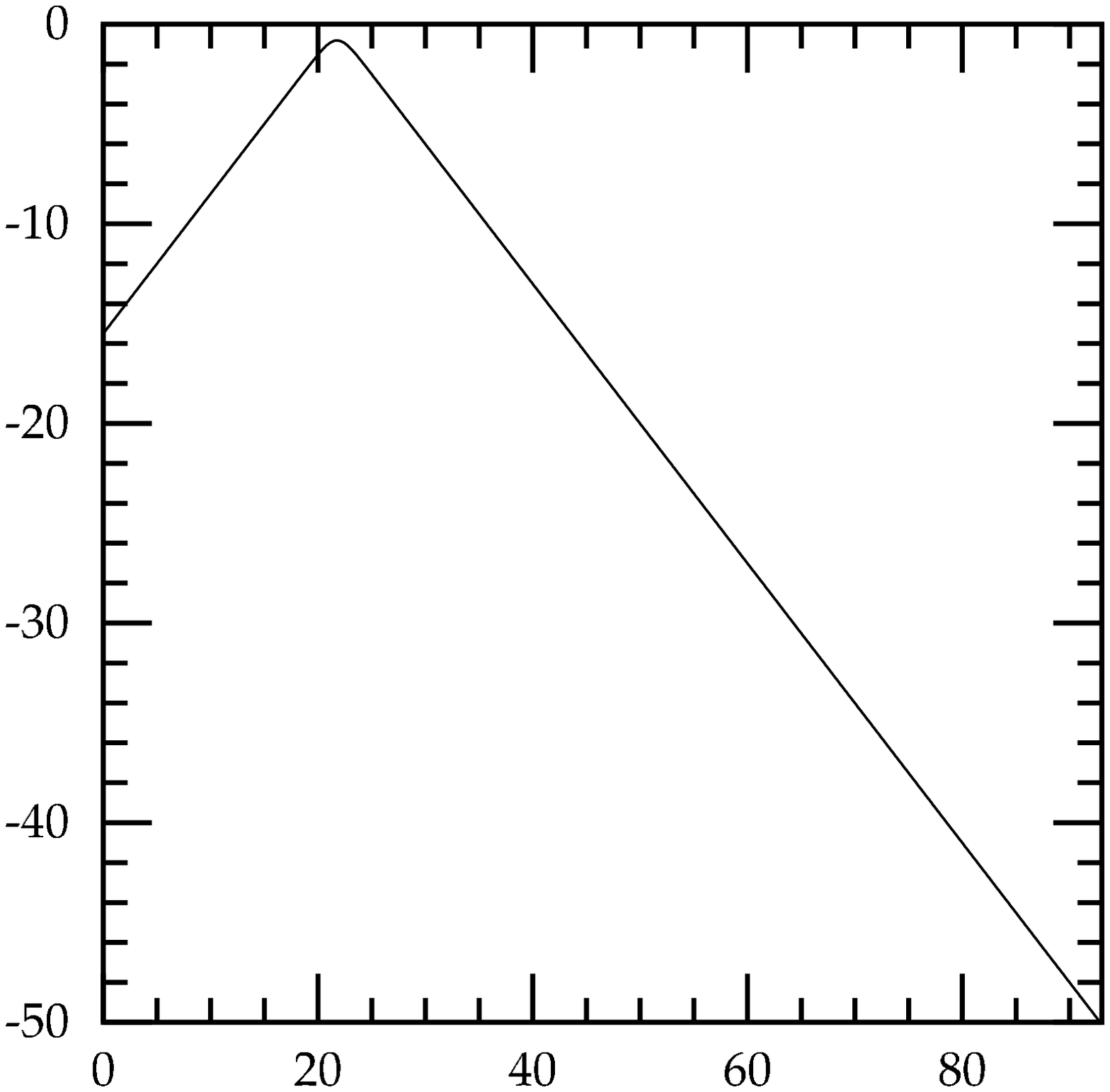}
	 \includegraphics[angle=0,width=4.5cm]{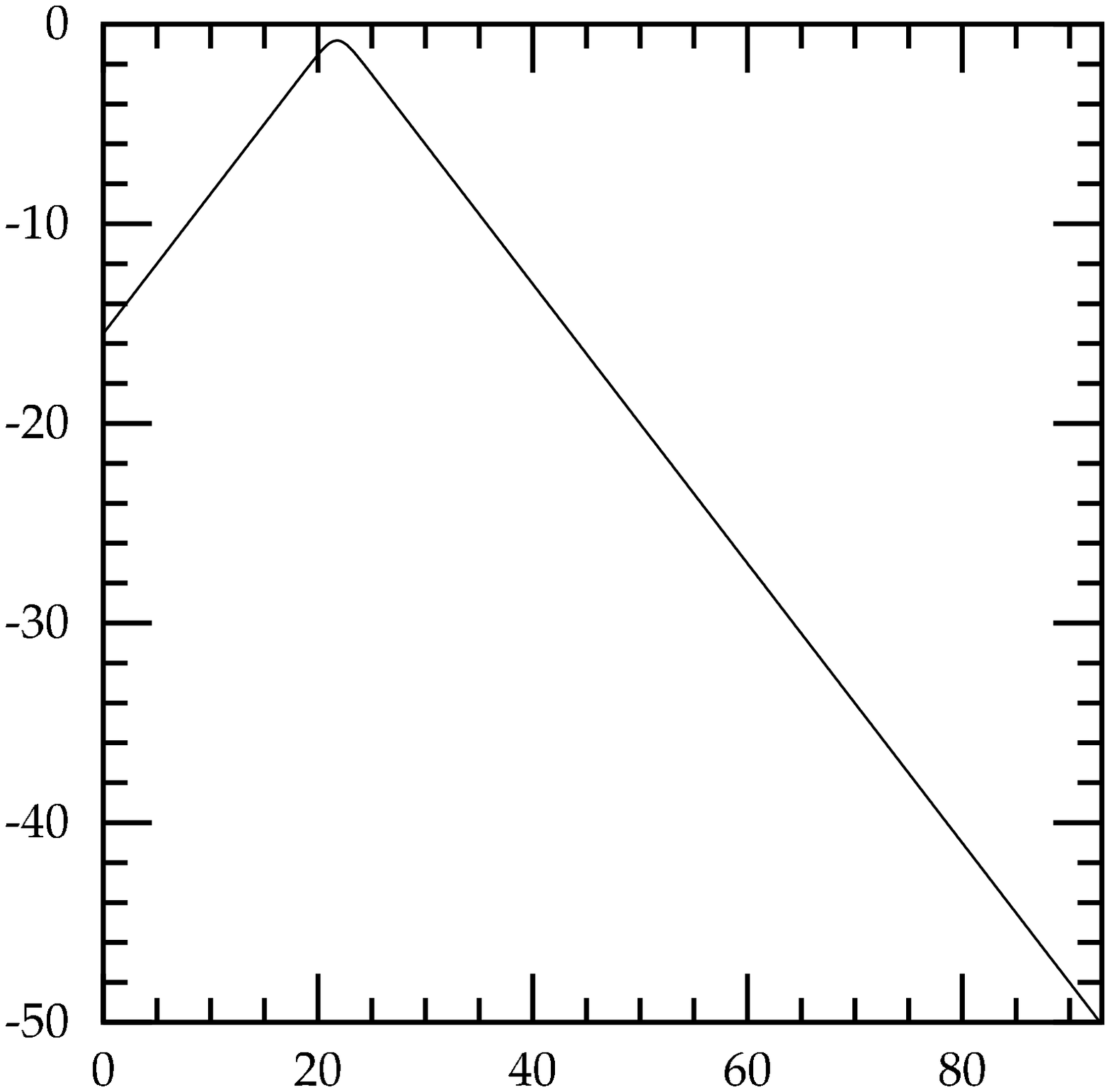}
    \begin{quote}
	\caption[AS]{  Trajectories: a)       {$n=2$}, b)       {$n=1.9$},} 
	\label{fig:Fig5}
    \end{quote}
\end{figure}

In both cases the kinks clearly 
come very close towards each other but then they repel and move away
from each other. 
At larger velocities they get closer before their repulsion always
sends them back. 
Similar results were obtained for other values of $d$, $n$ and their velocties.

Incidentally, the sine-Gordon model ({\it i.e.} the model with $n=2$)
possesses a solution 
describing two moving kinks and our results (for $n=2$) reproduce them very well
and, surprise, surprise, the model does not have any static solutions involving
more than one kink.

On the other hand, the sine-Gordon moving kinks solutions are known in an explicit 
form, and because these kinks are described by explicit functions it is
often said that the ''kinks pass through each other". This is clearly wrong
when one looks at the energy density of the moving kinks as they move
towards each other. In fact, one easily observes 
that the kinks never come on top of each other ({\it i.e.} the two peaks of the energy 
density never form a double peak); in practice, the functions which describe each kink switch
after the kinks' interaction.

We have also looked at the scattering of two kinks from the point of view 
of the integrability discussed in the previous section - {\it i.e.} from the point of view of the anomalies.

To do this we considered the scattering of two kinks for values of $n$ close to 2.
We looked at various positions of kinks and various velocities.
All results were qualitatively similar so here we present our results for
$v=.5$. The kinks were initially placed at $\pm 15.5$. 
 We performed many simulations of the dynamics of such systems.
In each case, as mentioned above, the kinks came close to each other, reflected 
and then moved to the boundaries with essentially the original velocity.
Thus the scattering was very elastic. Looking at the scattering in more detail
it was easy to see that, strictly speaking, there was also some radiation
emitted during the scattering and that the amount of this emitted radiation 
increased with the increase of $\vert\ve\vert =\vert n-2\vert$; however, even for $n=1$ this radiation constituted
less than 2\% of the total energy. Hence the scattering was very elastic.

We have also looked at the values of the first anomaly and its time integrated value
for these scatterings.

In fig 6. we present a representative selection of our results.
 Fig. 6a and 6b present the time dependence of the anomaly
and its time integrated form for $n=2$, as seen in our simulations. 
Of course we know that for $n=2$ the anomaly vanishes
so our results provide the test of our numerics. We note that our values of the anomaly are 
very small - {\it i.e.} consistent with zero. Next we looked at the values of $n\ne2$ for which
the anomaly does not vanish.
In fig 6c and 6d present our results for the anomaly and its integrated form for $n=1.99$
and fig 6e and 6f present similar results for $n=2.01$. Fig 6g and 6h refer to the case of $n=1.98$
while fig 6i and 6j  give the results for $n=2.1$ and fig 6k and 6l for $n=3.$

\begin{figure}[tbp]
    \centering
	 \includegraphics[angle=0,width=3.5cm]{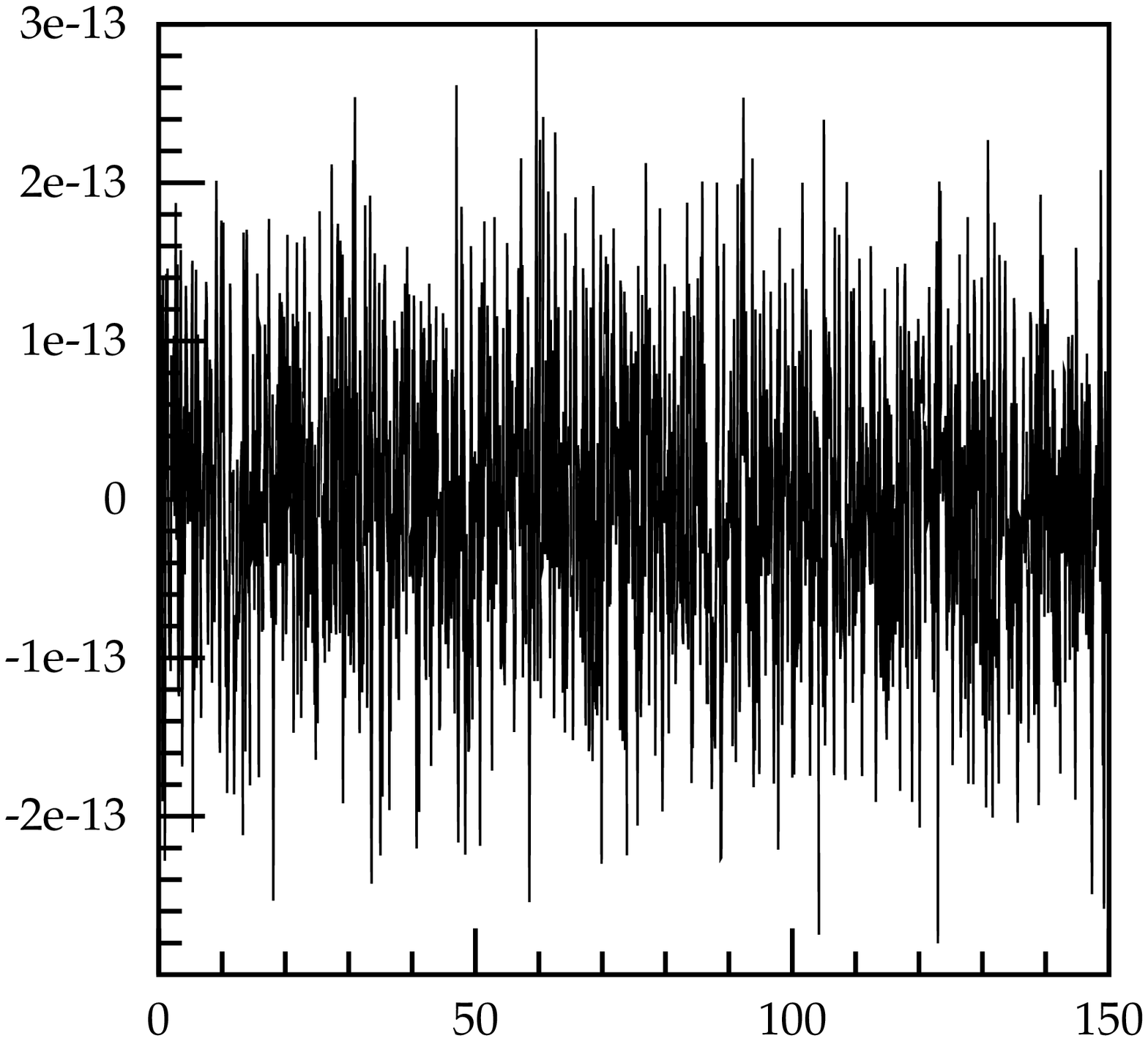}
	 \includegraphics[angle=0,width=3.5cm]{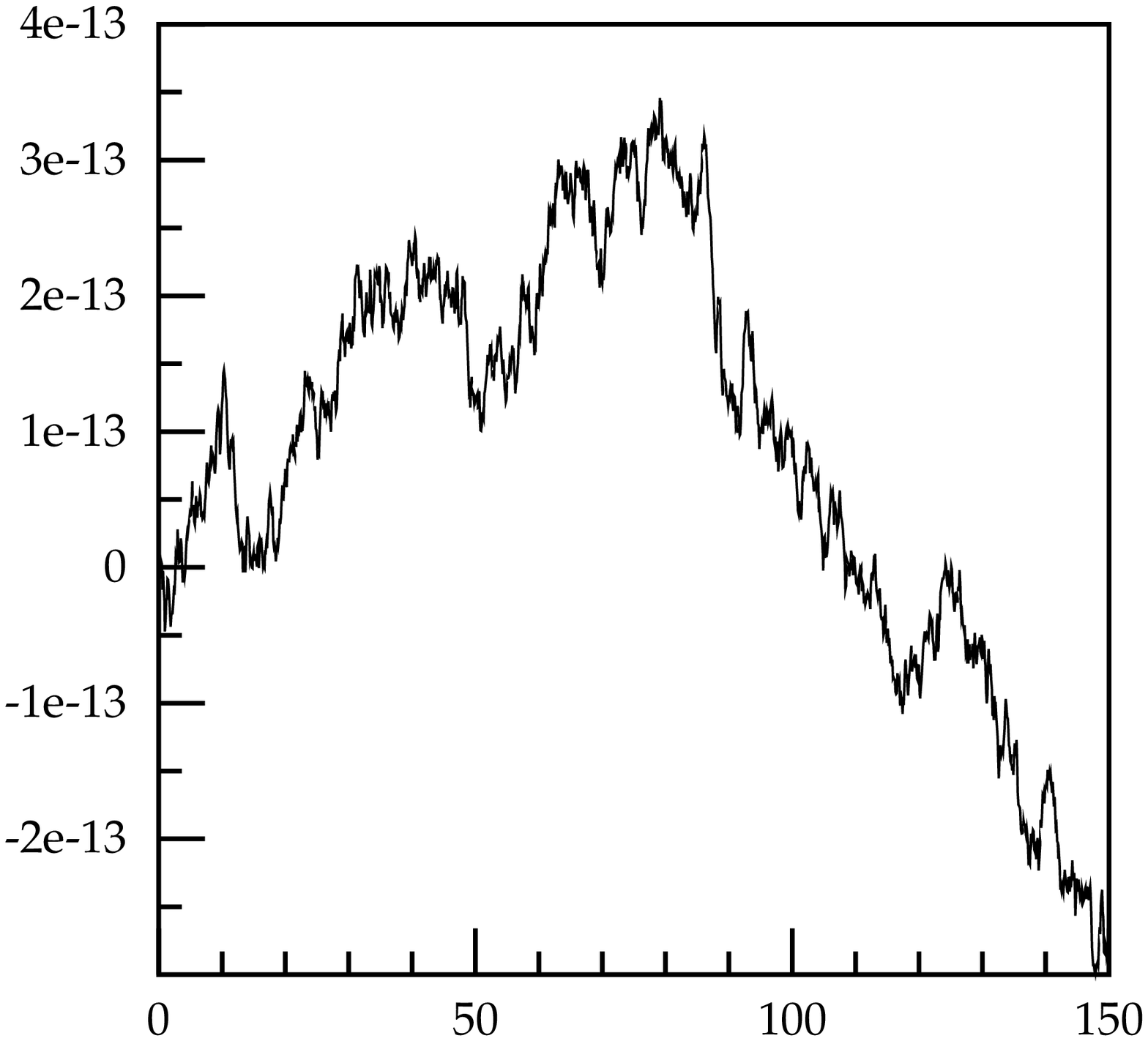}
	 \includegraphics[angle=0,width=3.5cm]{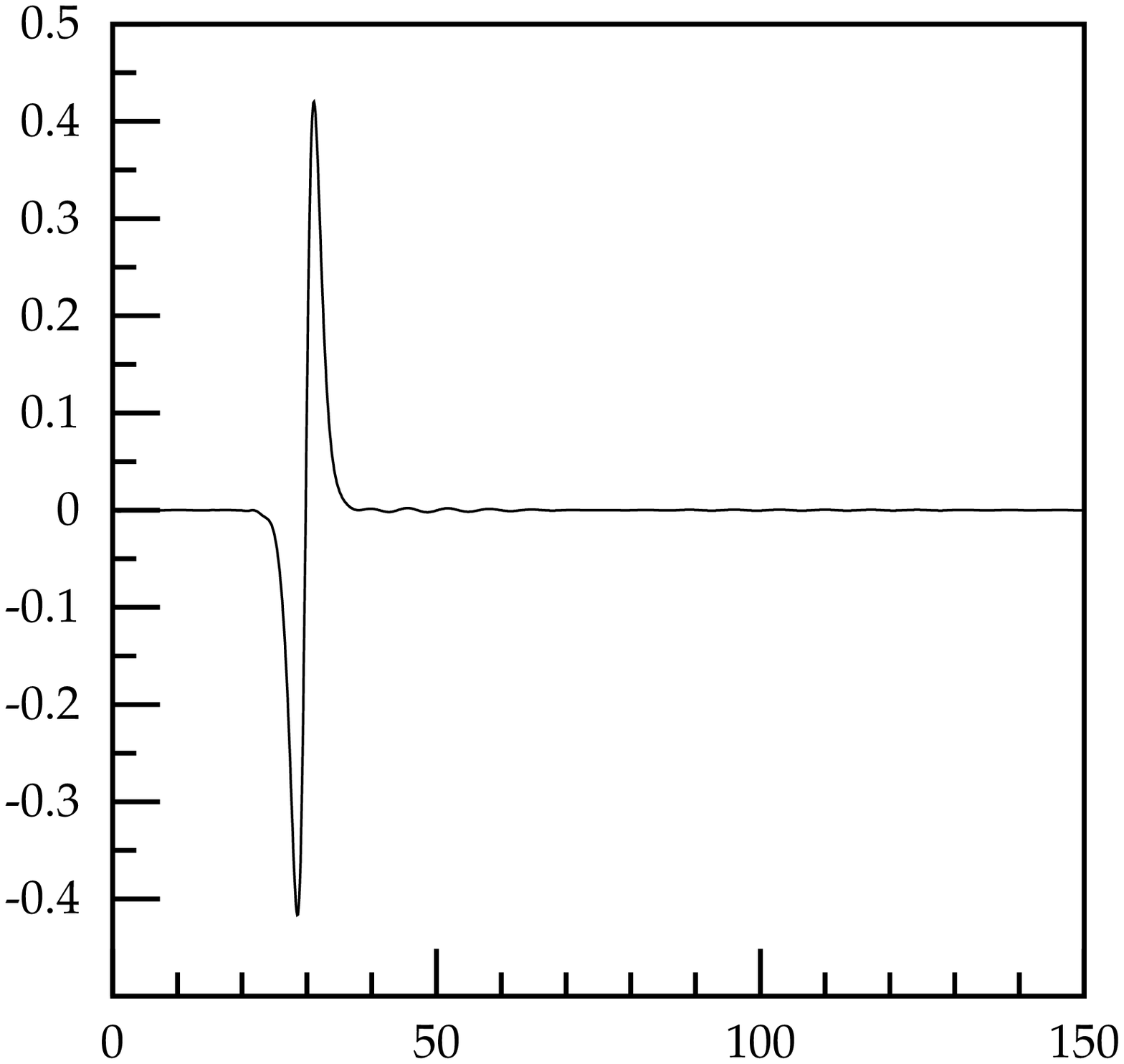}
	 \includegraphics[angle=0,width=3.5cm]{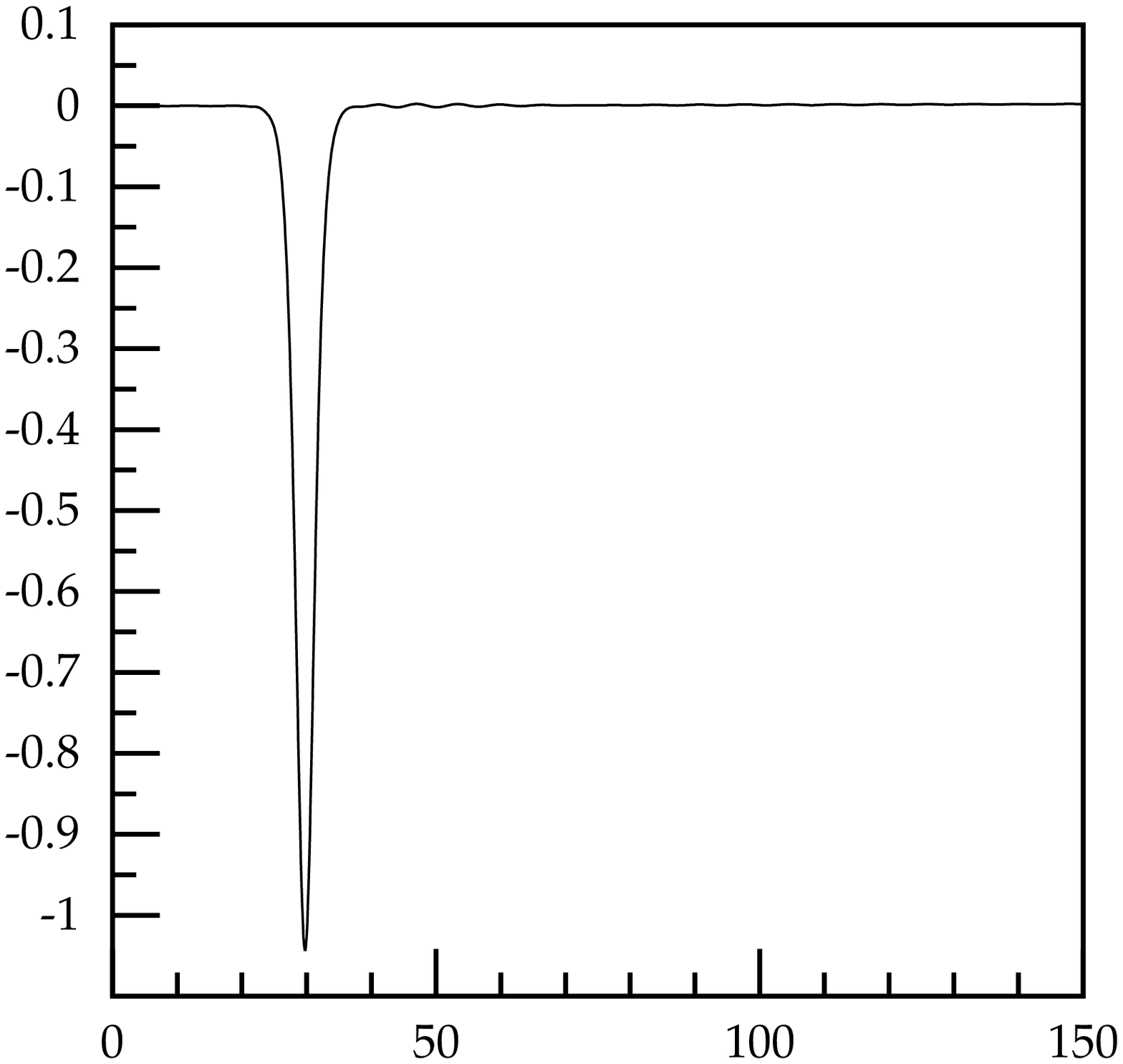}
	 \includegraphics[angle=0,width=3.5cm]{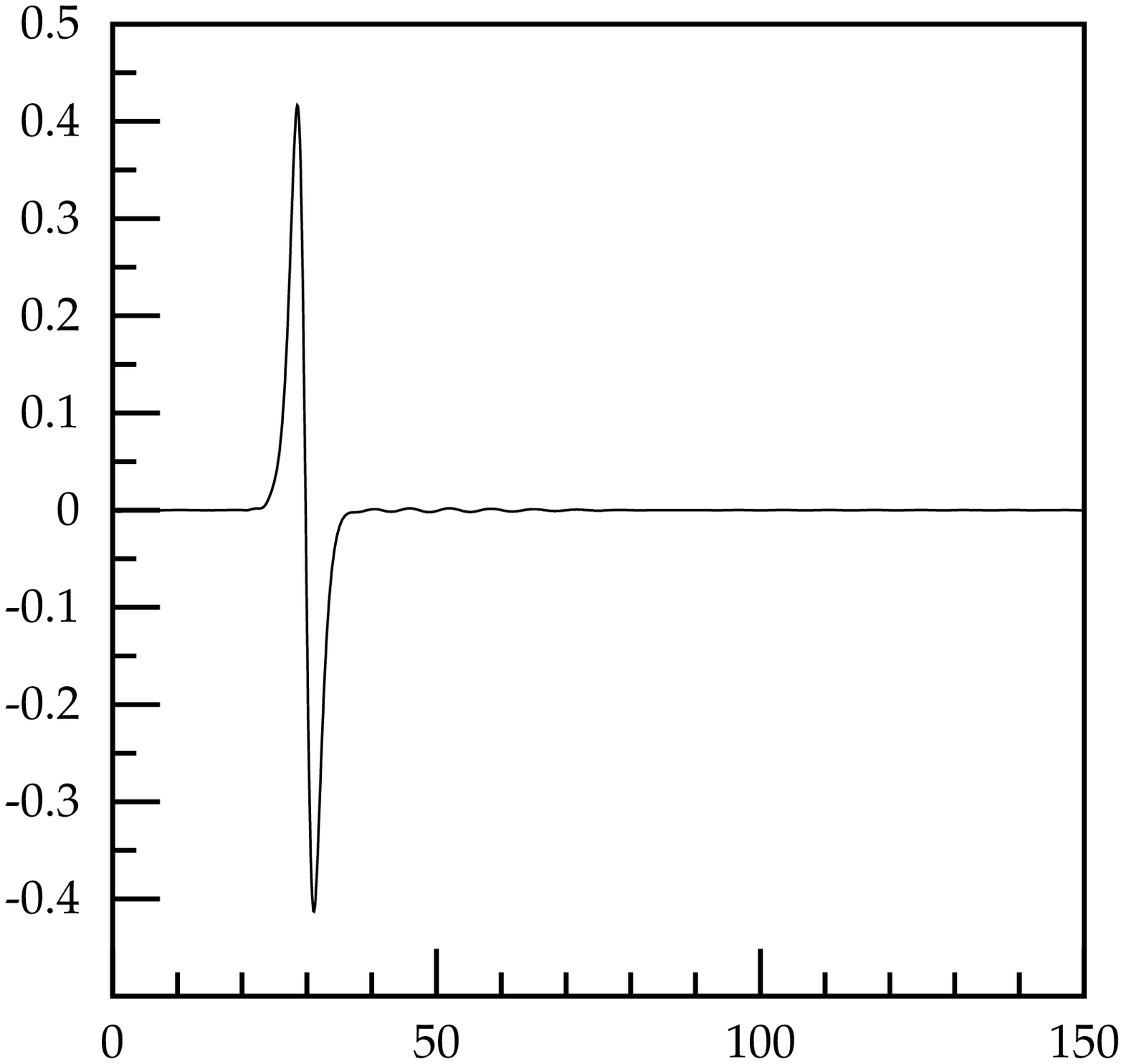}
	 \includegraphics[angle=0,width=3.5cm]{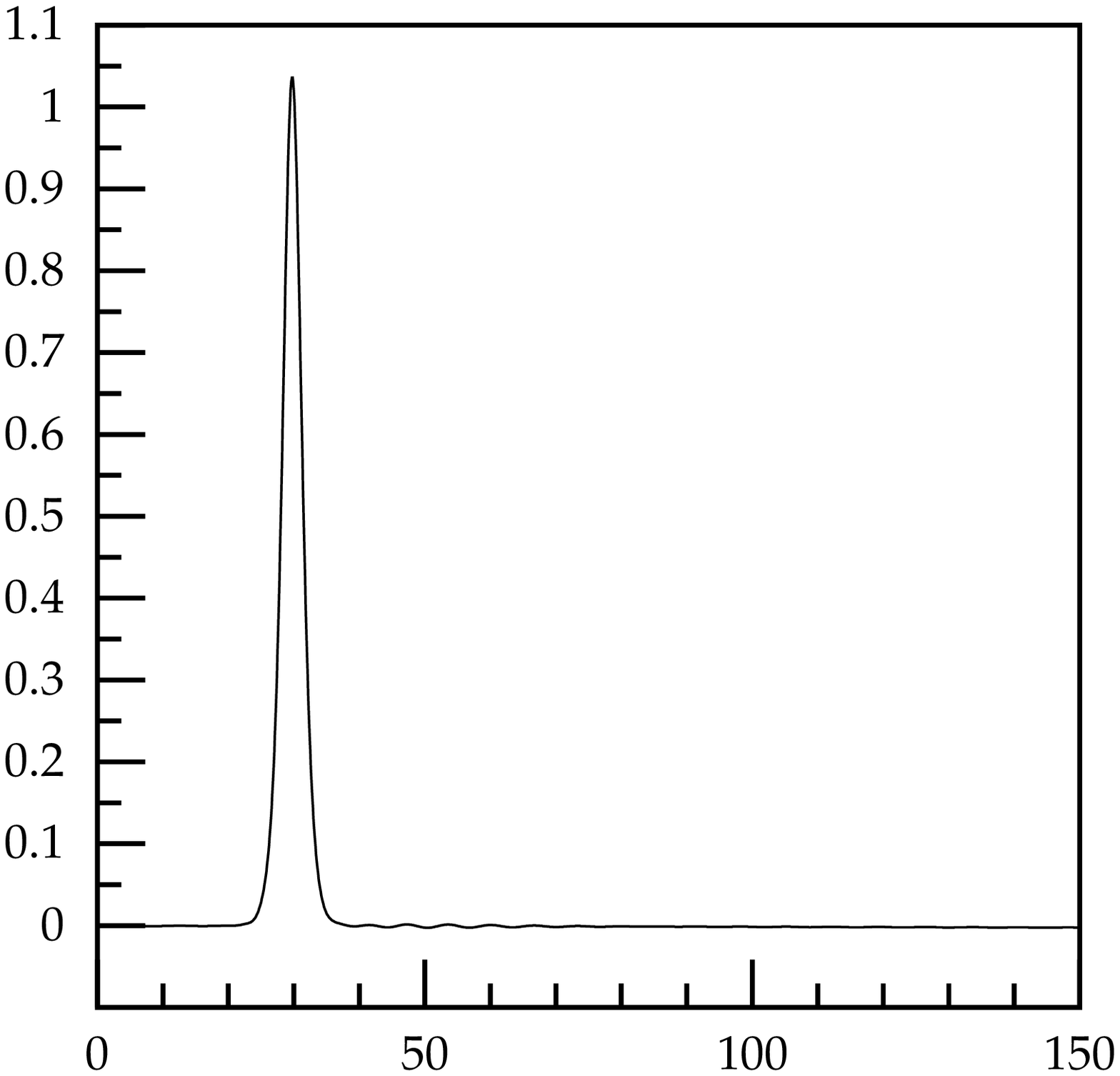}
	 \includegraphics[angle=0,width=3.5cm]{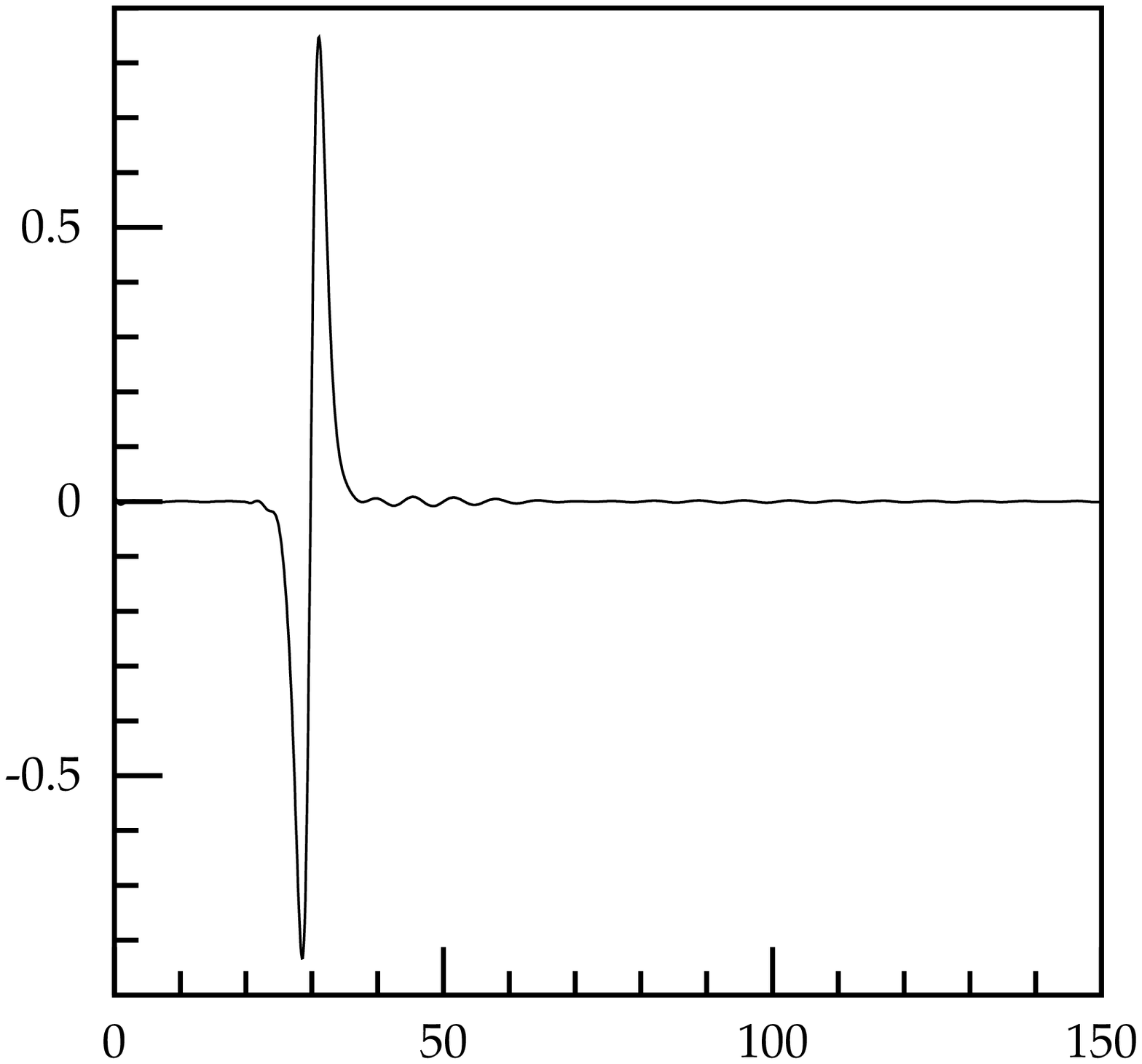}
	 \includegraphics[angle=0,width=3.5cm]{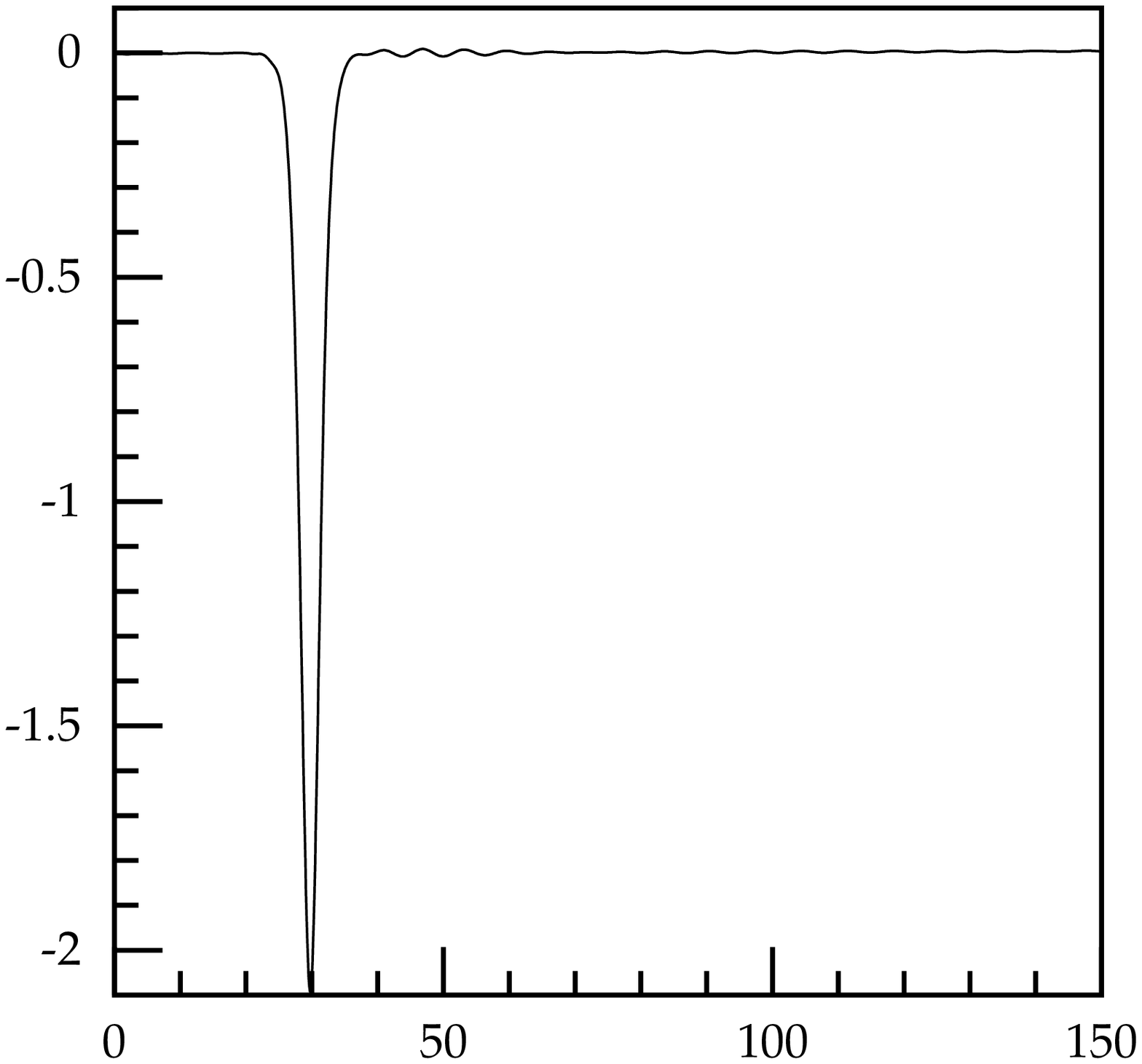}
	 \includegraphics[angle=0,width=3.5cm]{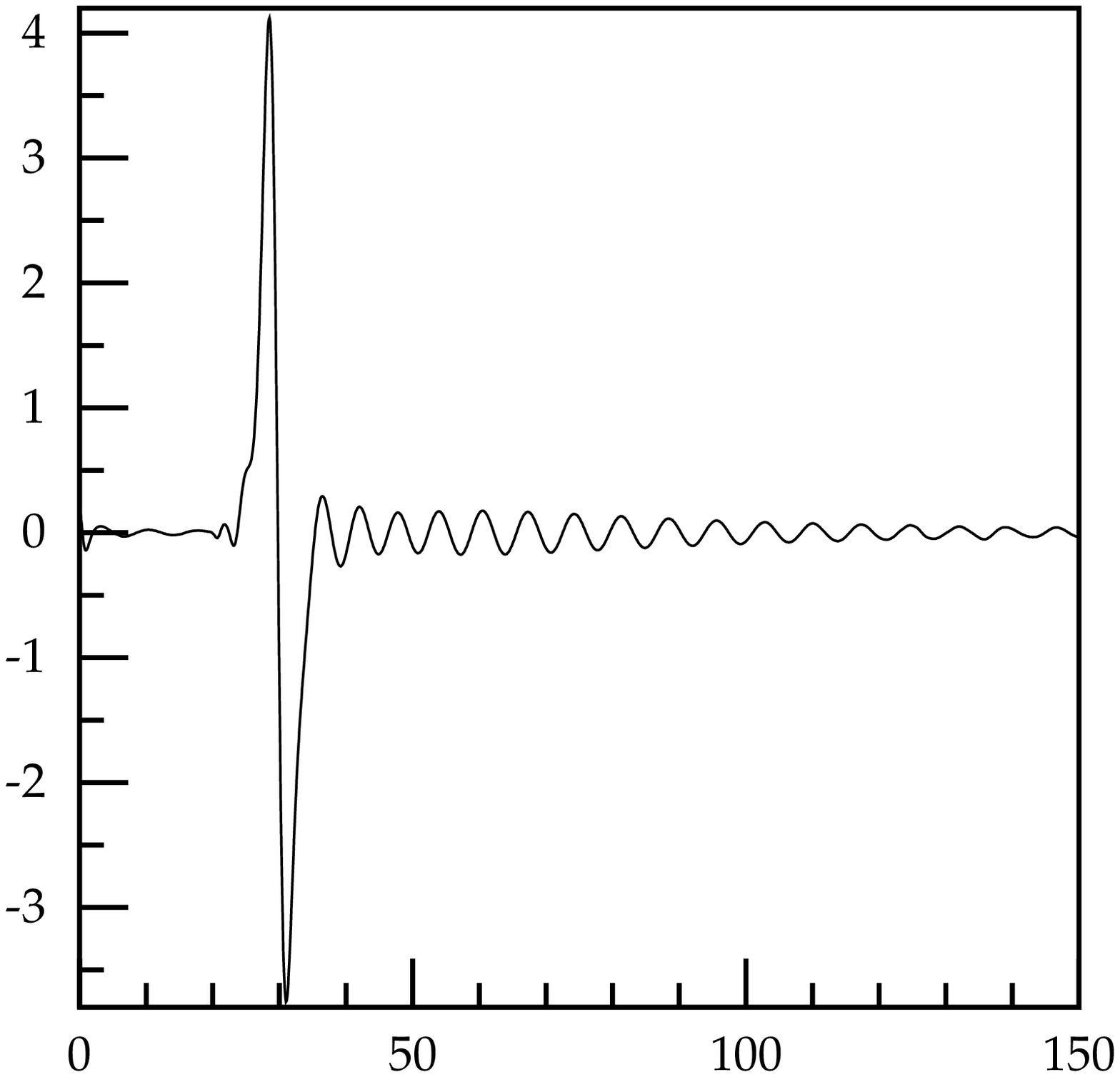}
	 \includegraphics[angle=0,width=3.5cm]{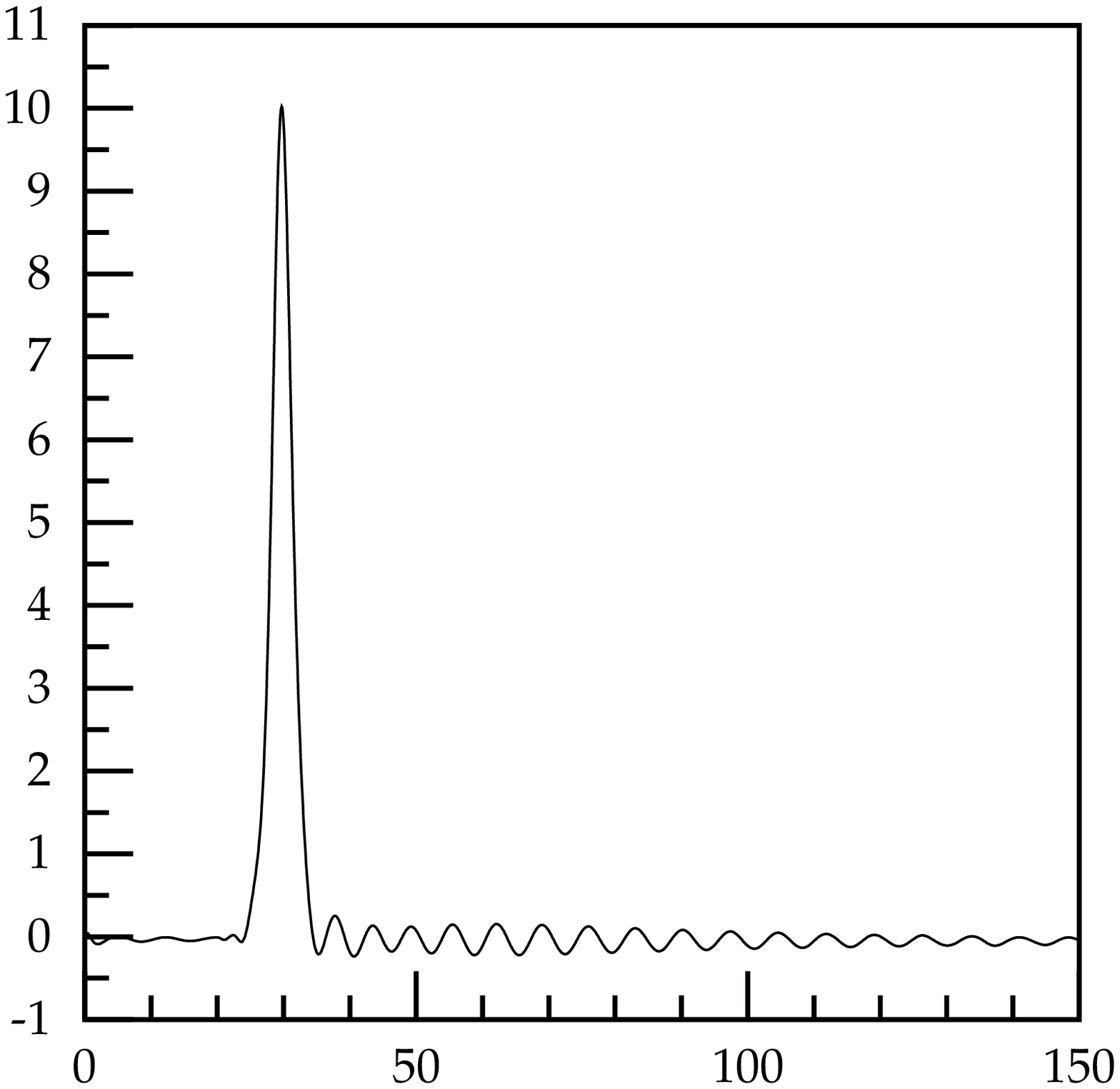}
         \includegraphics[angle=0,width=3.5cm]{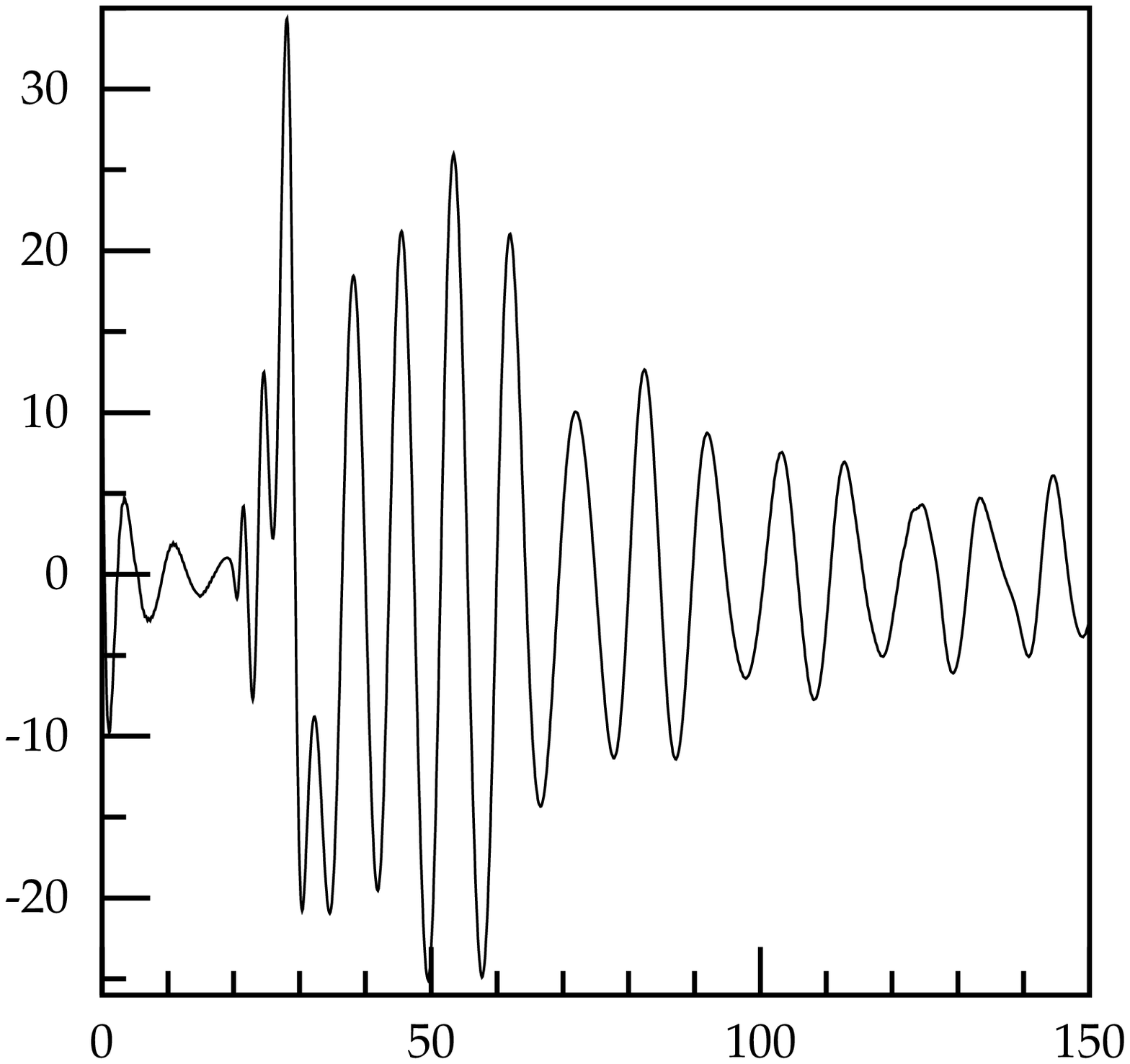}
	 \includegraphics[angle=0,width=3.5cm]{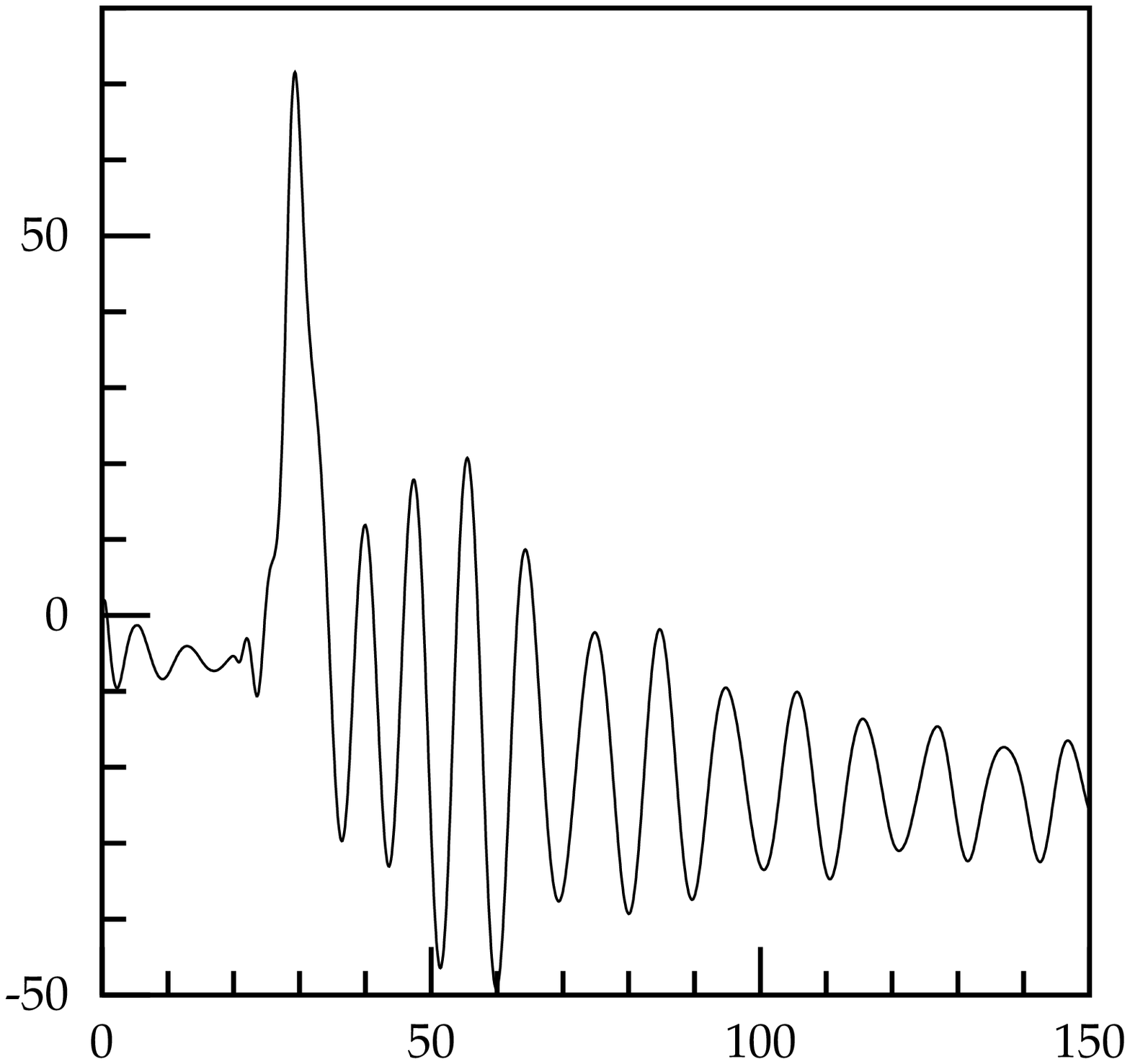}
    \begin{quote}
	\caption[AS]{Anomalies and the corresponding integrated anomalies (from left to right
and then down) for $n=2.0$,
$n=1.99$, $n=2.01$, $n=1.98$, $n=2.10$ and $n=3.00$} 
	\label{fig:Fig6a}
    \end{quote}
\end{figure}

From these results we see very clearly that for all values of $n$
(with the exception of $n=3$) the integrated anomaly is approximately
zero. This supports our analytical results and it shows 
that (for small $\ve$) the unintegrated total anomaly is
approximately proportional to $\ve=n-2$. This is supported
further by the observation that the anomaly  
changes sign as $\ve\rightarrow -\ve$. The second order (in
$\ve$)  
effects are comparable to those of the first order and the expansion in
$\ve$ clearly does not converge for $\ve\sim 1$. This last
point is very clear from the  
case of $n=3$ in which case $\ve=1$. Of course it would be nice
to understand why  
all the terms in the $\ve$ power series expansion are comparable
in magnitude; 
at this stage we have no understanding of this fact.

\subsection{Kink antikink scattering - {\bf quasi-breathers}}
\label{sec:kink-antikink}
\setcounter{equation}{0}

Next we have looked at the kink - antikink configurations and breathers.
In the sine-Gordon model we do have breathers and their analytical form 
is well known. They are in fact bound states of a kink and an antikink.
This is all well known; what is perhaps less known, is that one can generate
breathers by taking a kink and an antikink and placing them not too
close to each  
other and then let the configuration evolve in time. As the kink and the antikink attract
they move towards each other,
 alter their shape and, at the same time, emit some 
radiation and become a breather. Interestingly, they do not annihilate
but do form a breather. If we then absorb the energy at the boundaries
the system stabilises and essentially stops emitting further energy as
the fields have taken the shape of a breather which is a time dependent
solution of the model.

It is sometimes thought that the existence of breathers
and of other similar configurations (wobbles etc) is, at least in part,
 associated with the integrability of the sine-Gordon model. Actually, 
as we have stressed this before, the models of Bazeia et
 al \cite{Bazeia} do not appear to be integrable for any $n$ other than 2; so we have decided to apply 
our procedure  to look at configurations of a kink and an antikink for other values of $n$.
However, before we discuss some of our results obtained in such cases  let us first 
 present them for the sine-Gordon model.

 In fig. 7 we present the time dependence of the energy of the field configuration
which involved a kink and an antikink initially placed at $d=\pm 3$ (fig 7a)
and $d=\pm 5$ (fig 7b). In fig 7c we present the plot of the time dependence of the position 
of the kink (or antikink) for $x<0$. 

\begin{figure}[tbp]
    \centering
	\includegraphics[angle=0,width=4.5cm]{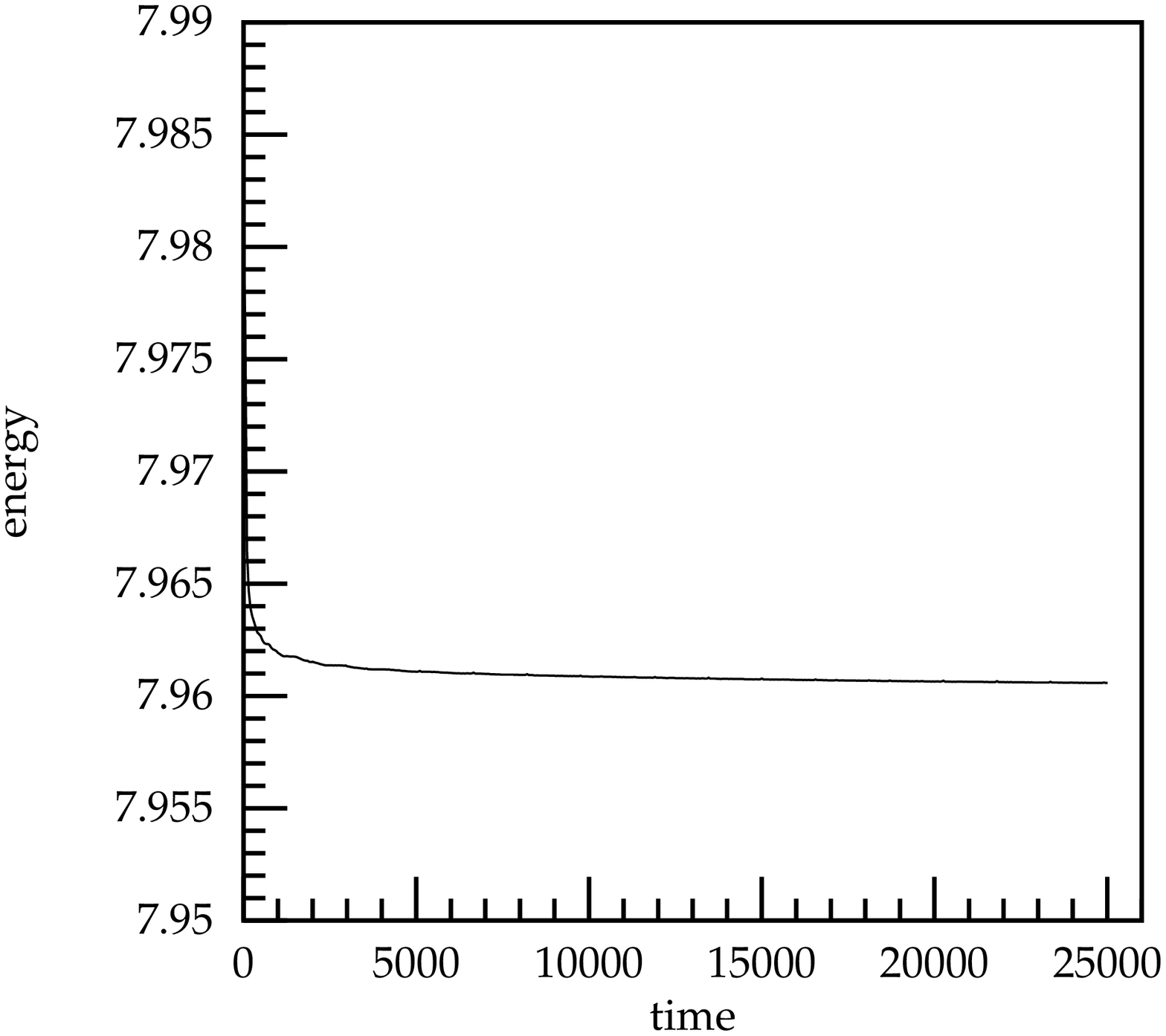}
	 \includegraphics[angle=0,width=4.5cm]{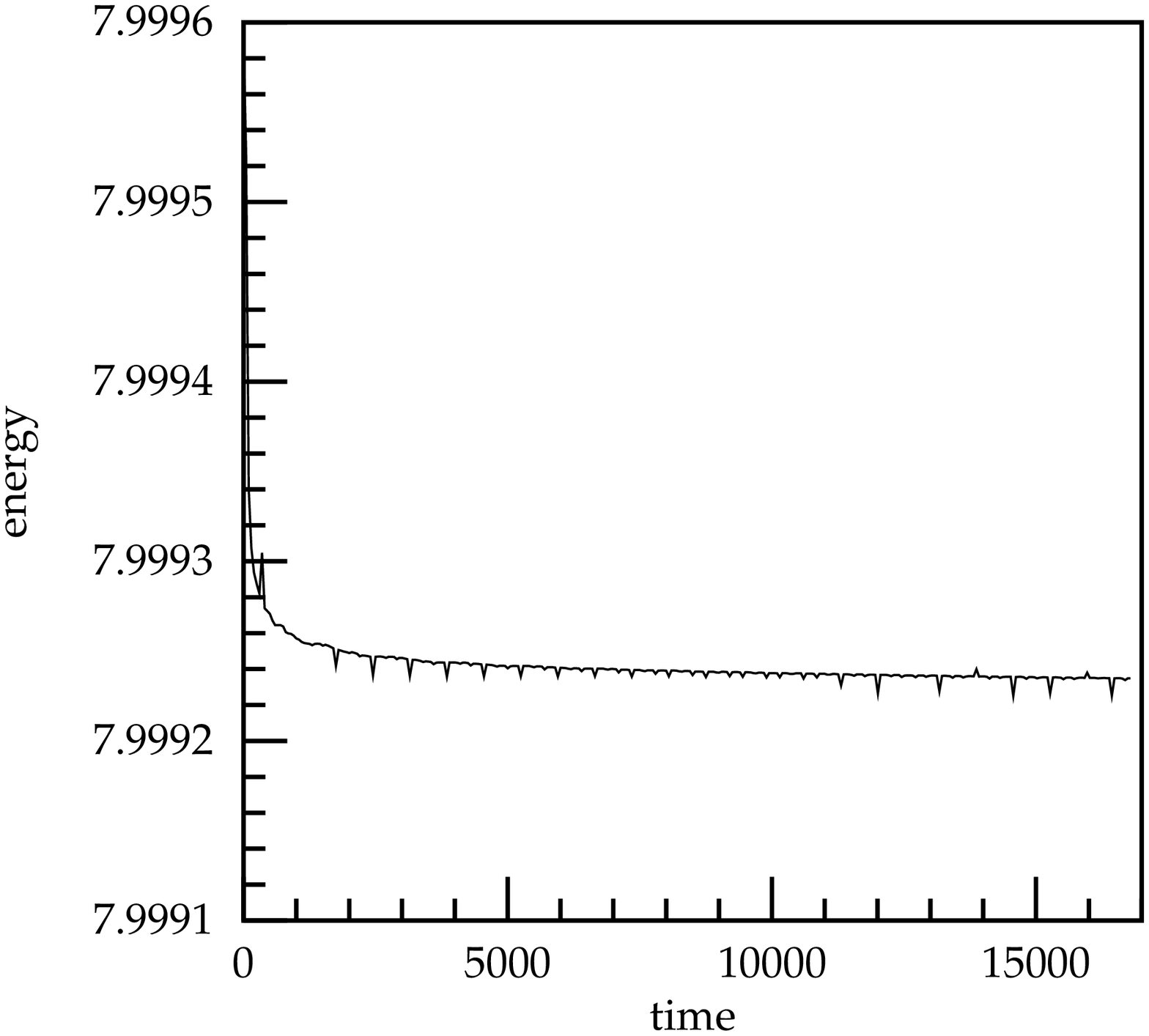}
	 \includegraphics[angle=0,width=4.5cm]{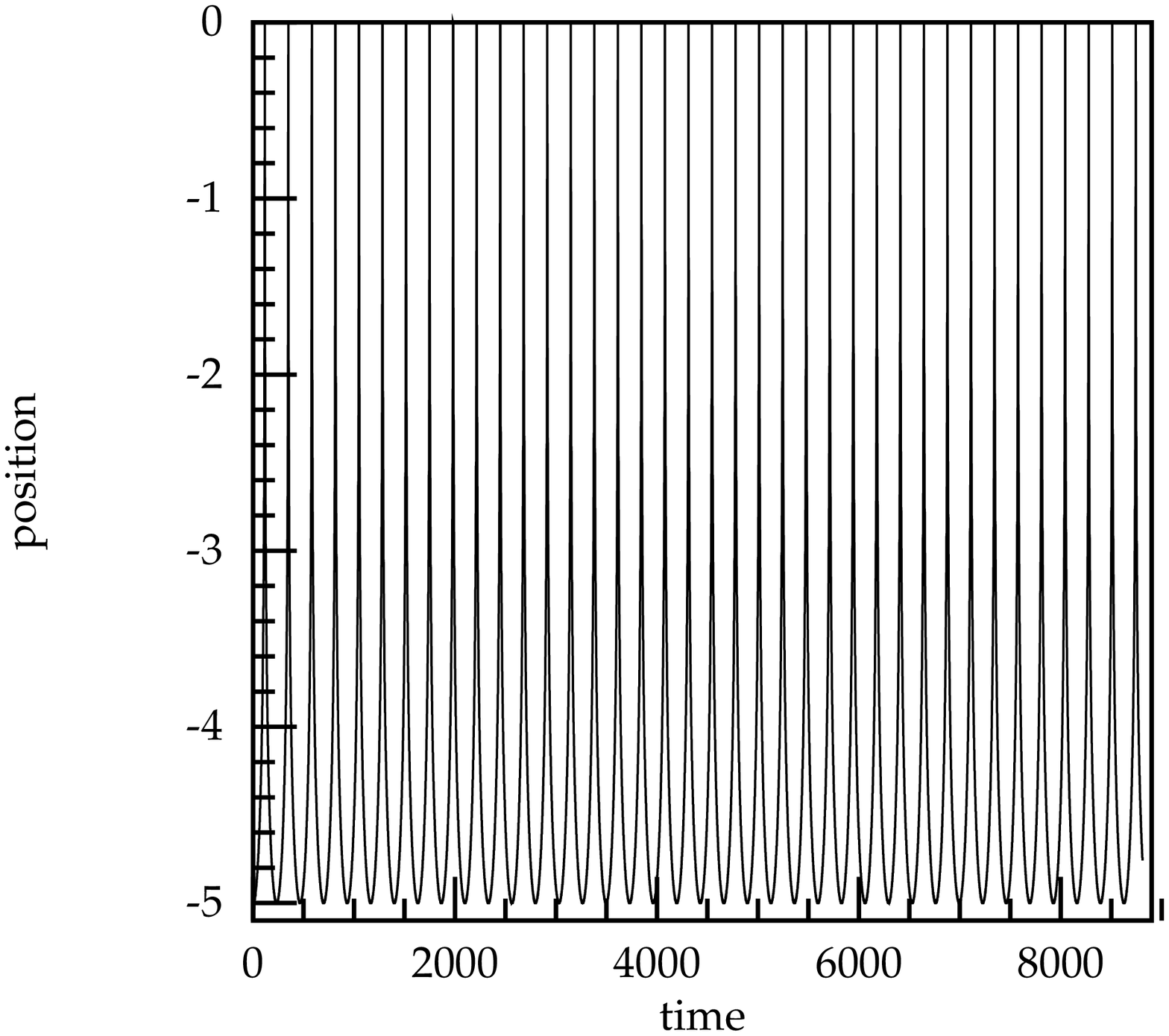}
    \begin{quote}
	\caption[AS]{(ab) Time dependence of total energies; (c). Position of the kink.} 
	\label{fig:Fig7}
    \end{quote}
\end{figure}

We note that in both cases (and also 
in other cases we have looked at, for which the results are not presented here) 
we have a very small initial drop of the total energy
of the configuration to the energy of the resultant breather.
The final energy of the breather is given by {$E=2E_0\sqrt{1-\omega^2}$}
and its frequency {$\omega$} is  related to the intial extend given by {$d$}.

Next, we repeated the same procedure of generating 
breathers for  field configurations corresponding to other values of $n$. 
In fig 8. we present our results for two values of $n$, namely $n=1$ and $n=3.1$,
in which the kink and the antikink were initially placed at $d=\pm4.0$. 
Our plots give the time dependence of the total energy of the configuration
(after the absorption at the boundaries has eliminated the radiation reaching
the boundaries).
\begin{figure}[tbp]
    \centering
	\includegraphics[angle=0,width=4.5cm]{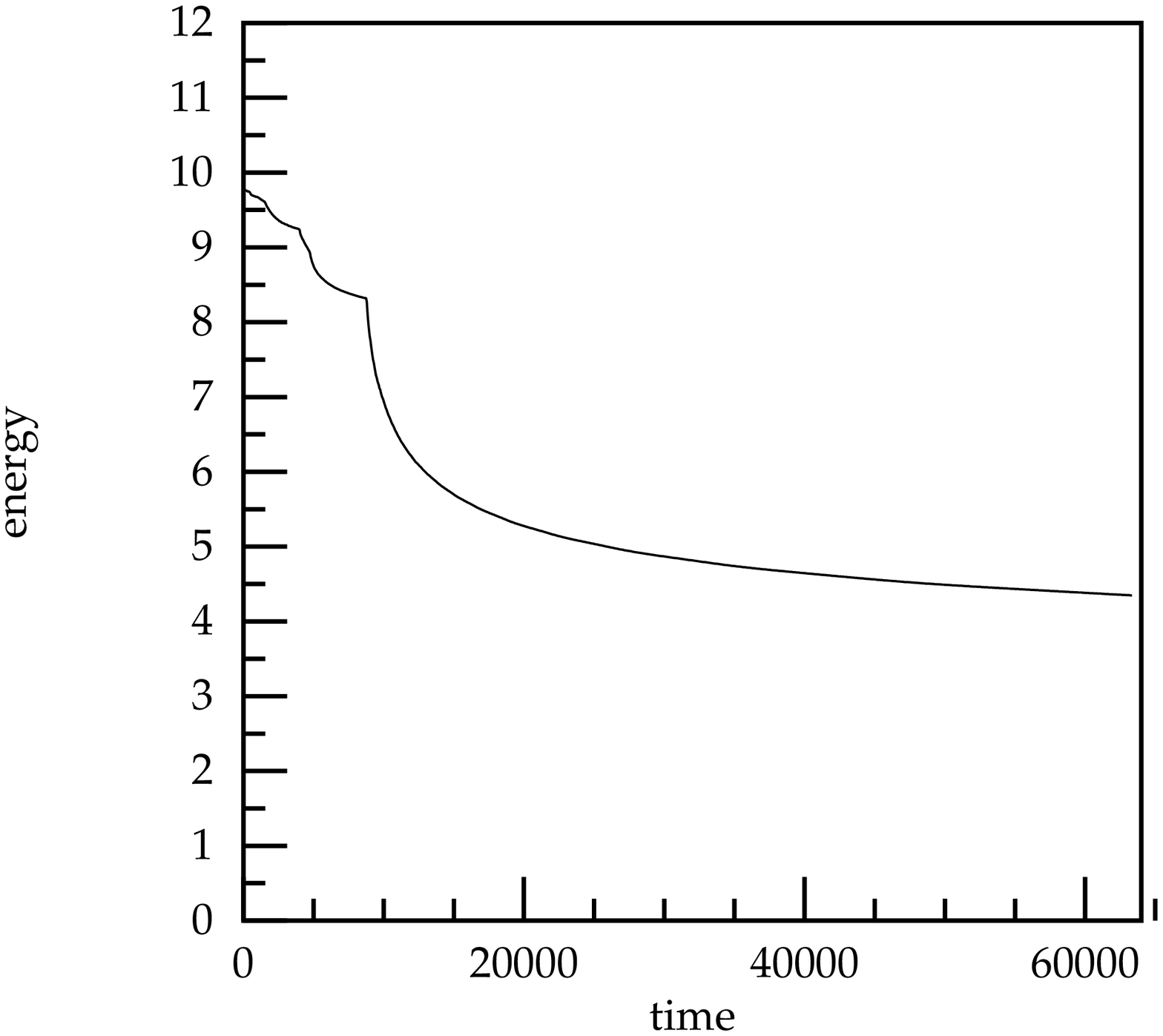}
	 \includegraphics[angle=0,width=4.5cm]{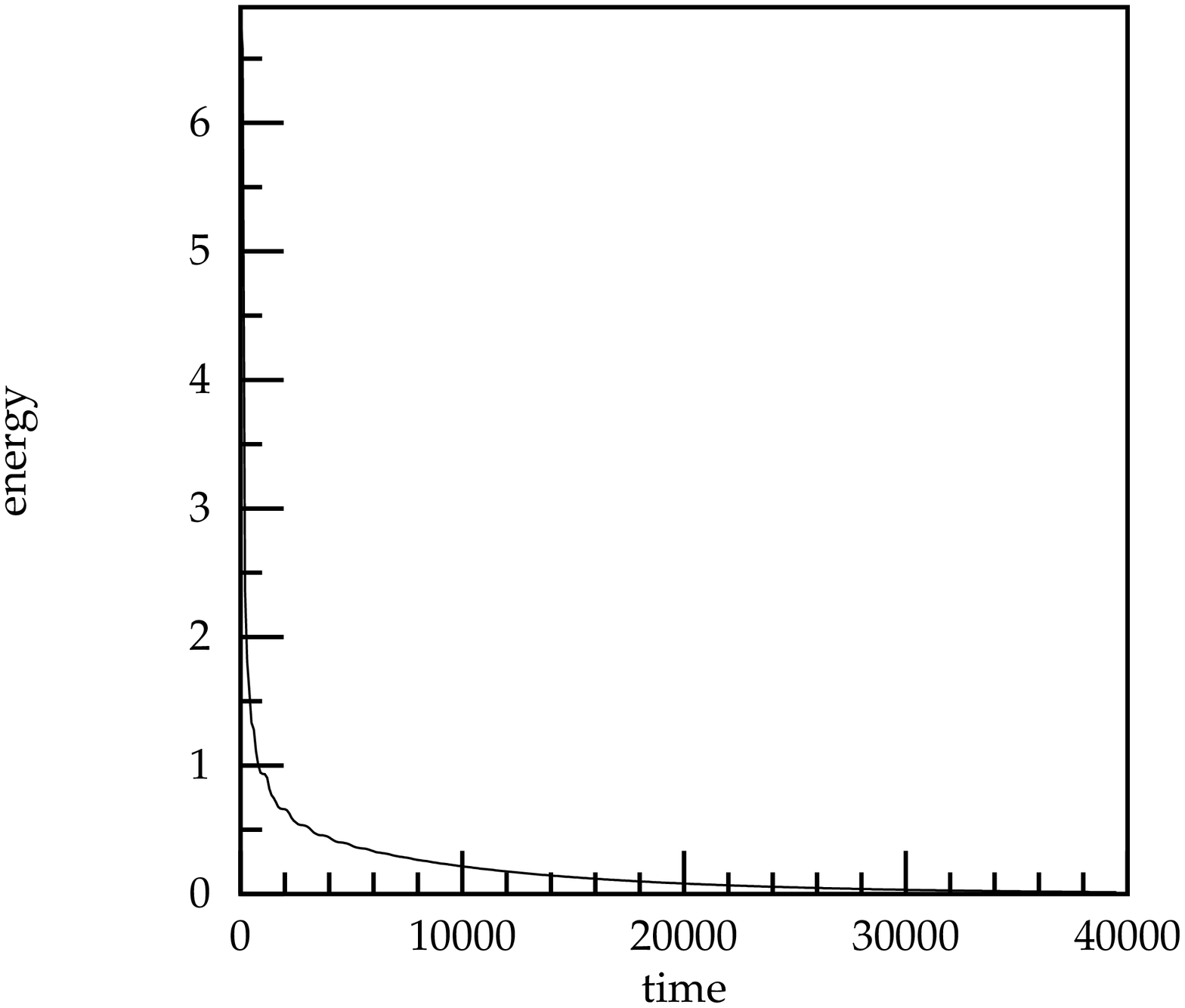}
    \begin{quote}
	\caption[AS]{Time dependence of the total energy (a) $n=1$, (b) $n=3.1$. } 
	\label{fig:Fig8}
    \end{quote}
\end{figure}

We note a fundamental difference; for $n=1$ the energy seems to `stabilise' 
around some finite nonzero value while for $n=3.1$ it quickly goes to zero.

We have performed many simulations (for different values of $n$
and for different distances between kinks and antikinks) running them for very long times. 
We have found that for some values of $n$ the fields annihilate very quickly;
while for the others the fields evolved towards breather-like configurations.
This was not much dependent on the distance between the initial kinks
but depended much more on $n$. In fact, as the distance $d$ increased the 
whole process, like for the sine-Gordon model, was slower, the initial
radiation was smaller and the generated breather was larger (and so its
oscillations were slower). Looking at the dependence on $n$ it was clear
that the closer $n$ was to 2 the more stable the breather was (this was true from about 
 $\sim 0.8$ to around $\sim 2.8$); and for some values 
of $n$ (very close to 2, like $2.01$ or so) the resultant configuration
was almost indistinguishable from a breather. In fact, in all such cases,
the energy kept decreasing but this decrease was infinitesimal.
Thus we could say that we had a {\bf quasi-breather}   ({\it i.e.} a long-lived 
breather). As the lifetime of such a quasi-breather could be counted in 
millions of units of time, such fields, for practical (but not purely mathematical)
reasons were not very different from a breather.

In fig. 9 we present some results of our simulations (for $n=2.01$) which demonstrate the existence
of our quasi-breathers. In fig 9ab we present the plots of the field configuration
for two values of $t$, namely $t=355500$ and $t=356300$. 
We see that the fields look very much like those of the $n=2$ breather.
\begin{figure}[tbp]
    \centering
	\includegraphics[angle=0,width=4.5cm]{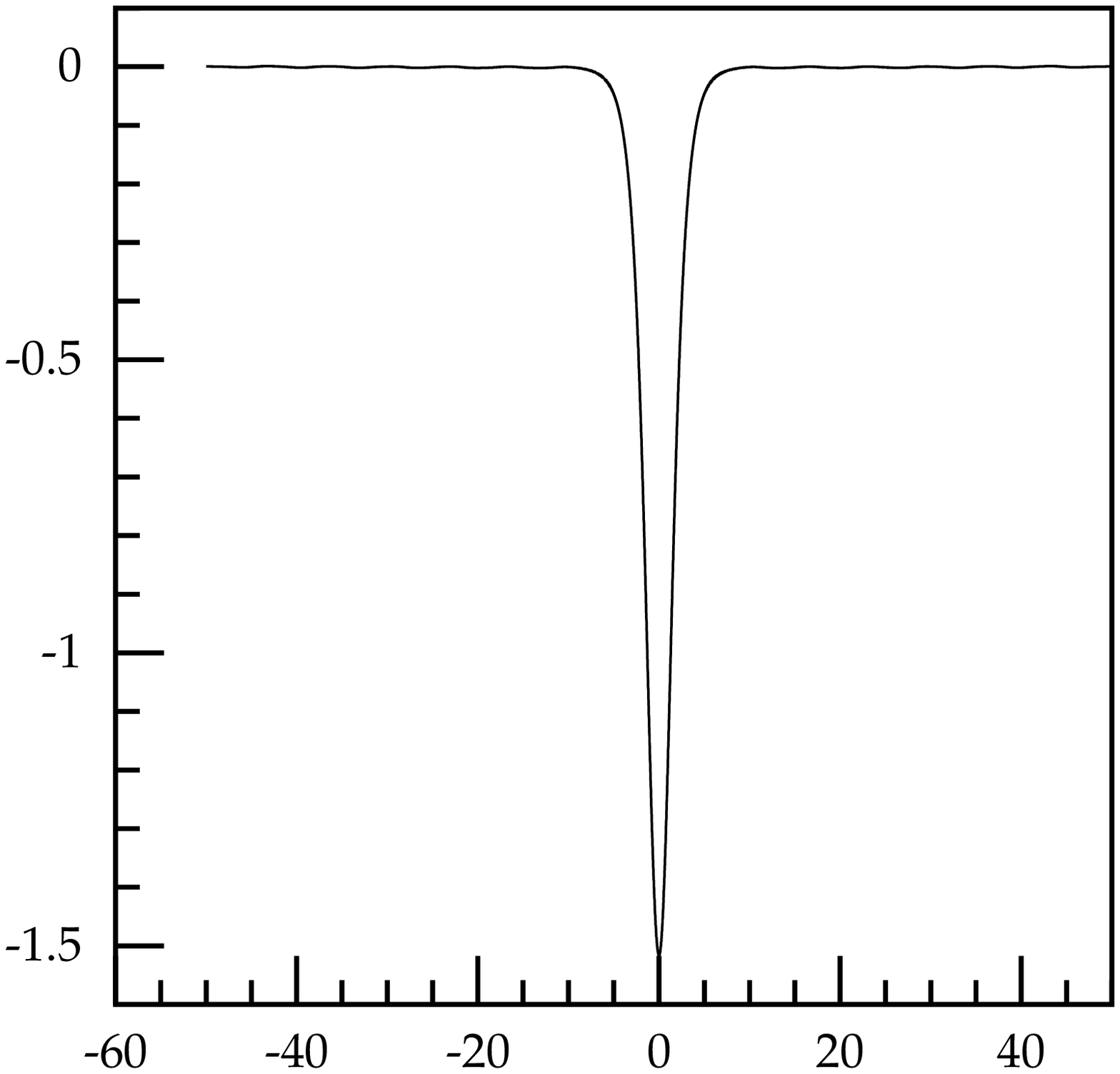}
	 \includegraphics[angle=0,width=4.5cm]{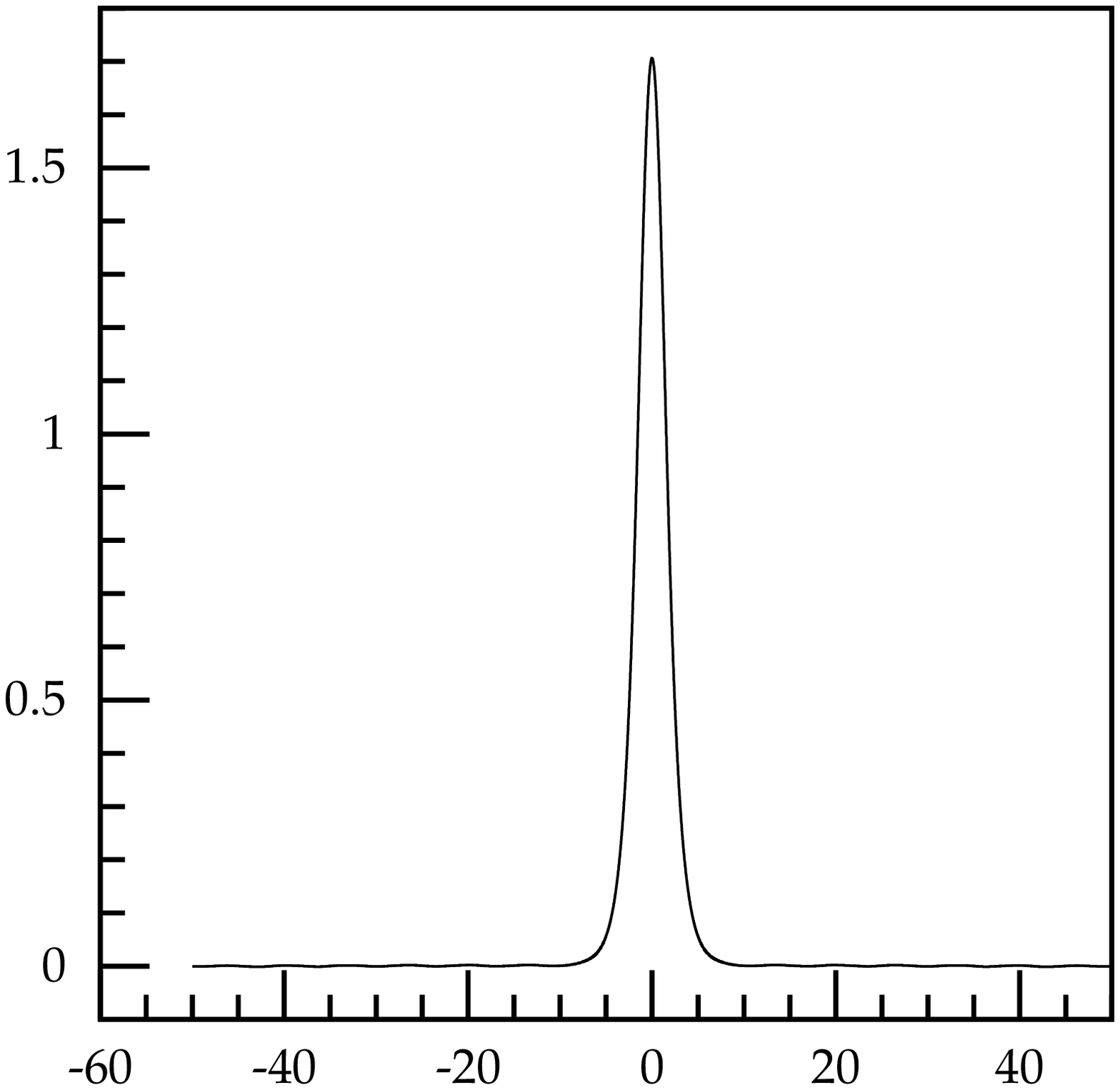}
    \begin{quote}
	\caption[AS]{Field configurations (of $n=2.01$) for (a)
          $t=355500$ and (b) $t=356300$}  
	\label{fig:Fig9}
    \end{quote}
\end{figure}

In fig. 10 ab and c we present the time dependence of the energy of
the configuration on $t$, 
a detail of this dependence at large $t$ and the time dependence of
the value of field 
at $x=0$. 
Note the extremely large values of $t$ in the plots of the energy
density. Note also the irregularity 
of the energy decrease. The energy gradually appears to decrease less
and less and then 
suddenly drops and changes its slope of decrease. It then continues in
the same way  
until the slope changes again etc. We do not understand these changes
but, in any case, 
the total decrease of the energy is still very modest and it is clear
that the quasi-breather 
is not going to ``die'' soon.

The plot in fig. 10c presents the time variation of the field at
$x=0$. As the field is symmetric around  
$x=0$ this plot demonstrates the frequency of the oscillation. 
\begin{figure}[tbp]
    \centering
	\includegraphics[angle=0,width=4.5cm]{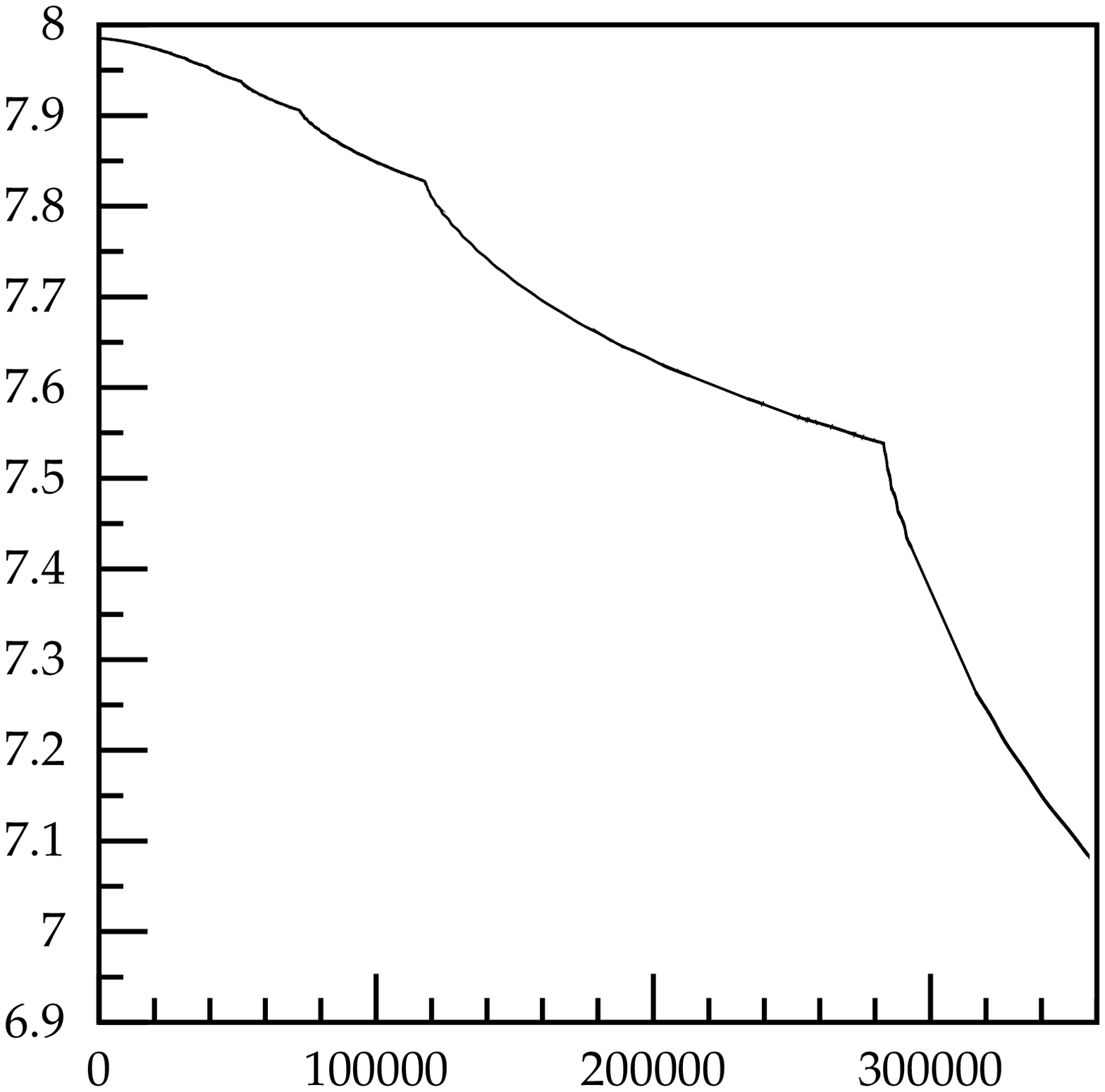}
	 \includegraphics[angle=0,width=4.5cm]{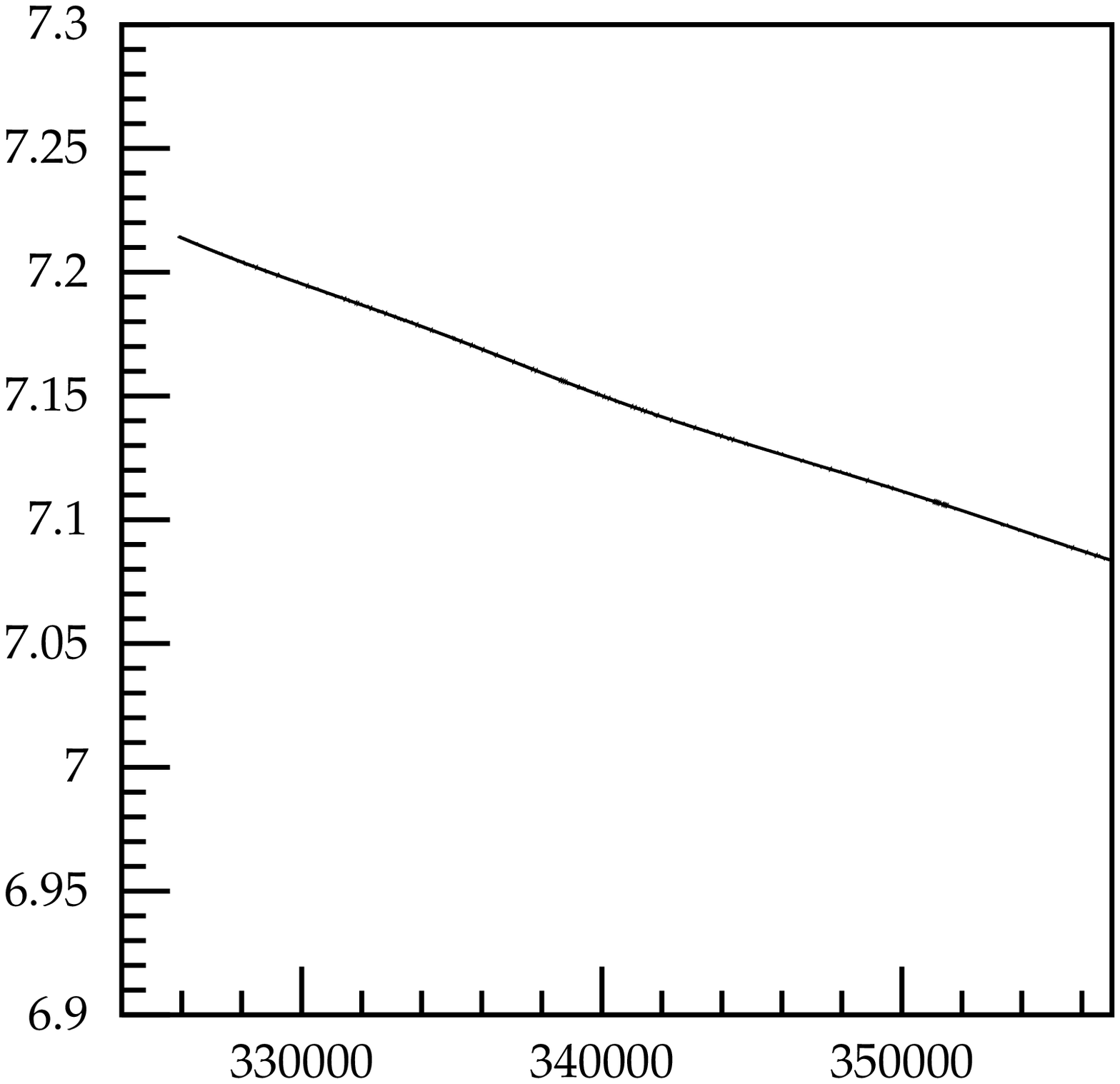}
	 \includegraphics[angle=0,width=4.5cm]{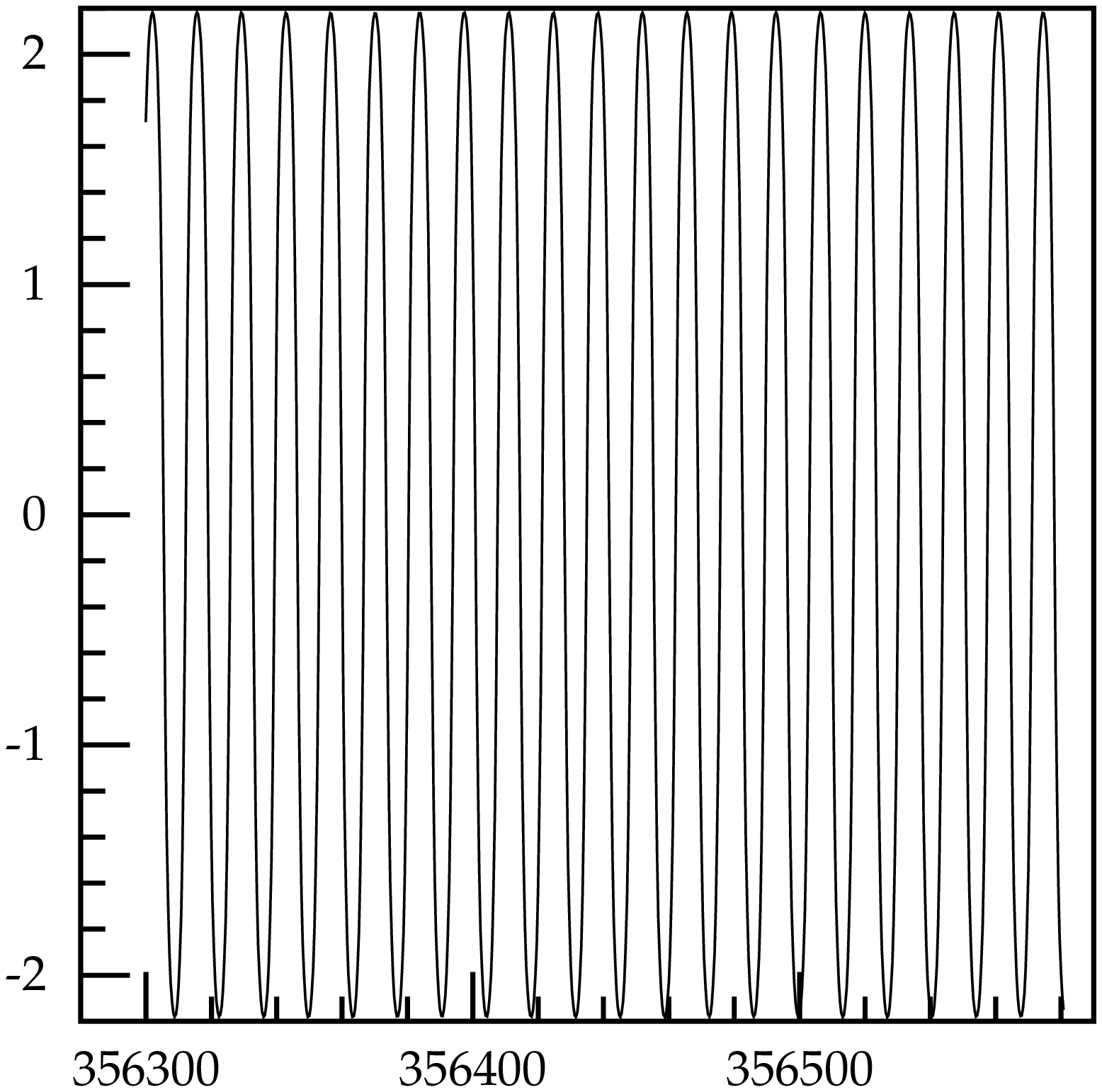}
    \begin{quote}
	\caption[AS]{The time dependence of the total energy (a), of the details of this dependence at the larger
values of $t$ (b) and of the values of the field at $x=0$ seen in a similation of a kink 
and an antikink in the $n=2.1$ model (c).} 
	\label{fig:Fig10}
    \end{quote}
\end{figure}

We have performed similar simulations starting with the initial kinks and antikinks 
at other distances from each other and for other values of $n$. The results were qualitatively
the same; the further the initial structures were - the slower was the decrease in energy ({\it i.e}
the longer the life-time of the breather). The same was true when we considered $n$ further away from 
$n=2$. This once again suggests that  the models for $n\ne 2$ (but close to 2) are quasi-integrable as discussed in the previous sections. 
Hence we have also looked at the behaviour of the anomaly for our quasi-breathers.

In fig. 11 ab we present the plots of the anomaly and the time integrated anomaly 
(at later times) of our $n=2.01$ quasi-breather and in fig. 11 cd the similar plots
for $n=2.7$.
\begin{figure}[tbp]
    \centering
	\includegraphics[angle=0,width=3.5cm]{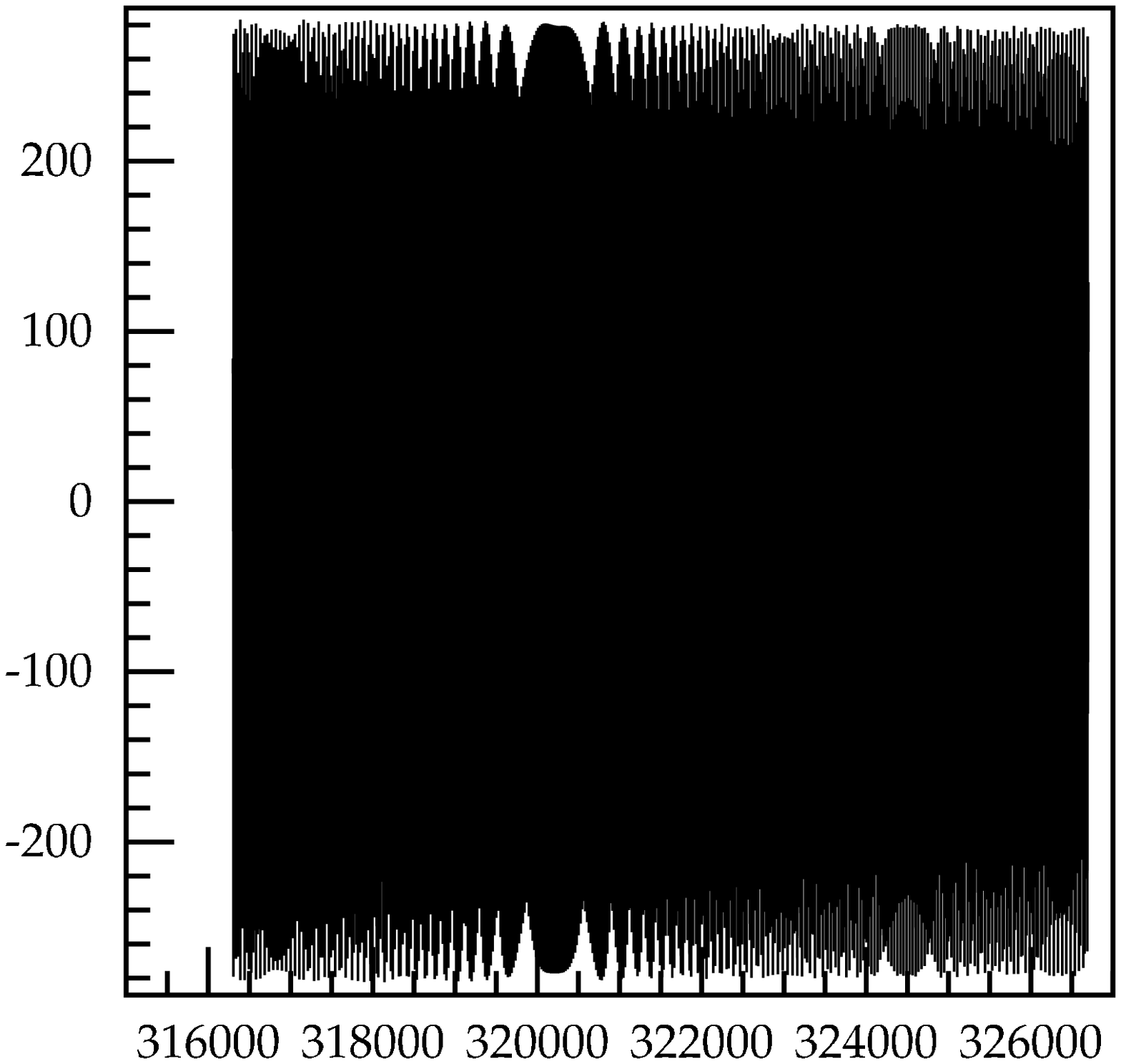}
	 \includegraphics[angle=0,width=3.5cm]{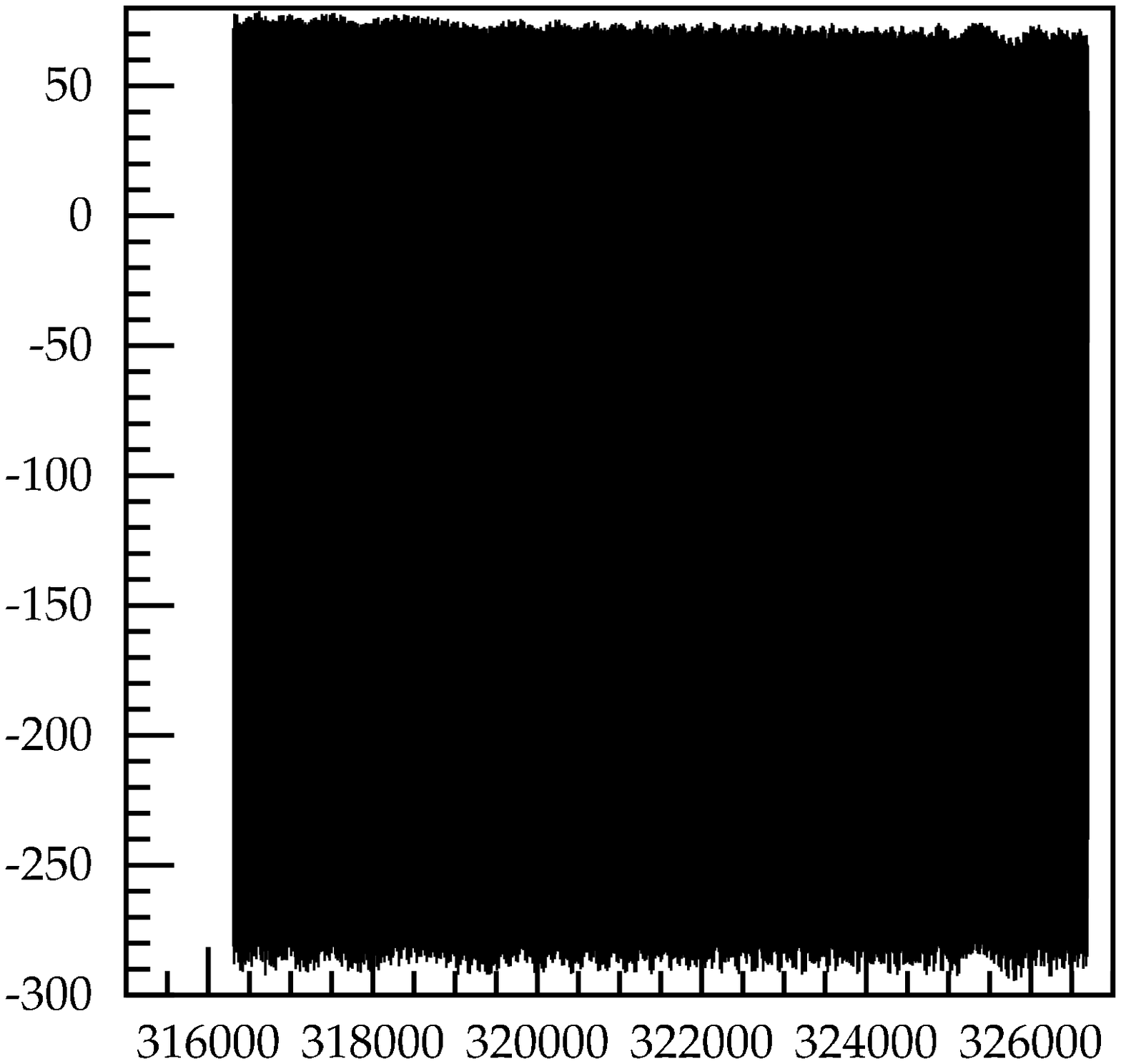}
	\includegraphics[angle=0,width=3.5cm]{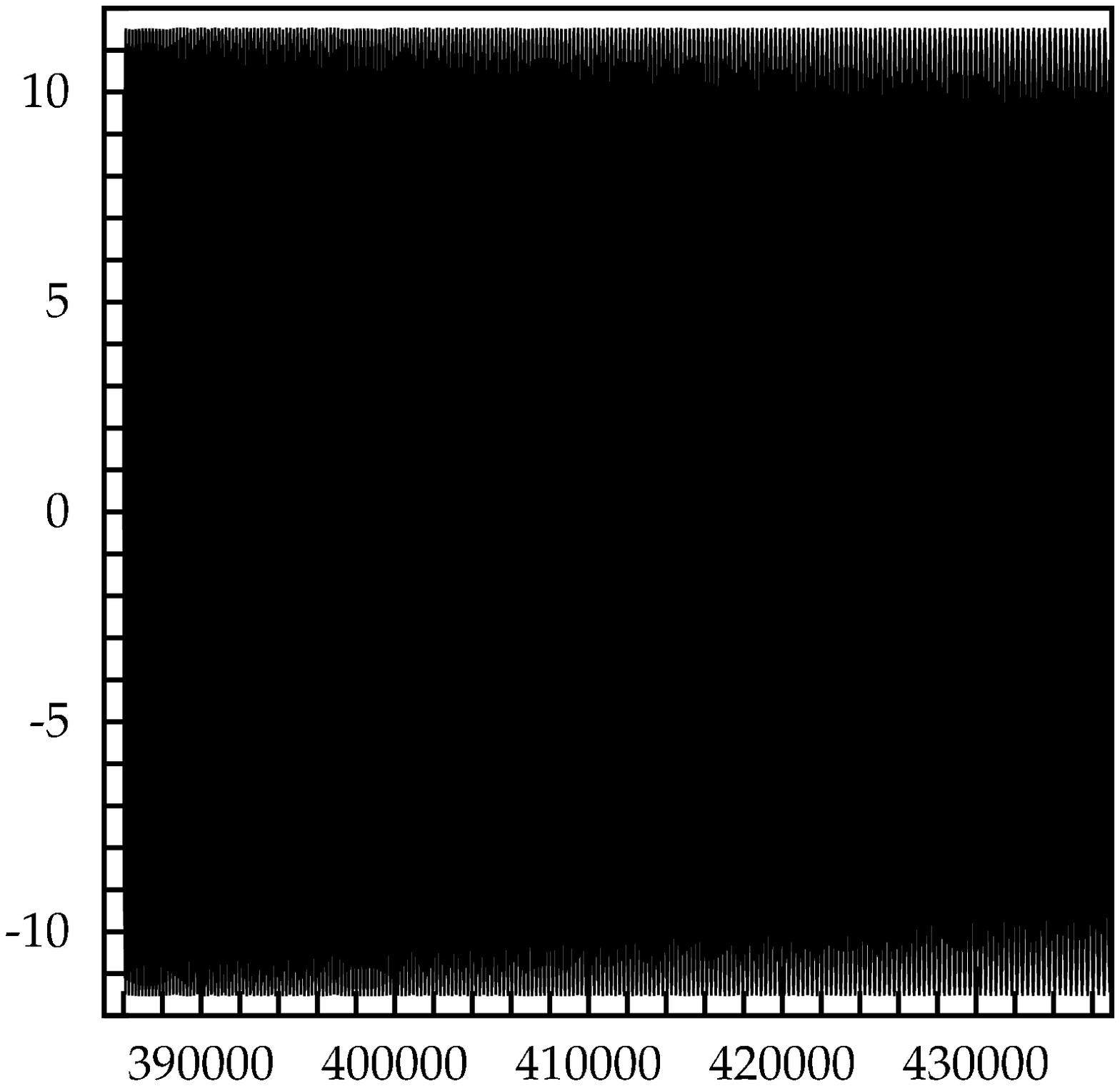}
	 \includegraphics[angle=0,width=3.5cm]{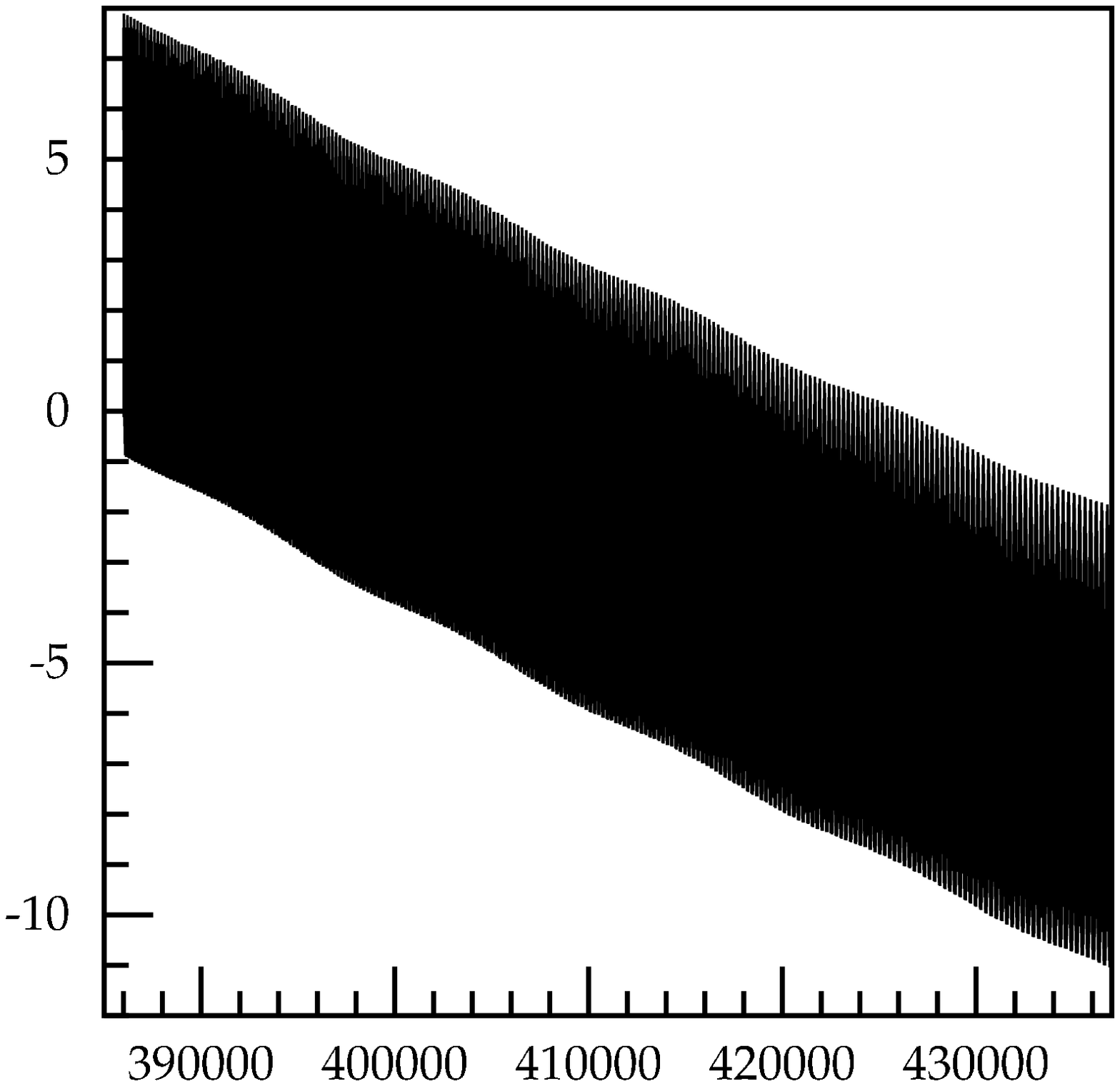}
    \begin{quote}
	\caption[AS]{The first anomaly, b) the integrated first anomaly for $n=2.01$,
c) and d) the same for $n=2.7$.} 
	\label{fig:Fig11}
    \end{quote}
\end{figure}

We see that, when compared to the two-kink case, the time integrated total anomaly is nonzero 
at all times (as the kink and antikink are bound into a pulsating quasi-breather) but when one 
looks at the time integrated anomaly over a multiple of the period of its oscillations its value is very very small ({\it i.e.} close to zero). For $n=2.01$ this value is so small that it is difficult to see that it changes at all,  for $n=2.7$ the value is small but nonzero and the value 
changes - thus the quantity corresponding to the anomaly is, strictly speaking, not conserved.

\subsection{Wobbles}
\label{sec:wobbles}
\setcounter{equation}{0}

Finally we looked at wobbles {\it i.e.}, fields involving a kink and a breather.
In the sine-Gordon model they are again well known and, in fact, one has
their analytical form. Of course, as before, we can generate them, numerically,
from field configurations involving an antikink and two kinks (or vice-versa).
However, as these configurations  have an excess of energy, which is emitted when an antikink
 a kink form a breather, this energy can be, in part, converted into
the motion of the remaining kink (or of the breather). Hence it is much harder, by comparison 
with  pure breathers (where one can exploit the symmetry
of the initial configuration), to generate non-moving wobbles. We have performed many simulations and the resultant fields sometimes were static but most of the time were moving. Clearly, the result of the simulation depends on the excess of energy - so further the initial structures were from each other the more likely there were to remain static. But this, in turn, slowed down the process of the generation of the breather. In addition, the futher $n$ was from $n=2$ the more radiation was sent out by the system and more likely it was that this radiation would set in motion 
the kink or  the breather. However, for $n$ close to 2 we did manage to obtain wobbles and in the plots given below we show some of our results.

First we present our results for $n=2$, {\it i.e.} for the sine-Gordon model.

In fig. 12 we plot the time dependence of the total and of the potential energy
seen in the simulation involving the kink, the antikink and the kink originally
located at -15.5, 0 and +15. In the following figure we exhibit the field configurations
for three values of $t$, namely for $t=0$, $t=6400$ and $t=12800$.
We note a fast decrease of the total energy over the initial period and then stability. The potential energy is virtually always close to 12 and then it decreases 
to just over 4 when the breather is `breathing', {\it i.e.} when almost all its energy
is kinetic. Note that, for the breather, the flow of energy between the kinetic and the potential 
energies is very uneven; most of the time the breather's energy is mainly potential and the 
periods over which the kinetic energy dominates are relatively short. 
\begin{figure}[tbp]
    \centering
	\includegraphics[angle=0,width=4.5cm]{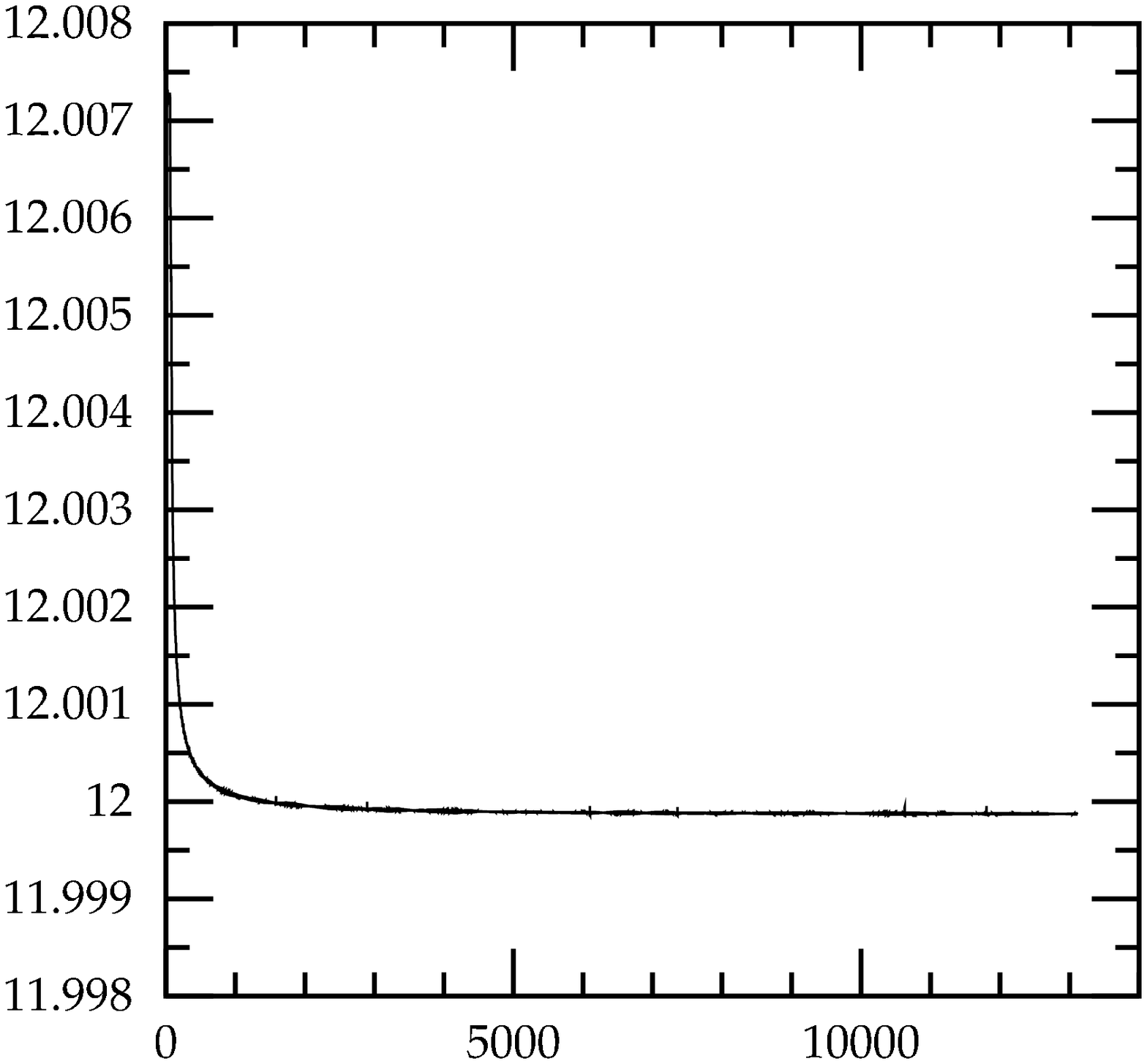}
	 \includegraphics[angle=0,width=4.2cm]{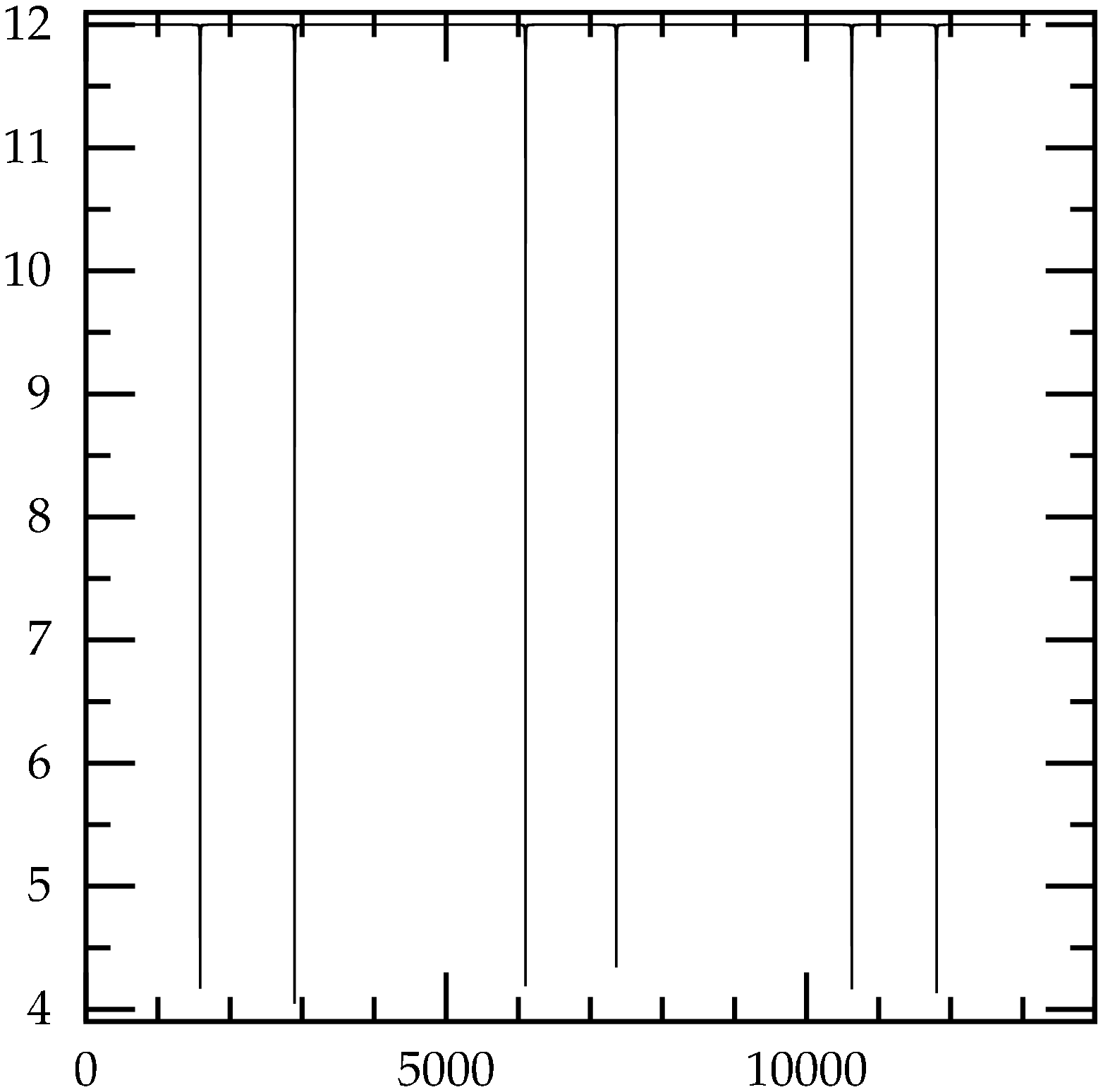}
    \begin{quote}
	\caption[AS]{a) Total energy, b) Potential energy} 
	\label{fig:Fig12}
    \end{quote}
\end{figure}

\begin{figure}[tbp]
    \centering
	\includegraphics[angle=0,width=4.5cm]{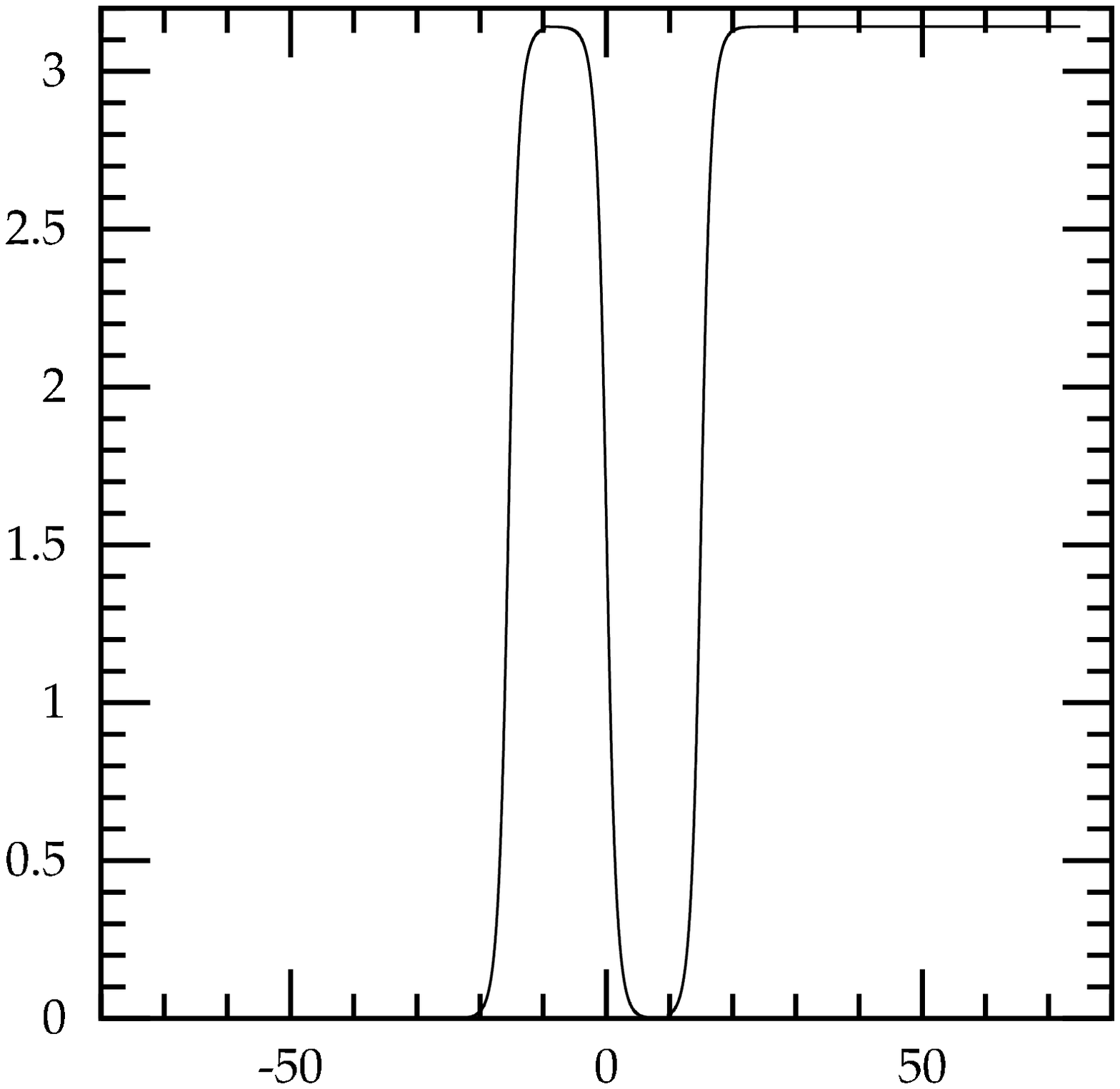}
	 \includegraphics[angle=0,width=4.3cm]{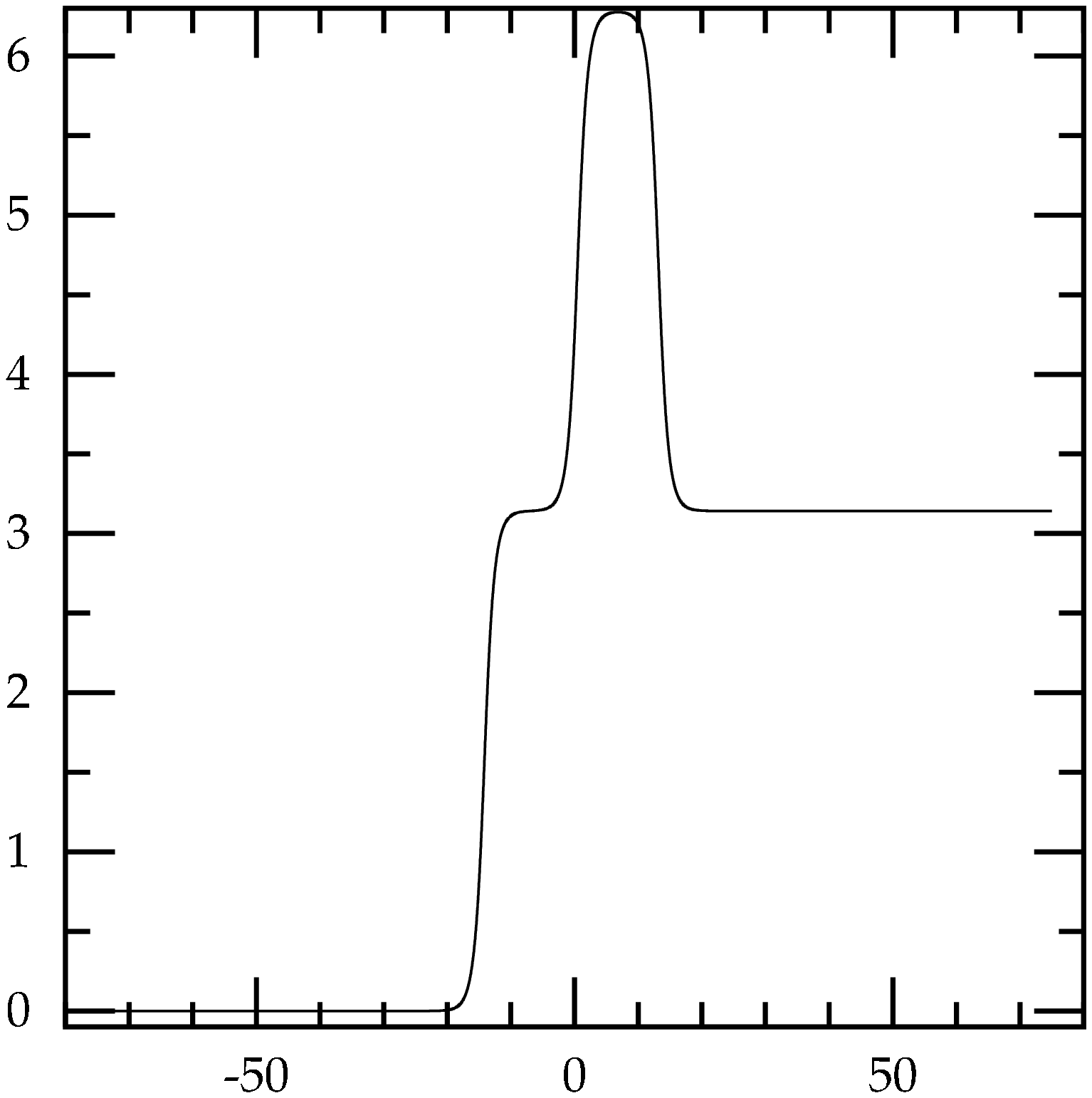}
	 \includegraphics[angle=0,width=4.5cm]{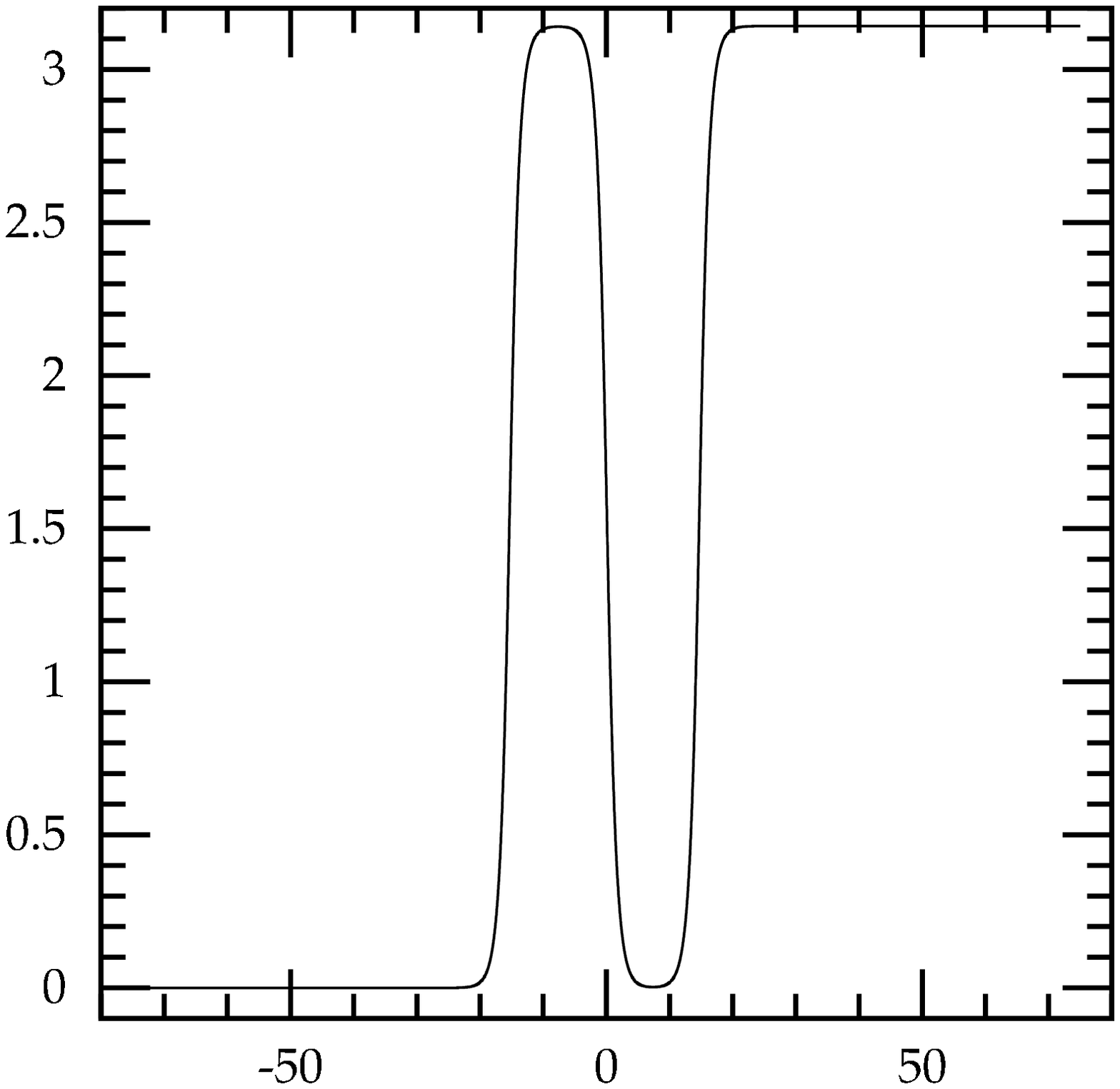}
    \begin{quote}
	\caption[AS]{a) $t=0$, b) $t=6400$, c) $t=12800$} 
	\label{fig:Fig13}
    \end{quote}
\end{figure}

Next we present our results for the case of $n\ne2$ {\it i.e.} $n=2.01$.
In this case the energy continues to decrease but this decrease is very slow.
In fig 14c we present the details of the plot of  the total energy 
for larger values of $t$. We clearly see the decrease - hence the breather
is slowly dying but its decay is very slow indeed. And fig 15 
shows the fields at some representative  values of time (for $n=2.01$). 
 It is clearly very difficult to see any fundamental difference between the wobbles
in these two systems (for $n=2$ and $n=2.01$), although as fig 14b shows, the time dependences of the breather oscillations in both cases are very different (much shorter 
for the $n=2.01$ wobble and slowly decreasing when compared to that
of $n=2.00$).

\begin{figure}[tbp]
    \centering
	\includegraphics[angle=0,width=4.5cm]{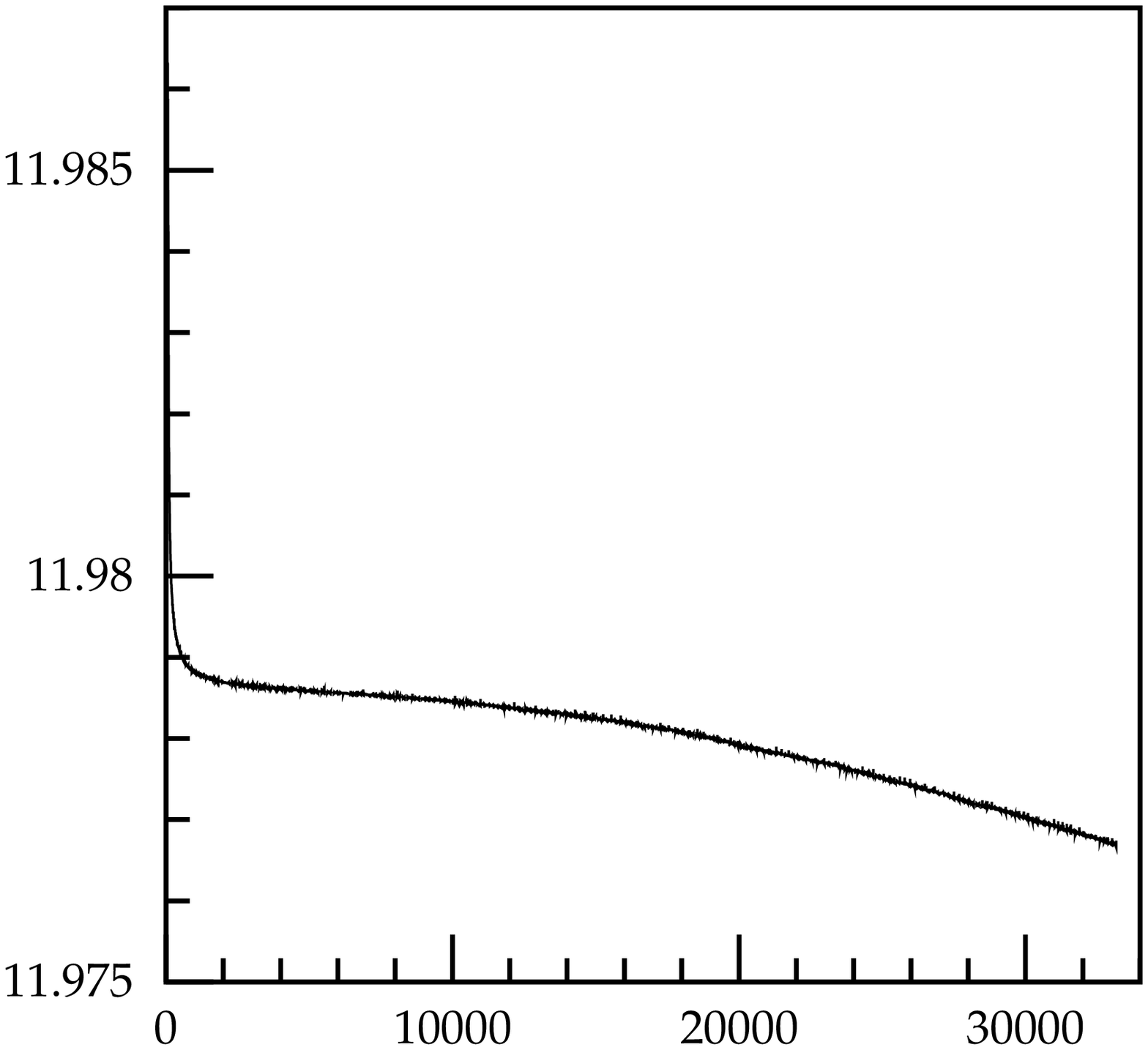}
	 \includegraphics[angle=0,width=4.5cm]{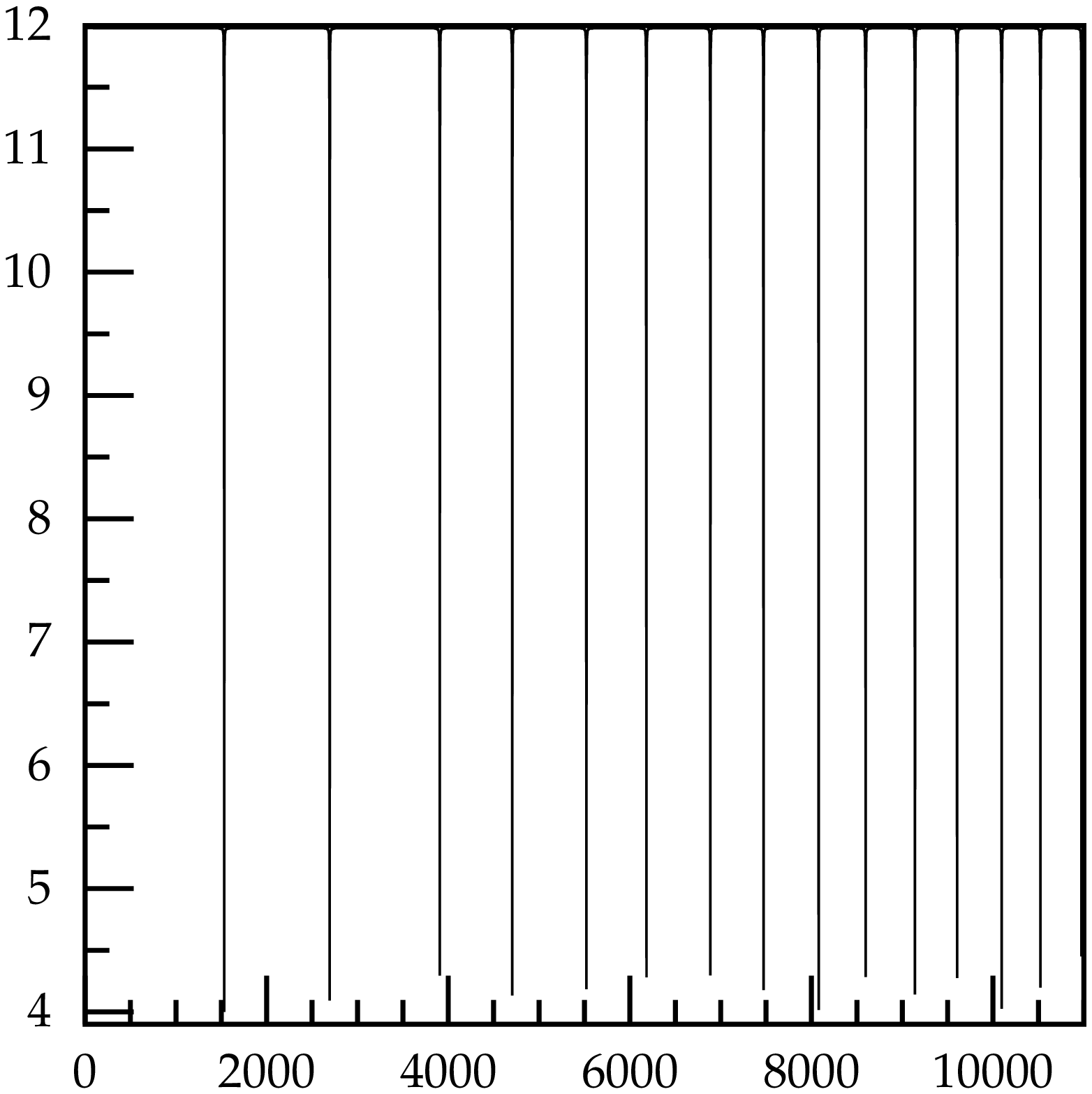}
	 \includegraphics[angle=0,width=4.5cm]{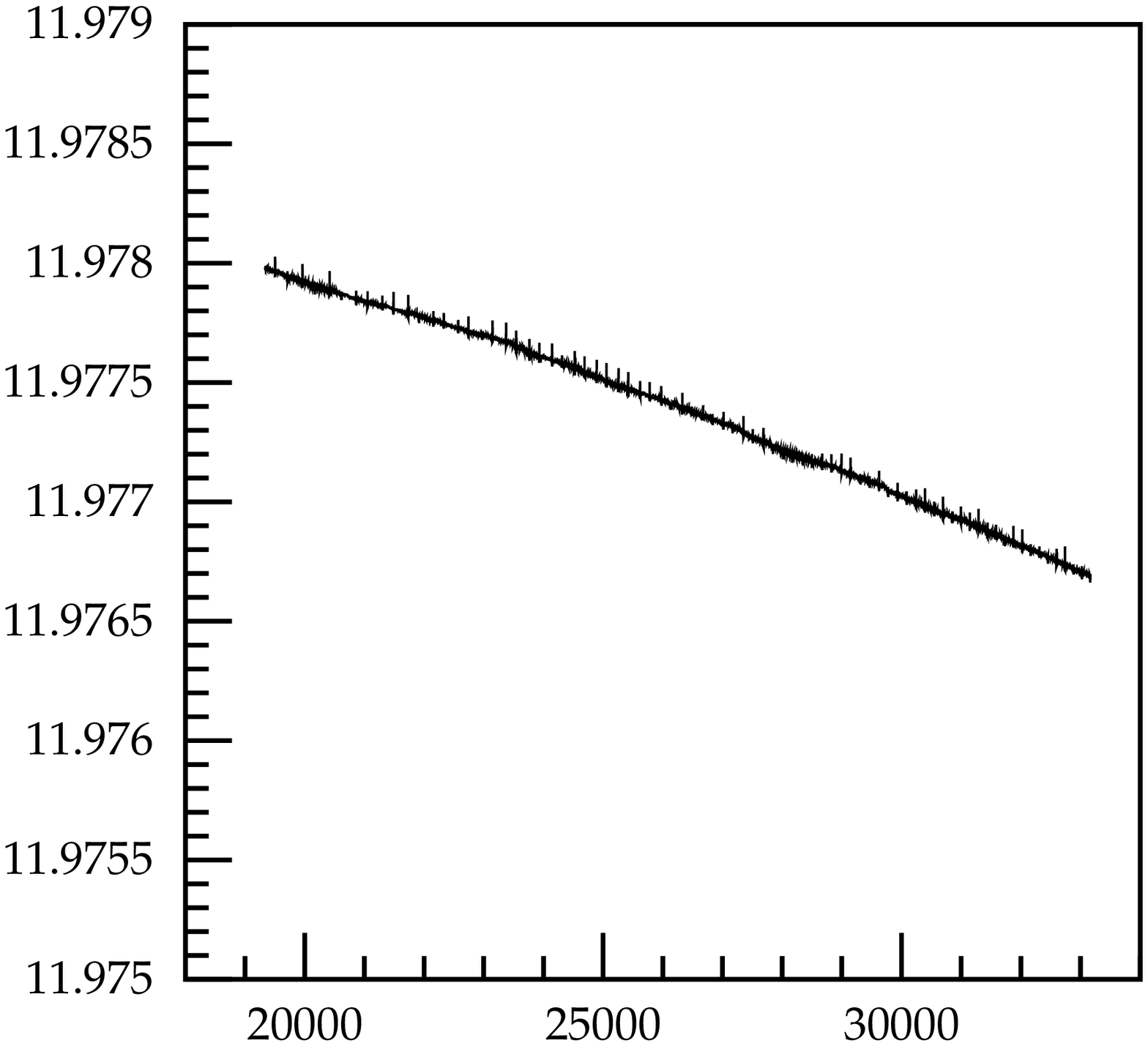}
    \begin{quote}
	\caption[AS]{a) Total energy, b) Potential energy, c) Total energy for large values of $t$ ($n=2.01$)} 
	\label{fig:Fig14}
    \end{quote}
\end{figure}

\begin{figure}[tbp]
    \centering
	\includegraphics[angle=0,width=4.5cm]{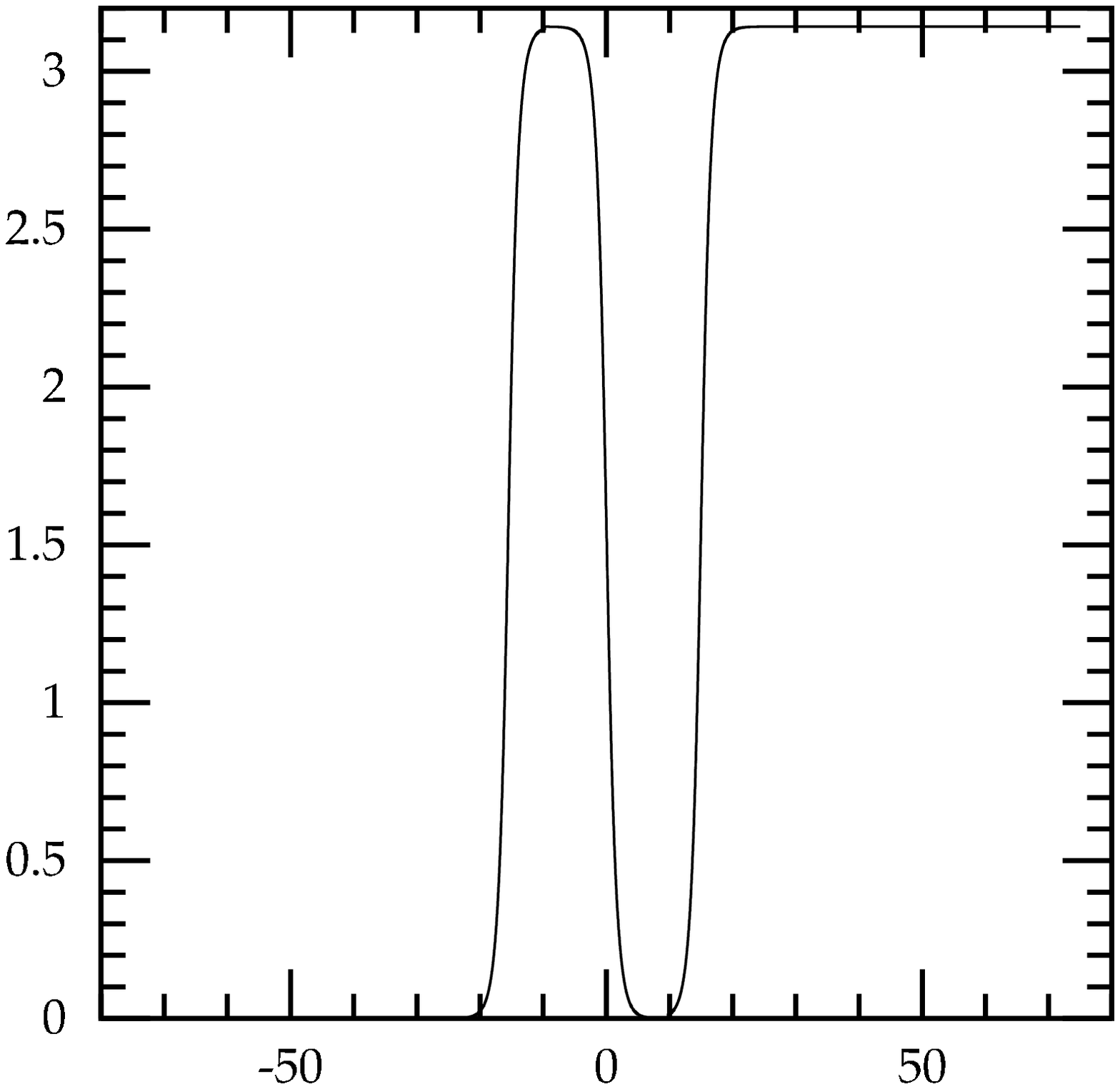}
	 \includegraphics[angle=0,width=4.5cm]{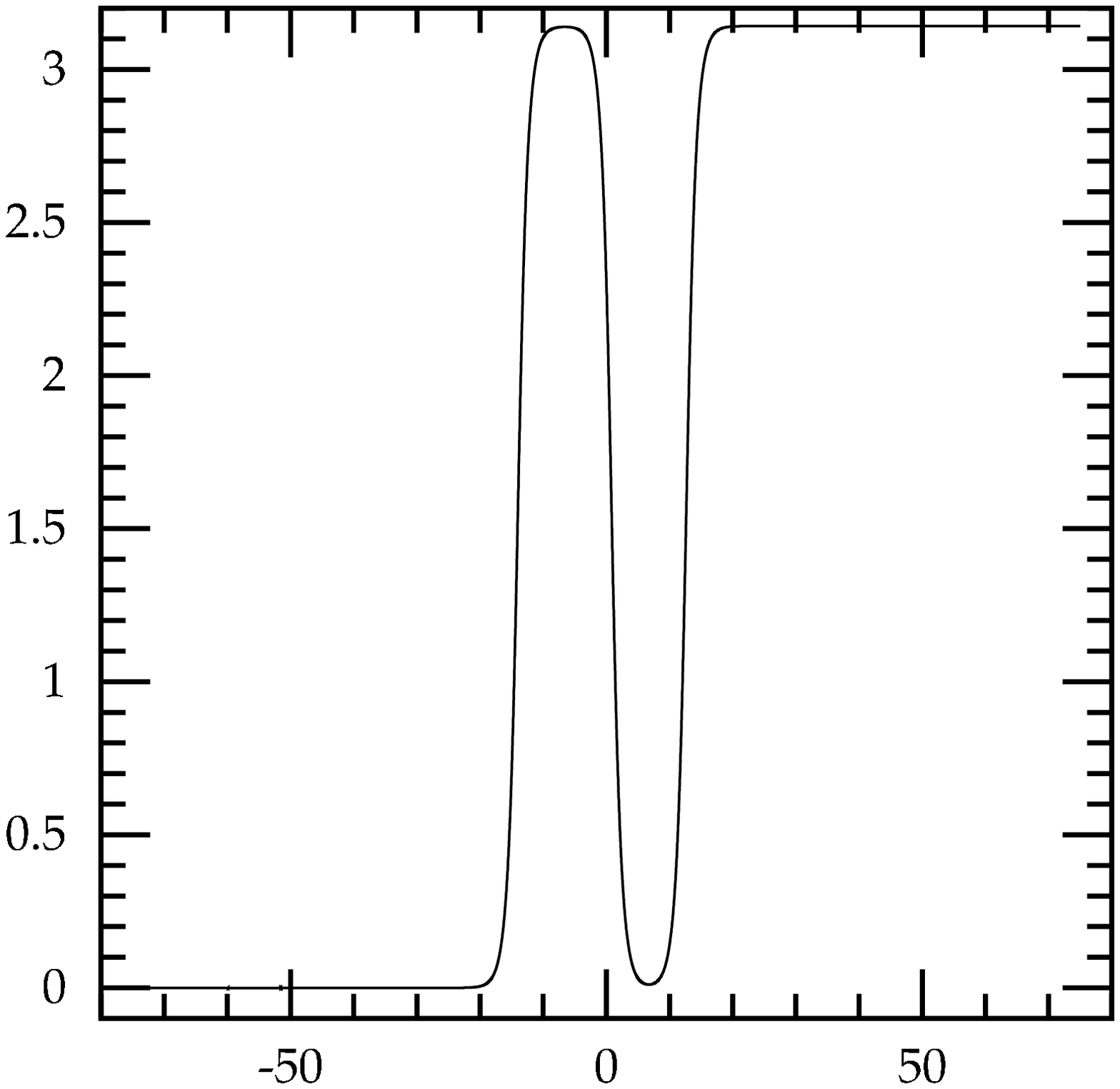}
	 \includegraphics[angle=0,width=4.4cm]{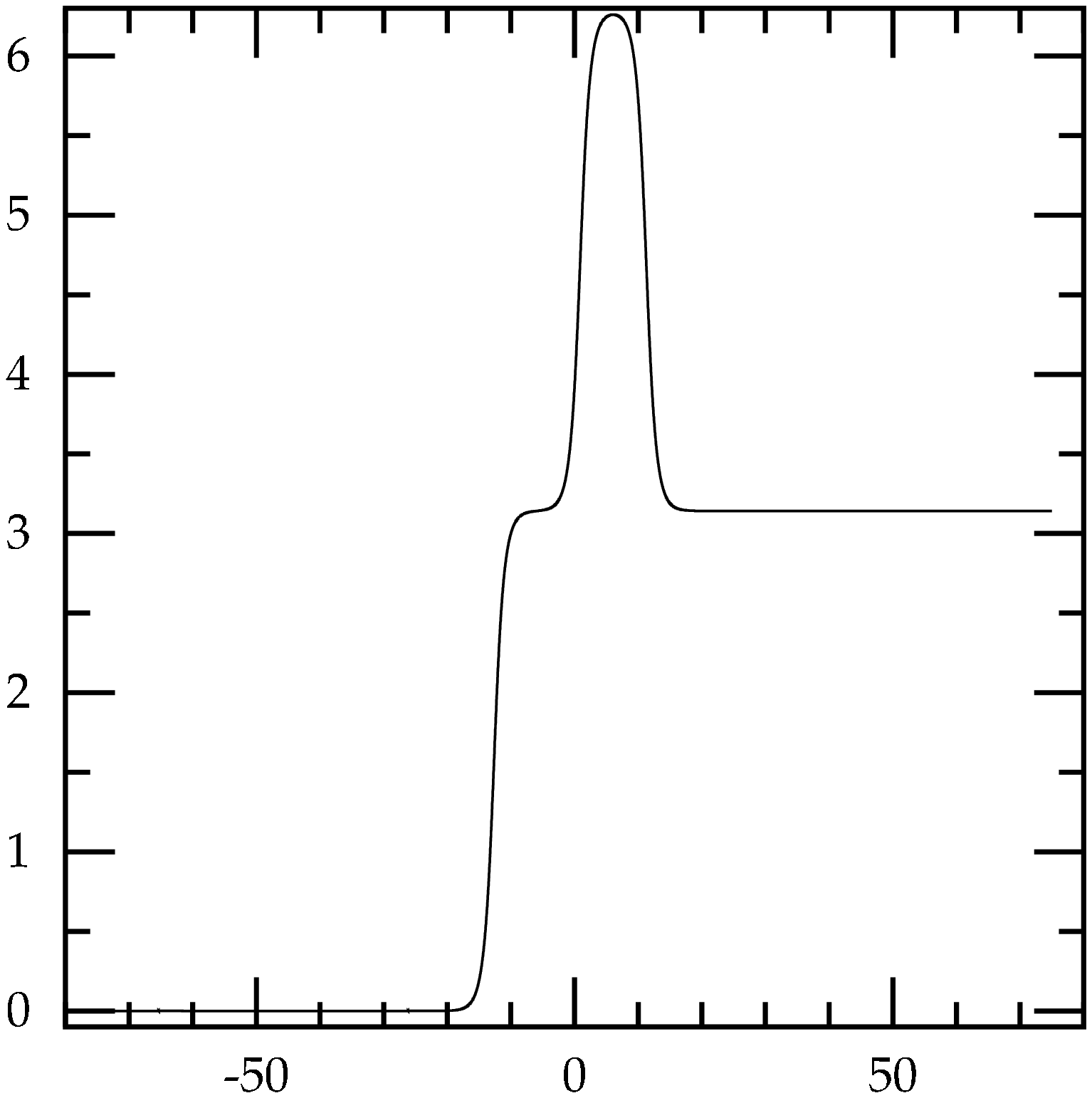}
    \begin{quote}
	\caption[AS]{Fields at 3 values of $t$; a) $t=0$, b) $t=6400$,
          c) $t=12800$}  
	\label{fig:Fig15}
    \end{quote}
\end{figure}

As we have mentioned before, in our studies we have also seen simulations
in which the breather and the kink which form the wobble move relative
to each other. In fig 16. we show the field 
configurations for one of such cases. This case corresponds to $n=1.9$.

\begin{figure}[tbp]
    \centering
	\includegraphics[angle=0,width=4.5cm]{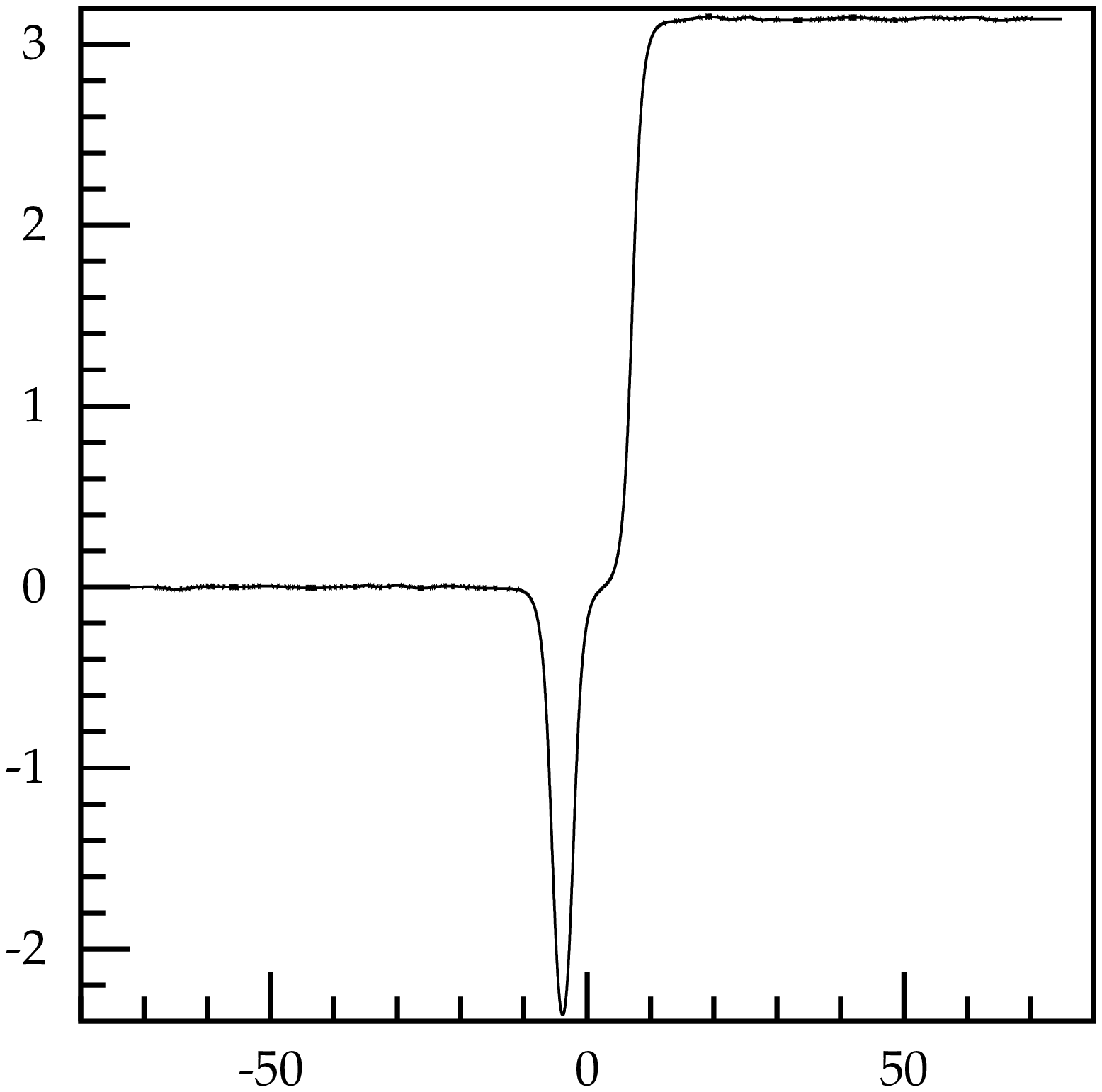}
	 \includegraphics[angle=0,width=4.5cm]{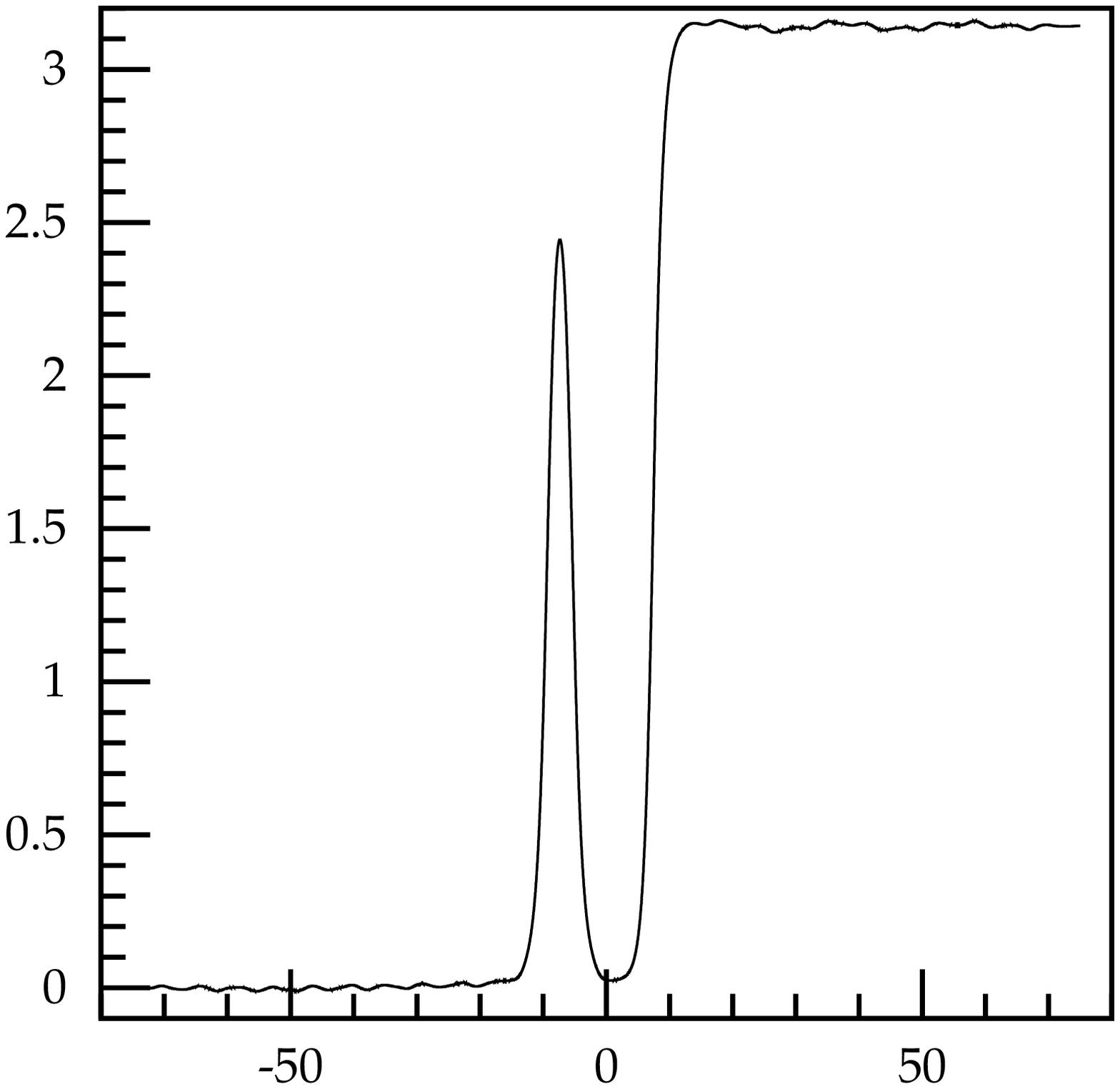}
	 \includegraphics[angle=0,width=4.5cm]{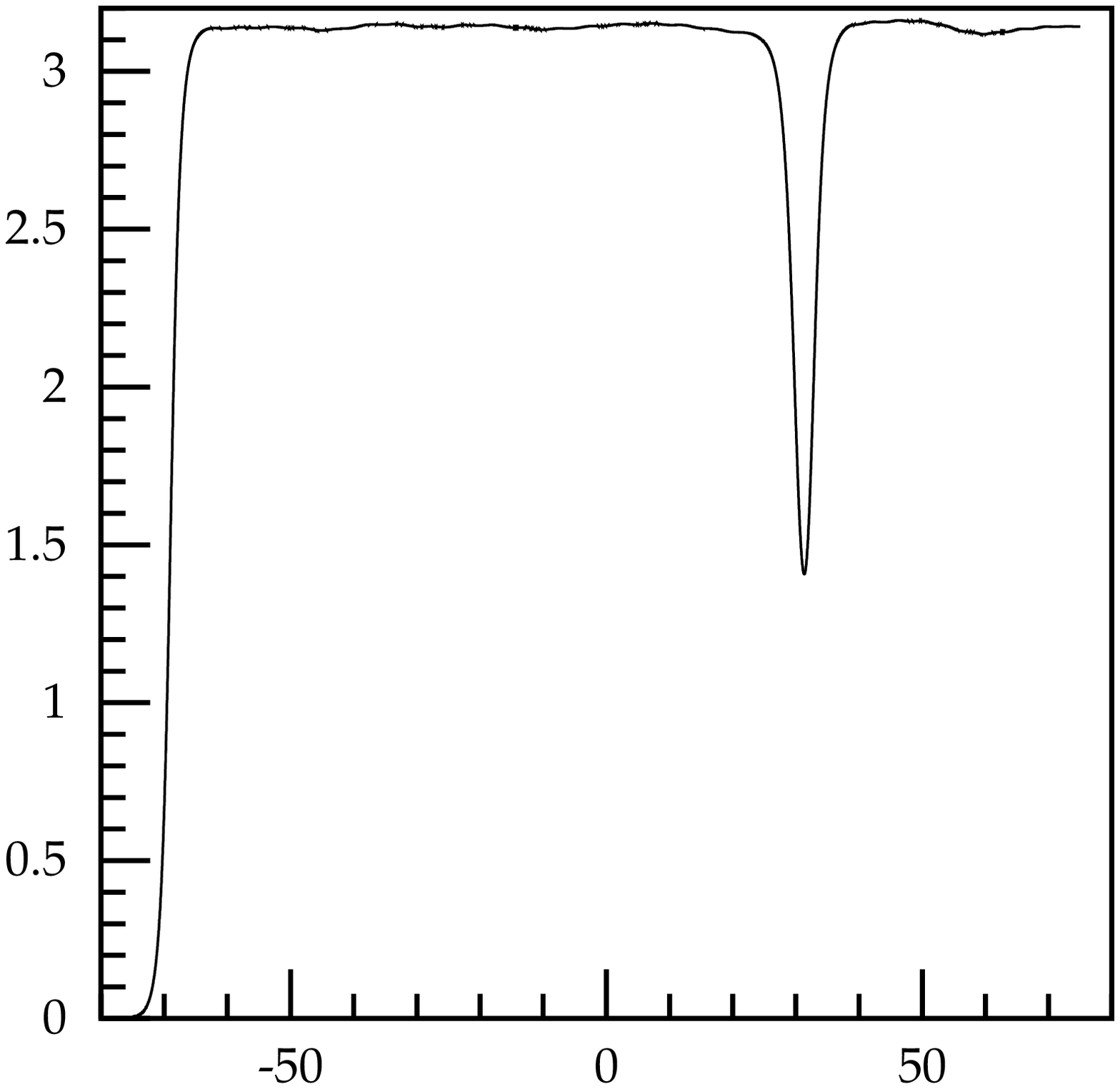}
    \begin{quote}
	\caption[AS]{Fields at 3 values of $t$ (for {$n=1.9$}):   a)
          $t=500$, b) $t=2000$, c) $t=5000$.}  
	\label{fig:Fig16}
    \end{quote}
\end{figure}

 \section{Summary}
\label{sec:summary}
\setcounter{equation}{0}

In this paper we have made the first steps to introduce the concept of
quasi-integrability and discussed it 
on the example of the models of Bazeia et al \cite{Bazeia}. We showed
that when the models are close 
to being intergrable and so can be compared with them then one can
introduce many quantities, 
which in the integrable case are conserved and which, in the
non-integrable case, 
are not conserved. One can then calculate their {\em anomalies} 
that are responsible for this nonconservation, in a power series of the
difference  
of the quasi-integrable models from their integrable neighbours. 
In the models of Bazeia et al \cite{Bazeia} this difference is provided
by $\ve = n-2$. 

We can then calculate these anomalies for various field configurations. 
We have shown that the anomalies are very small  for
many such configurations 
and are only significant when the fields describe strongly
time-dependent tightly bound objects; 
{\it i.e.} the field configurations like those of breathers or wobbles.
And, conveniently, the models of Bazeia et al do possess such configurations.

In fact, the models of Bazeia et al, which depend 
on a parameter $n$ (which when $n=2$ reduce to the integrable
sine-Gordon models) have many very similar
properties and can be used
to discuss the concept of quasi-integrability. All models ({\it i.e.}
for any $n$) have one kink solutions and their scattering properties are very
similar. Moreover, no other analytic solutions of these models 
(when $n\ne2$) are known. 

However, the models can be studied numerically. 
When we studied these models for $n\ne2$ but close to 2 we have found that
the models do, indeed,  possess long-lived breather-like field configurations;
{\it i.e.} when we have constructed breather-like field configurations and
let them evolve they gradually emitted some energy but this process was extremely slow; and so
we can claim that these models (for $n$ close to 2) possess
'very long-lived' breather-like solutions. Their life-time is closely
related to how close $n$ is to 2 and when $n<0.8$ or $n>2.8$ this decay
was relatively  fast so that the `existence' of these states cannot be taken too seriously.

We have also looked at wobble-like states ({\it i.e.} states involving
a breather and a kink) and 
the situation was found to be similar although the range of $n$ for
which such states appeared to be long lived 
was smaller. This is partly related to our construction of such states; 
we generated them all by taking initial configurations consisting 
of kinks and antikinks and then evolving them and absorbing, at the boundaries,
any energy emitted by the configuration. For the wobble-like states, as the
configuration was less symmetric the energy was emitted non-symmetrically 
and this often lead to more perturbation of the resultant (wobble-like) 
field configuration.

Thus, in addition to supporting our studies of quasi-integrability,
 our numerical results demonstrated also the existence of long-lived
breather-like  
and wobble-like states.

\newpage

\noindent {\bf Acknowledgements:} LAF and WJZ thank the Royal Society
for a grant 
which set up their collaboration on the topics of this paper.
 LAF is also partially supported by a CNPq
grant while WJZ acknowledges a FAPESP grant which supported his visit
to IFSC/USP.    

\appendix

\section{The algebra}
\label{sec:appendix-algebra}
\setcounter{equation}{0}

We consider the $sl(2)$ algebra
\be
\sbr{T_3}{T_{\pm}}=\pm\, T_{\pm}, \qquad\qquad \sbr{T_{+}}{T_{-}}=2\, T_3.
\ee
We take the following basis for the corresponding loop algebra
\be
b_{2m+1}=\lambda^{m}\(T_{+}+\lambda\, T_{-}\),\qquad 
F_{2m+1}=\lambda^{m}\(T_{+}-\lambda\, T_{-}\),\qquad
F_{2m}=2\,\lambda^m\, T_3.
\ee
The algebra is
\br
\sbr{b_{2m+1}}{b_{2n+1}}&=&0,\nonumber\\
\sbr{F_{2m+1}}{F_{2n+1}}&=&0,\nonumber\\
\sbr{F_{2m}}{F_{2n}}&=&0,\nonumber\\
\sbr{b_{2m+1}}{F_{2n+1}}&=&-2\, F_{2(m+n+1)},\nonumber\\
\sbr{b_{2m+1}}{F_{2n}}&=&-2\, F_{2(m+n)+1},\nonumber\\
\sbr{F_{2m+1}}{F_{2n}}&=&-2\, b_{2(m+n)+1}.\nonumber
\er

We have a grading operator
\be
d= T_3+ 2\,\lambda \frac{d\;}{d\lambda}
\lab{gradingop}
\ee
such that
\be
\sbr{d}{b_{2m+1}}=\(2m+1\)\,b_{2m+1},\qquad\qquad 
\sbr{d}{F_{m}}=m\,F_{m}.
\ee

\section{The gauge transformation \rf{gaugeminus}}
\label{sec:appendix-gauge}
\setcounter{equation}{0}

The first six parameters $\zeta_n$ of the gauge transformation
\rf{gaugeminus}, determined through
\rf{determinezeta}  are given by 
\begin{eqnarray}
\zeta_1&=&
\frac{1}{2} i \omega \varphi^{(0,1)}, 
\nonumber\\
\zeta_2&=&
\frac{1}{2} i \omega \varphi^{(0,2)}, 
\nonumber\\
\zeta_3&=&
\frac{1}{6} i \left(\omega^3 (\varphi^{(0,1)})^3+3 \omega \varphi^{(0,3)} \right),
\nonumber\\
\zeta_4&=&
\frac{1}{6} i \left(4 (\varphi^{(0,1)})^2 \varphi^{(0,2)}  \omega^3+3
  \varphi^{(0,4)}  
   \omega\right),
\nonumber\\
\zeta_5&=&
\frac{1}{30} i \left(3 \omega^5 (\varphi^{(0,1)})^5+30 \omega^3
   \varphi^{(0,3)}  (\varphi^{(0,1)})^2+40 \omega^3
   (\varphi^{(0,2)})^2 \varphi^{(0,1)} +15 
   \omega \varphi^{(0,5)} \right),
\nonumber\\
\zeta_6&=&
\frac{1}{30} i \left(23 (\varphi^{(0,1)})^4 \varphi^{(0,2)}  \omega^5+
40 (\varphi^{(0,2)})^3 \omega^3+145 \varphi^{(0,1)}  \varphi^{(0,2)}
\varphi^{(0,3)}  \omega^3
\right. \nonumber\\
&+&\left. 35 (\varphi^{(0,1)})^2 \varphi^{(0,4)} 
   \omega^3+15 \varphi^{(0,6)}  \omega\right),
\nonumber
\end{eqnarray}
where $\varphi^{(0,n)}\equiv \partial_{-}^n \varphi$.

The first few components of transformed gauge potentials introduced in
\rf{gaugeminus}  are given by
\br
a_{-}&=& \frac{1}{2} \, b_{-1}
\nonumber\\
&+& b_{1}\, \left[-\frac{1}{4} \omega^2  (\partial_{-}\varphi)^2\right]
\nonumber\\
&+&b_{3} 
\left[-\frac{1}{16} \omega^4 (\partial_{-}\varphi)^4-\frac{1}{4} \omega^2 
\partial_{-}^3\varphi \partial_{-}\varphi\right]\nonumber\\
&+&b_{5} \left[-\frac{1}{32} \omega^6
  (\partial_{-}\varphi)^6-\frac{7}{16} \omega^4  
\partial_{-}^3\varphi 
(\partial_{-}\varphi)^3-\frac{11}{16} \omega^4 
(\partial_{-}^2\varphi)^2 
(\partial_{-}\varphi)^2-\frac{1}{4} \omega^2 
\partial_{-}^5\varphi 
\partial_{-}\varphi\right]\,+\,....
\nonumber
\er
and
\br
a_{+}&=&
b_{1}\, \left[\frac{1}{2} \, \(\omega^2\, V-m\)\right]
\nonumber\\
&+& b_{3} \,
\left[\frac{1}{4}\omega^2\, \partial_{-}^2\varphi \, 
\frac{d\, V}{d\, \varphi}\, 
-\frac{1}{2} i \omega \partial_{-}\varphi X\right]
\nonumber\\
&+& b_{5} \,
\left[-\frac{3}{8} i \omega^3 (\partial_{-}\varphi)^3 
X+\frac{5}{16} \omega^4 \partial_{-}^2\varphi 
(\partial_{-}\varphi)^2 \, \frac{d\, V}{d\, \varphi}\,
-\frac{1}{2} i \omega \partial_{-}\varphi 
\partial_{-}^2 X+\frac{1}{2} i \omega 
\partial_{-}^2\varphi \partial_{-}X
\right. 
\nonumber\\
&-&\left. \frac{1}{2} i \omega 
\partial_{-}^3\varphi X+\frac{1}{4} \omega^2\,
\partial_{-}^4\varphi \frac{d\,V}{d\,\varphi}\right]
\nonumber\\
&+& F_{2} \, X
\nonumber\\
&+& F_{3} \, \partial_{-}X
\nonumber\\
&+&
F_{4} \left[\frac{1}{2} \omega^2 (\partial_{-}\varphi)^2 
X+\partial_{-}^2 X\right]
\nonumber\\
&+& F_{5} 
\left[\omega^2 (\partial_{-}\varphi)^2 
\partial_{-}X+\frac{1}{2} \omega^2 
\partial_{-}\varphi \partial_{-}^2\varphi 
X+\partial_{-}^3 X\right]
\nonumber\\
&+& F_{6} 
\left[\frac{3}{8} \omega^4 (\partial_{-}\varphi)^4 
X+\frac{3}{2} \omega^2 (\partial_{-}\varphi)^2 
\partial_{-}^2 X+\frac{5}{2} \omega^2 
\partial_{-}^2\varphi \partial_{-}\varphi 
\partial_{-}X+\omega^2 \partial_{-}^3\varphi 
\partial_{-}\varphi X+\partial_{-}^4 X\right] \,+\,...
\nonumber
\er

The anomalous terms of the gauge transformation \rf{gaugeminus}, {\it i.e.},
those that do no vanish due to the anomaly $X$ introduced in
\rf{xdef}, are given by
\br
X\, g\, F_1\, g^{-1} &-& \sbr{a_{+}}{a_{-}}=
\nonumber\\
&& b_{3} \,\left[i \omega 
\partial_{-}^2\varphi X\right]
\nonumber\\
&+&
b_{5}\, \left[\frac{3}{2} i \omega^3 (\partial_{-}\varphi)^2 
\partial_{-}^2\varphi X+i \omega 
\partial_{-}^4\varphi X\right]
\nonumber\\
&+& F_{2}\,\left[- \partial_{-}X\right]
\nonumber\\
&+& F_{3}\, \left[- \partial_{-}^2 X\right]
\nonumber\\
&+& F_{4} \,
\left[-\frac{1}{2} \omega^2 (\partial_{-}\varphi)^2 
\partial_{-}X-\omega^2 \partial_{-}\varphi 
\partial_{-}^2\varphi 
X-\partial_{-}^3 X\right]
\nonumber\\
&+& F_{5}\, 
\left[-\frac{5}{2} \omega^2 \partial_{-}\varphi 
\partial_{-}^2\varphi \partial_{-}X-\omega^2 
(\partial_{-}\varphi)^2 \partial_{-}^2 X-\frac{1}{2} 
\omega^2 (\partial_{-}^2\varphi)^2 X-\frac{1}{2} 
\omega^2 \partial_{-}\varphi \partial_{-}^3\varphi 
X-\partial_{-}^4 X\right]\nonumber\\
&+&\, ... \nonumber
\nonumber
\er

\section{The gauge transformation \rf{gaugeplus}}
\label{sec:appendix-gauge2}
\setcounter{equation}{0}

The first six parameters $\zeta_{-n}$ of the gauge transformation
\rf{gaugeplus} and \rf{gaugeplus2}  are given by
\begin{eqnarray}
\zeta_{-1}&=&
\frac{1}{2} i \omega  \varphi^{(1,0)}, 
\nonumber\\
\zeta_{-2}&=&
\frac{1}{2} i \omega  \varphi^{(2,0)}, 
\nonumber\\
\zeta_{-3}&=&
\frac{1}{6} i \left(\omega^3 (\varphi^{(1,0)})^3+3
   \omega  \varphi^{(3,0)} \right),
\nonumber\\
\zeta_{-4}&=&
\frac{1}{6} i \left(4 (\varphi^{(1,0)})^2 \varphi^{(2,0)}  \omega ^3+
3 \varphi^{(4,0)}  \omega \right),
\nonumber\\
\zeta_{-5}&=&
\frac{1}{30} i \left(3 \omega ^5 (\varphi^{(1,0)})^5+30
   \omega ^3 \varphi^{(3,0)}  (\varphi^{(1,0)})^2+
40 \omega ^3 (\varphi^{(2,0)})^2 \varphi^{(1,0)} +15
   \omega  \varphi^{(5,0)} \right),
\nonumber\\
\zeta_{-6}&=&
\frac{1}{30} i \left(23 (\varphi^{(1,0)})^4 \varphi^{(2,0)}  \omega ^5+
40 (\varphi^{(2,0)})^3 \omega ^3+145 \varphi^{(1,0)}  \varphi^{(2,0)}  
\varphi^{(3,0)}  \omega ^3
\right. \nonumber\\
&+&\left. 35 (\varphi^{(1,0)})^2 \varphi^{(4,0)} 
   \omega ^3+15 \varphi^{(6,0)}  \omega \right),
\nonumber
\end{eqnarray}
where $\varphi^{(n,0)} \equiv \partial_{+}^n\varphi$.

The first few components of transformed gauge potentials introduced in
\rf{gaugeplus} are given by
\br
{\tilde a}_{+}&=& \frac{1}{2} b_{1}
\nonumber\\
&+& b_{-1}\left[-\frac{1}{4} \omega ^2  (\partial_{+}\varphi)^2\right]
\nonumber\\
&+& b_{-3} \left[-\frac{1}{16} \omega ^4 (\partial_{+}\varphi)^4-
\frac{1}{4} \omega ^2 \partial_{+}^3\varphi 
\partial_{+}\varphi\right]
\nonumber\\
&+& b_{-5} \left[-\frac{1}{32} \omega ^6
    (\partial_{+}\varphi)^6-
\frac{7}{16} \omega ^4 \partial_{+}^3\varphi 
(\partial_{+}\varphi)^3-\frac{11}{16} \omega ^4 
(\partial_{+}^2\varphi)^2 
(\partial_{+}\varphi)^2-\frac{1}{4} \omega ^2 
\partial_{+}^5\varphi 
\partial_{+}\varphi\right]\,+\,...
\nonumber
\er
and 
\br
{\tilde a}_{-}&=& 
b_{-1}\left[\frac{1}{2}  \(\omega^2\, V-m\)\right]
\nonumber\\
&+&b_{-3} \left[\frac{1}{4} \, \omega^2\,
\partial_{+}^2\varphi \frac{d\, V}{d\,\varphi}+\frac{1}{2} i
    \omega  \partial_{+}\varphi
    {\tilde X}\right]
\nonumber\\
&+&
b_{-5} \left[\frac{5}{16} \omega^4 \partial_{+}^2\varphi
    (\partial_{+}\varphi)^2 
\frac{d\, V}{d\,\varphi}+\frac{1}{4}\,\omega^2\, \partial_{+}^4\varphi
    \frac{d\, V}{d\,\varphi}+\frac{3}{8} i \omega ^3 
(\partial_{+}\varphi)^3 {\tilde X}+\frac{1}{2} i
    \omega  \partial_{+}\varphi
    \partial_{+}^2{\tilde X}
\right. \nonumber\\
&-& \left. \frac{1}{2} i \omega  
\partial_{+}^2\varphi \partial_{+}{\tilde X}+\frac{1}{2} i
    \omega  \partial_{+}^3\varphi
    {\tilde X}\right]
\nonumber\\
&+& F_{-2}\left[- {\tilde X}\right]
\nonumber\\
&+& F_{-3}\left[ - \partial_{+}{\tilde X}\right]
\nonumber\\
&+& F_{-4} \left[-\frac{1}{2} \omega ^2 
(\partial_{+}\varphi)^2
    {\tilde X}-\partial_{+}^2{\tilde X}\right]
\nonumber\\
&+& F_{-5}
    \left[\omega ^2 \left(-(\partial_{+}\varphi)^2\right)
    \partial_{+}{\tilde X}-\frac{1}{2} \omega ^2 
\partial_{+}\varphi \partial_{+}^2\varphi
    {\tilde X}-\partial_{+}^3 {\tilde X}\right]
\nonumber\\
&+& F_{-6}
    \left[-\frac{3}{8} \omega ^4 (\partial_{+}\varphi)^4
    {\tilde X}-\frac{3}{2} \omega ^2 
(\partial_{+}\varphi)^2 \partial_{+}^2{\tilde X}-\frac{5}{2}
    \omega ^2 \partial_{+}^2\varphi 
\partial_{+}\varphi \partial_{+}{\tilde X}-\omega ^2
    \partial_{+}^3\varphi \partial_{+}\varphi {\tilde X}
-\partial_{+}^4 {\tilde X}\right]\nonumber\\
&+&\, ... \nonumber
\nonumber
\er

The anomalous terms of the gauge transformation \rf{gaugeplus}, {\it i.e.},
those that do no vanish due to the anomaly ${\tilde X}$ introduced in
\rf{ydef}, are given by
\br
&&{\tilde X}\, {\tilde g}\, F_{-1}\, {\tilde g}^{-1} 
- \sbr{{\tilde a}_{+}}{{\tilde a}_{-}}=
\nonumber\\
&&b_{-3}\left[i \omega \partial_{+}^2\varphi {\tilde X}\right]
\nonumber\\
&+& b_{-5} \left[\frac{3}{2} i \omega ^3 (\partial_{+}\varphi)^2
    \partial_{+}^2\varphi {\tilde X}+i \omega  
\partial_{+}^4\varphi {\tilde X}\right]
\nonumber\\
&+& F_{-2}\left[ - \partial_{+}{\tilde X} \right]
\nonumber\\
&+& F_{-3}\left[ - \partial_{+}^2{\tilde X}\right]
\nonumber\\
&+& F_{-4} \left[-\frac{1}{2} \omega ^2 (\partial_{+}\varphi)^2
    \partial_{+}{\tilde X}-\omega ^2 
\partial_{+}\varphi \partial_{+}^2\varphi
    {\tilde X}-\partial_{+}^3 {\tilde X}\right]
\nonumber\\
&+& F_{-5} \left[-\frac{5}{2} \omega ^2 \partial_{+}\varphi 
\partial_{+}^2\varphi \partial_{+}{\tilde X}-\omega ^2
    (\partial_{+}\varphi)^2
    \partial_{+}^2{\tilde X}-\frac{1}{2} \omega ^2 
(\partial_{+}^2\varphi)^2 {\tilde X}-\frac{1}{2} \omega
    ^2 \partial_{+}\varphi \partial_{+}^3\varphi
    {\tilde X}-\partial_{+}^4 {\tilde X}\right]\nonumber\\
&+&\, ... \nonumber
\nonumber
\er

\section{The $\ve$-expansion}
\label{sec:appendix-expansion}
\setcounter{equation}{0}

\br
V &=& V\mid_{\ve=0} + \frac{d\,V}{d\,\ve}\mid_{\ve=0}\, \ve +\ldots
\nonumber\\
&=&
  V\mid_{\ve=0} + \left[\frac{\partial V}{\partial \ve}+
\frac{\partial V}{\partial \vp}\,\frac{\partial \vp}{\partial
  \ve}\right]_{\ve=0} \, \ve \nonumber\\
&+& \left[\frac{\partial^2 V}{\partial \ve^2}+
2\,\frac{\partial^2 V}{\partial \ve \partial \vp}\,\frac{\partial \vp}{\partial
  \ve}+\frac{\partial V}{\partial \vp}\,\frac{\partial^2 \vp}{\partial
  \ve^2}+\frac{\partial^2 V}{\partial \vp^2}\,\(\frac{\partial \vp}{\partial
  \ve}\)^2\right]_{\ve=0}\, \ve^2
+\ldots
\er
Analogously, we have
\br
\frac{\partial\,V}{\partial\,\vp} &=& 
\frac{\partial\,V}{\partial\,\vp} \mid_{\ve=0} +
\left[\frac{d\,}{d\,\ve}\(\frac{\partial\,V}{\partial\,\vp}
  \)\right]_{\ve=0}\, \ve +\ldots\nonumber\\
&=& 
\frac{\partial\,V}{\partial\,\vp} \mid_{\ve=0} +
 \left[\frac{\partial^2 V}{\partial \ve\partial\vp}+
\frac{\partial^2 V}{\partial \vp^2}\,\frac{\partial \vp}{\partial
  \ve}\right]_{\ve=0} \, \ve \nonumber\\
&+&
 \left[\frac{\partial^3 V}{\partial \ve^2\partial\vp}
+
2\,\frac{\partial^3 V}{\partial\ve\partial \vp^2}\,\frac{\partial \vp}{\partial
  \ve}
+
\frac{\partial^2 V}{\partial \vp^2}\,\frac{\partial^2 \vp}{\partial
  \ve^2}
+
\frac{\partial^3 V}{\partial \vp^3}\,\(\frac{\partial \vp}{\partial
  \ve}\)^2\right]_{\ve=0} \, \ve^2 
+\ldots
\er
and
\br
\frac{\partial^2\,V}{\partial\,\vp^2} &=& 
\frac{\partial^2\,V}{\partial\,\vp^2} \mid_{\ve=0} +
\left[\frac{d\,}{d\,\ve}\(\frac{\partial^2\,V}{\partial\,\vp^2}
  \)\right]_{\ve=0}\, \ve +\ldots\nonumber\\
&=& 
\frac{\partial^2\,V}{\partial\,\vp^2} \mid_{\ve=0} +
 \left[\frac{\partial^3 V}{\partial \ve\partial\vp^2}+
\frac{\partial^3 V}{\partial \vp^3}\,\frac{\partial \vp}{\partial
  \ve}\right]_{\ve=0} \, \ve 
\nonumber\\
&+&
 \left[\frac{\partial^4 V}{\partial \ve^2\partial\vp^2}
+
2\,\frac{\partial^4 V}{\partial\ve\partial \vp^3}\,\frac{\partial \vp}{\partial
  \ve}
+\frac{\partial^3 V}{\partial \vp^3}\,\frac{\partial^2 \vp}{\partial
  \ve^2}
+\frac{\partial^4 V}{\partial \vp^4}\,\(\frac{\partial \vp}{\partial
  \ve}\)^2\right]_{\ve=0} \, \ve^2 
+\ldots
\er
Calculating we have
\br
V\mid_{\ve=0}&=& \frac{1}{8}\,\sin^2 \(2\,\vp_0\)=
\frac{1}{16}\left[1-\cos\(4\,\vp_0\)\right]\nonumber\\
\frac{\partial\,V}{\partial\,\vp} \mid_{\ve=0}&=&
\frac{1}{4}\,\sin\(4\,\vp_0\)\nonumber\\
\frac{\partial^2\,V}{\partial\,\vp^2} \mid_{\ve=0}&=&
\cos\(4\,\vp_0\)\nonumber\\
\frac{\partial^3 V}{\partial \vp^3}\mid_{\ve=0}&=&-4\,\sin\(4\,\vp_0\)
\nonumber\\
\frac{\partial^4 V}{\partial \vp^4}\mid_{\ve=0}&=&-16\,\cos\(4\,\vp_0\)
\er
and
\br
\frac{\partial\,V}{\partial\ve} \mid_{\ve=0}&=&
-\frac{1}{2} \sin ^2(\vp_0) \left[2\,\sin ^2(\vp_0) \, \log \mid\sin
   (\vp_0)\mid+\cos^2\(\vp_0\)\right]\\
\frac{\partial^2\,V}{\partial\,\vp\partial\ve} \mid_{\ve=0}&=&
-\frac{1}{4} \sin (2 \vp_0) 
\left[8\,\sin^2\vp_0\, \log \mid\sin (\vp_0)\mid
+\cos (2 \vp_0) +1\right]\nonumber\\
\frac{\partial^3\,V}{\partial\,\vp^2\partial\ve} \mid_{\ve=0}&=&
-\frac{1}{2} \left[
4 \(\cos (2 \vp_0)-\cos (4 \vp_0)\)\log \mid\sin (\vp_0)\mid
+\cos (2 \vp_0) +1\right]
\nonumber\\
\frac{\partial^4\,V}{\partial\,\vp^3\partial\ve} \mid_{\ve=0}&=&
\sin (2 \vp_0) \left[-(-4 \log \mid\sin (\vp_0)\mid+4 \cos (2 \vp_0) (4 \log
   \mid\sin (\vp_0)\mid+1)+1)\right]
\nonumber
\er
and
\br
\frac{\partial^2\,V}{\partial\ve^2} \mid_{\ve=0}&=&
\frac{1}{4} \tan ^2(\vp_0) \left[\sin ^4(\vp_0) 
\left(8 \log ^2\mid\sin(\vp_0)\mid
-8 \log \mid\sin (\vp_0)\mid+3\right)
\right.\nonumber\\
&+&\left.\sin ^2(\vp_0) \left(-4
   \log ^2\mid\sin (\vp_0)\mid+8 \log \mid\sin
   \vp_0\mid-6\right)+3\right]
\nonumber\\
\frac{\partial^3\,V}{\partial \vp\partial\ve^2} \mid_{\ve=0}&=&
\frac{1}{32} \tan (\vp_0) \sec ^2(\vp_0) \left[24 \log ^2\mid\sin
   (\vp_0)\mid+16 \log \mid\sin (\vp_0)\mid
\right. \nonumber\\
&+&\left.8 \cos (6 \vp_0) \log
   ^2\mid\sin (\vp_0)\mid
\right. \\
&+&\left.\cos (2 \vp_0) \left(-40 \log
   ^2\mid\sin (\vp_0)\mid+4 \log \mid\sin (\vp_0)\mid+23\right)
\right. \nonumber\\
&+&\left. 8
   \cos (4 \vp_0) (\log \mid\sin (\vp_0)\mid-1)^2-4 \cos (6 \vp_0) \log
   \mid\sin (\vp_0)\mid+\cos (6 \vp_0)+16\right]
\nonumber\\
\frac{\partial^4\,V}{\partial \vp^2\partial\ve^2} \mid_{\ve=0}&=&
\frac{1}{16} \sec ^4(\vp_0) \left[44 \log ^2\mid\sin (\vp_0)\mid+16 \log
   \mid\sin (\vp_0)\mid
\right. \nonumber\\
&+&\left. 10 \cos (6 \vp_0) \log ^2\mid\sin
   (\vp_0)\mid+4 \cos (8 \vp_0) \log ^2\mid\sin (\vp_0)\mid
\right. \nonumber\\
&+&\left. 2 \cos (2 \vp_0)
   \left(-29 \log ^2\mid\sin (\vp_0)\mid+\log \mid\sin
   (\vp_0)\mid+6\right)
\right. \nonumber\\
&-&\left. 2 \cos (6 \vp_0) \log \mid\sin (\vp_0)\mid+\cos (4 \vp_0)
   (3-16 \log \mid\sin (\vp_0)\mid)+9\right]
\nonumber
\er

\end{document}